\RequirePackage{ifpdf}
\documentclass[12pt,letterpaper]{article}
\pdfoutput=1
\usepackage{jheppub}
\usepackage{epsfig}
\usepackage{enumitem}
\usepackage{bbm,amsfonts}
\usepackage{rotating,graphicx}
\usepackage{amssymb,amsmath,amsfonts,mathtools}
\usepackage{fancybox,diagbox}
\usepackage{enumerate}
\usepackage{dsfont}
\usepackage{verbatim}
\usepackage{wrapfig}
\usepackage{slashed}
\usepackage{braket}
\usepackage{pdflscape}
\usepackage{accents}
\usepackage{afterpage}
\usepackage{breqn}

\author[a]{Marco S. Bianchi}
 
\affiliation[a]{Facultad de Ingeniería, Universidad San Sebastián, Santiago, Chile}

\emailAdd{marco.bianchi@uss.cl}  
  
\abstract{The perturbative expansion of two-point functions of lowest dimension supersymmetric operators in $\mathcal{N}=4$ SYM and ABJM theory exhibits uniform transcendental weight. Inspired by this, we construct an explicit basis of uniformly transcendental master integrals for these correlators, through four loops in four and three loops in three dimensions.
In terms of these bases, the two-point functions simplify to rational linear combinations. Conversely, such explicit bases of uniformly transcendental integrals can be useful for other applications.}

\title{Uniformly transcendental bases for protected two-point functions}

\keywords{$\mathcal{N}=4$ SYM, ABJM, transcendentality, perturbation theory}

\numberwithin{equation}{section}

\begin{document}

\maketitle
\allowdisplaybreaks

\section{Introduction}

Transcendentality plays an important role in the analytic structure of perturbative Quantum Field Theory \cite{Henn:2014qga}.
It governs the organization of multi-loop Feynman integrals in terms of transcendental weight, helping to streamline their expansion in dimensional regularization \cite{Henn:2013pwa}.

In four-dimensional $\mathcal{N}=4$ super-Yang-Mills (SYM) theory scattering amplitudes \cite{Bern:1994zx,Bern:1997nh,Anastasiou:2003kj,Bern:2004cz,Bern:2005iz,Bern:2008ap,Goncharov:2010jf}, Wilson loops \cite{Drummond:2007bm,Anastasiou:2009kna,DelDuca:2009au,DelDuca:2010zg}, and form factors \cite{vanNeerven:1985ja,Bork:2010wf,Gehrmann:2011xn,Brandhuber:2012vm,Brandhuber:2014ica,Banerjee:2016kri,Boels:2017ftb,Brandhuber:2018xzk,Ahmed:2019yjt,Huber:2019fxe,Lin:2020dyj,Agarwal:2021zft,Lee:2021lkc} often exhibit uniform transcendentality, a feature conjecturally tied to the underlying integrable structure of the theory \cite{Beisert:2010jr,Arkani-Hamed:2012zlh}.
In this context, transcendentality has provided insights into hidden symmetries and structures \cite{Bern:2005iz,Drummond:2006rz,Drummond:2007aua,Drummond:2008vq,Brandhuber:2008pf,Drummond:2009fd,Arkani-Hamed:2012zlh}, dualities \cite{Drummond:2006rz,Drummond:2007cf,Drummond:2007au,Brandhuber:2007yx}, and integrability \cite{Kotikov:2002ab,Kotikov:2003fb,Kotikov:2004er,Eden:2006rx,Beisert:2006ez,Marboe:2014sya,Marboe:2016igj,Kniehl:2021ysp}.   
Furthermore, uniform transcendentality can be leveraged to calculate scattering amplitudes \cite{Dixon:2011pw,Dixon:2011nj,Dixon:2013eka,Drummond:2014ffa,Caron-Huot:2016owq,Dixon:2016nkn,Drummond:2018caf,Caron-Huot:2019vjl,Dixon:2020cnr} and form factors \cite{Brandhuber:2012vm,Dixon:2020bbt,Dixon:2022rse,Dixon:2022xqh} via bootstrap strategies.

Similarly, the three-dimensional ABJM model \cite{Aharony:2008ug,Aharony:2008gk} shares many of these hidden properties \cite{Klose:2010ki}. In parallel, uniform transcendentality has been detected in its scattering amplitudes \cite{Chen:2011vv,Bianchi:2011dg,Bianchi:2011aa,Caron-Huot:2012sos,Bianchi:2014iia}, Wilson loops \cite{Henn:2010ps,Bianchi:2013pva} and form factors \cite{Brandhuber:2013gda,Young:2013hda,Bianchi:2013iha,Bianchi:2013pfa} as well. 

In previous work \cite{Bianchi:2023llc,Bianchi:2024nah}, we pointed out that two-point functions of the lowest-dimension supersymmetric operators in both theories exhibit uniform transcendentality when expanded in dimensional regularization in a scheme that preserves supersymmetry.
This was established from explicit perturbative evaluations up to three and two loops, respectively, for $\mathcal{N}=4$ SYM \cite{Bianchi:2023llc} and ABJM \cite{Bianchi:2024nah}, and conjectured to hold for the whole perturbative series. Based on this assumption we also proposed a conjecture for re-summing the perturbative expansion of the first non-trivial order in the dimensional regulator expansion in the case of $\mathcal{N}=4$ SYM \cite{Bianchi:2023llc}.

This work extends these ideas. 
Previously, uniform transcendentality was verified only a posteriori, i.e.~after expanding and combining all required momentum master integrals. This finding suggests that there should exist a basis of uniformly transcendental master integrals in terms of which the correlator would display uniform transcendentality manifestly.
In this work we identify such uniformly transcendental bases. 
In four dimensions this is a straightforward task for two-and three-loop integrals. At four loops, most master integrals can be expanded in a uniformly transcendental way via multiplication by a suitable rational function of the dimensional regulator $\epsilon$ \cite{Lee:2011jt}. 
We extend these results to all master integrals, and we identify the relevant rational prefactors as arising from modified denominator or numerator powers generated by the integration-by-parts (IBP) reduction \cite{Tkachov:1981wb,Chetyrkin:1981qh}. 
We checked that our basis is uniformly transcendental up to weight twelve, by using the explicit expansions in dimensional regularization of \cite{Baikov:2010hf,Lee:2011jt}. 
Past that order uniform transcendentality is conjectured.
We remark that, at least in principle, the present results could be derived from earlier uniformly transcendental bases constructed for form-factor integrals \cite{Henn:2016men,Boels:2017ftb,Lee:2021lkc,Lee:2023dtc} by means of IBP reduction. Nevertheless, our approach offers a direct and minimal realization, albeit more heuristic, adapted specifically to the correlators considered here.

In three dimensions fewer results are known analytically \cite{Lee:2015eva}. We were able to define uniformly transcendental bases for up to three-loop integrals, which is the relevant order for the two-point function calculations in \cite{Bianchi:2024nah}. Again, uniform transcendentality can only be checked up to some order in the $\epsilon$ regulator expansion, corresponding to up to transcendental weight ten in this case, and conjectured to extend to the whole series.

Using these bases the correlators can be expressed in compact analytic form up to three loops in $\mathcal{N}=4$ SYM and two loops in ABJM, where only simple rational coefficients appear, in contrast to other generic bases and theories, where such coefficients would be rational functions of the regulator $\epsilon$. This is the hallmark of the hidden simplicity of $\mathcal{N}=4$ SYM and ABJM, applied to their protected two-point functions.

Beyond exposing this hidden simplicity, uniformly transcendental bases are useful more generally.
The propagator integrals considered here are ubiquitous in perturbative Quantum Field Theory, for instance when analyzing renormalization group properties, so their range of applicability extends well beyond superconformal theories.
Uniformly transcendental integrals are simpler to expand in $\epsilon$, since their reduced functional basis allows to reconstruct the coefficients from numerical data much more straightforwardly. Moreover, they naturally serve as input for differential equations in canonical form \cite{Henn:2013pwa}. 


\section{Two-point functions of short protected operators}

We consider gauge-invariant local operators consisting of two scalars of the form
\begin{equation}
O(x) = \mathrm{Tr}\!\left(\Phi \Phi\right)(x)
\end{equation}
where $\Phi$ denotes a complex scalar field of the corresponding $\mathcal{N}=4$ SYM or ABJM theory. Their flavor indices are chosen appropriately, such that the operators belong to protected BPS multiplets. 
We compute their momentum-space two-point functions
\begin{equation}
\langle O(p)\, \bar O(-p) \rangle 
= \sum_{L=0}^{\infty} \lambda^{L}\, a^{(L)}(p^2)
\end{equation}
where $a^{(L)}(p^2)$ denotes the $L$-loop contribution.
For $\mathcal{N}=4$, the coupling constant $\lambda$ coincides with the 't Hooft coupling $\lambda\equiv \frac{g^2N}{16\pi^2}$.
We do not assume the planar limit in this work; however, no non-planar corrections arise through the perturbative order we consider, namely three loops, although they are expected to appear at higher orders.

In three dimensions we work in the ABJM model with different gauge group ranks $N_1$ and $N_2$ \cite{Aharony:2008gk} and the coupling constant is generically $\lambda = \frac{N_i}{k}$, where $k$ is the Chern-Simons coupling, which is assumed to be a large integer. At each loop order different color structures may contribute; we review these below.

We work in dimensional regularization, $d=4-2\epsilon$ for $\mathcal{N}=4$ SYM and $d=3-2\epsilon$ for ABJM theory. 
Other choices of regulators, while theoretically conceivable, would not allow for the high-order calculations required here. Moreover, transcendentality properties become more transparent within dimensional regularization.
We further adopt the dimensional-reduction (DRED) scheme \cite{Siegel:1979wq}, ensuring vanishing perturbative $\beta$ functions up to the required order \cite{Velizhanin:2008rw}. 
The uniform transcendentality properties rely crucially on this choice, which is natural since DRED preserves supersymmetry.

A deep conceptual connection between supersymmetry and uniform transcendentality is not yet fully understood, but suggestive evidence has existed since the seminal observation of maximal transcendentality in the comparison between QCD and $\mathcal{N}=4$ SYM anomalous dimensions \cite{Kotikov:2002ab,Kotikov:2003fb}. 
While supersymmetry may not be strictly necessary for uniform transcendental behavior in general, it appears to play a crucial organizing role in suppressing lower-weight contributions. 
In this sense, it is natural that a supersymmetry-preserving scheme such as DRED helps maintain the uniform transcendental structure of the two-point functions considered here, by avoiding regulator-induced artifacts that would otherwise spoil it. 
Conversely, our findings further support the intuition that uniform transcendentality is closely linked to the supersymmetric organization of perturbation theory.

In addition, relaxing dimensional reduction would generate spurious divergences to the two-point functions considered here, which we want to avoid in order to keep the two-point functions perturbatively protected.

In dimensional regularization, each of the coefficients $a^{(l)}$ is itself an $\epsilon$-dependent function. While individual higher-order terms in the $\epsilon$ expansion are scheme dependent and therefore not directly physical, they carry non-trivial structural information. 
In particular, such terms would be required for consistent higher-loop renormalization and for insertions into more complicated observables. 
Moreover, uniform transcendentality is a property of the full $\epsilon$ expansion, making control over higher orders is technically important, even if individual terms are not directly physical observables.

At each loop order $L$, $a^{(L)}(p^2)$ are expressed in terms of $(L+1)$-loop momentum integrals. This slight mismatch in loop counting is a standard feature and we clarify it here.
These integrals can be decomposed onto a finite basis of master integrals $I_i^{(L+1)}$:
\begin{equation}
a^{(L)}(p^2) = \sum_i b_i^{(L)}(\epsilon)\, I_i^{(L+1)}(p^2)
\end{equation}
with rational-function coefficients $b_i^{(L)}(\epsilon)$ determined via IBP reductions.

For these reductions we employed \textsc{Forcer} \cite{Ruijl:2017cxj} scripts executed in \textsc{Form} \cite{Vermaseren:2000nd,Tentyukov:2007mu,Ruijl:2017dtg}, which are tailored to this class of propagator-type integrals.


\subsection{Results}
\paragraph{$\mathcal{N}=4$ SYM.}
In $\mathcal{N}=4$ SYM, we computed the two-point function up to three loops in \cite{Bianchi:2023llc}. Related perturbative multi-loop computations of correlation functions and closely connected observables, including higher-loop anomalous dimensions of composite operators \cite{Velizhanin:2008jd,Fiamberti:2008sh,Eden:2012fe}, multi-point correlation functions of protected operators \cite{Eden:1998hh,Eden:1999kh,Eden:2000mv,Dolan:2004mu,Alday:2010zy,Eden:2011we,Eden:2012tu,Bourjaily:2016evz,Chicherin:2015edu,Chicherin:2018avq,Coronado:2018cxj,Bourjaily:2025iad,Bargheer:2022sfd,Bercini:2024pya,Bargheer:2025uai} and form factors \cite{Bork:2010wf,Gehrmann:2011xn,Brandhuber:2012vm,Boels:2012ew,Yang:2016ear,Boels:2017ftb,Boels:2017skl,Huber:2019fxe,Lin:2021qol,Lee:2021lkc,Henn:2024pki}, have been carried out extensively in $\mathcal{N}=4$ SYM. 
These works provide important benchmarks for the multi-loop technology employed here.

The perturbative expansion of the two-point function can be written as
\begin{equation}\label{eq:perturbative}
\left\langle O(p) \bar O(-p) \right\rangle 
= \frac{2\,(N^2-1)}{(4\pi)^{2-\epsilon}}\,
\left( n^{(0)} + n^{(1)}\lambda + n^{(2)}\lambda^2 + n^{(3)}\lambda^3 + O(\lambda^4) \right)
\end{equation}
The 't Hooft coupling is defined as $\lambda \equiv \frac{g^2 N}{16\pi^2}$ and suffices up to three loops, since no non-planar corrections contribute through this order.
The tree-level result reads
\begin{equation}
n^{(0)} = \vcenter{\hbox{\includegraphics[scale=0.13]{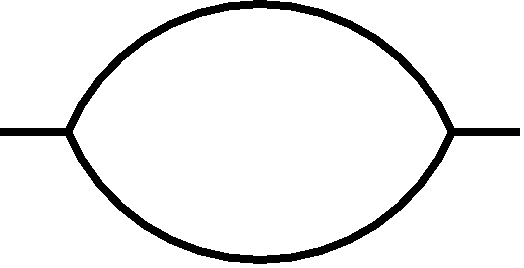}}}
\end{equation}
where the one-loop momentum integral is defined as
\begin{equation}\label{eq:1loop4d}
\vcenter{\hbox{\includegraphics[scale=0.1]{figures/M1L1.png}}}\,
\equiv \int \frac{d^{4-2\epsilon}k}{\pi^{2-\epsilon}}\frac{1}{k^2(k-p)^2}
= \frac{\Gamma (1-\epsilon )^2 \Gamma (\epsilon )}{\Gamma (2-2 \epsilon )}
\end{equation}
with this normalization chosen to eliminate $(4\pi)$-factors from the integrals expansions.
The coefficients $n^{(L)}$ encode the loop corrections.
To streamline the expressions, we set $p^2 = 1$; the full momentum dependence can be restored by dimensional analysis.
We further absorb residual factors of $e^{\gamma_E \epsilon}$ and $(4\pi)^\epsilon$ into the definition of $\lambda$.
With these conventions, the one-loop coefficient, normalized by the tree-level contribution, expands in dimensional regularization as \cite{Bianchi:2023llc}:
\begin{align}\label{eq:1L}
&\frac{n^{(1)}}{n^{(0)}} = -12 \zeta _3 \epsilon-18 \zeta _4 \epsilon ^2+\left(6 \zeta _2 \zeta _3-84 \zeta _5\right) \epsilon ^3+\left(64 \zeta _3^2-\tfrac{657 }{4}\zeta _6\right) \epsilon ^4+\left(\tfrac{741}{4} \zeta _3 \zeta _4+42 \zeta _2 \zeta _5
\right.\nonumber\\&\left.
-588 \zeta _7\right) \epsilon ^5
+\left(-32 \zeta _2 \zeta _3^2+\tfrac{3872}{5} \zeta _5 \zeta _3-\tfrac{18285}{16} \zeta _8\right) \epsilon ^6
+ O(\epsilon^7)
\end{align}
at two loops
\begin{align}\label{eq:2L}
&\frac{n^{(2)}}{n^{(0)}} =100 \zeta _5 \epsilon+\left(244 \zeta _3^2+250 \zeta _6\right) \epsilon ^2+\left(732 \zeta _3 \zeta _4-100 \zeta _2 \zeta _5+1718 \zeta _7\right) \epsilon ^3+
    \left(\tfrac{7288}{3} \zeta _5 \zeta _3
\right.\nonumber\\&\left.    
    -\tfrac{1296}{5} \zeta_{5,3}-244 \zeta _2 \zeta _3^2+\tfrac{179647}{30} \zeta _8\right)\epsilon ^4 
    + O(\epsilon^5)
\end{align}
and at three loops
\begin{align}\label{eq:3L}
&\frac{n^{(3)}}{n^{(0)}}=-980 \zeta _7 \epsilon+\left(-5560 \zeta _3 \zeta _5-3430 \zeta _8\right) \epsilon ^2    
+\left(-\tfrac{17560}{3} \zeta _3^3-13900 \zeta _6 \zeta _3-8340 \zeta _4 \zeta _5
\right.\nonumber\\&\left.    
+1470 \zeta _2 \zeta _7-\tfrac{292220}{9} \zeta _9\right) \epsilon ^3
+ O(\epsilon^4)
\end{align}
where $\zeta$ denotes the Riemann zeta function.
For compactness we display only the first few transcendental orders. 
The complete results up to transcendental weight twelve can be found in \cite{Bianchi:2023llc}. 
The key point is that these coefficients exhibit uniform transcendentality: this is verified explicitly up to three loops in perturbation theory and weight twelve, and conjectured to persist to all orders in the 't~Hooft coupling $\lambda$ (including, potentially, its non-planar corrections) and in the dimensional regulator $\epsilon$.
In momentum space, it is crucial to normalize quantum corrections by the tree-level result, since the integral \eqref{eq:1loop4d} that produces the tree-level contribution is not uniformly transcendental. Only after this normalization does the structure of uniform transcendentality in the loop corrections become manifest.
Alternatively, upon Fourier transforming to position space, which is the more natural setting for two-point functions, the correlator displays uniform transcendentality without the need to factor out the tree-level contribution. 
However, since our goal is to construct a uniformly transcendental basis of momentum-space integrals, we will work directly in momentum space and therefore adhere to \eqref{eq:perturbative}.

In \cite{Bianchi:2023llc}, explicit expressions were given in terms of a basis of master integrals, namely the one used by \textsc{Forcer}. 
In that representation, each coefficient appears as a linear combination of master integrals with rational functions of the regulator $\epsilon$.
In the following sections we show that these expansions simplify dramatically once a basis of integrals with uniform transcendental weight is adopted.

\paragraph{ABJM.}

In \cite{Bianchi:2024nah} we carried out the analogous analysis for two-point functions in ABJM theory in three dimensions, again finding clear evidence for uniform transcendentality (this was already detected at finite order in $\epsilon$ in \cite{Minahan:2009wg,Young:2014lka,Bianchi:2020cfn} at two loops and to all orders in a specific color limit \cite{Bianchi:2016rub}). Up to two-loop order
\begin{equation}\label{eq:norm3d}
    \langle O(p) \bar O(-p)\rangle = \frac{2 N_1N_2}{(4\pi)^{3/2-\epsilon}} \left(c_0 +\frac{1}{k^2}\left( \left(N_1^2+N_2^2\right) c_{N_1^2} + N_1N_2 c_{N_1N_2} + c_1 \right) \right) + O\left(k^{-4}\right)
\end{equation}
The tree-level result reads
\begin{equation}
c_0 = \vcenter{\hbox{\includegraphics[scale=0.13]{figures/M1L1.png}}}
\end{equation}
where the one-loop momentum integral is defined as
\begin{align}\label{eq:1loop3d}
\vcenter{\hbox{\includegraphics[scale=0.1]{figures/M1L1.png}}}\,\, 
&\equiv \int \frac{d^{3-2\epsilon}k}{\pi^{3/2-\,\epsilon}}\frac{1}{k^2(k-p)^2}
= \frac{\Gamma \!\left(\frac12-\epsilon \right)^2 \Gamma \!\left(\frac12+\epsilon \right)}{\Gamma (1-2\epsilon )} \\[2mm]
&= \pi^{3/2}\,4^{\epsilon}\left(1+\frac{5 \zeta _2 \epsilon ^2}{2}-\frac{\zeta _3 \epsilon ^3}{3}
+\frac{241 \zeta _4 \epsilon^4}{16}
+\left(-\frac{5}{6}\zeta _2 \zeta _3-\frac{\zeta _5}{5}\right)\epsilon^5
+O(\epsilon^6)\right)
\nonumber
\end{align}
with a normalization chosen to match \eqref{eq:1loop4d}. We recall that we suppress $\gamma_E$ factors everywhere.
This convention introduces half-integer powers of $\pi$ in intermediate expressions; we factor out such overall factors whenever convenient.

The one-loop correction vanishes identically due to the antisymmetry properties of the Levi-Civita tensors ubiquitous in Chern-Simons perturbation theory. 
The same mechanism eliminates all odd-loop contributions to these two-point functions.

At two loops, the correction to the two-point function again displays uniform transcendentality for each of its three color structures, namely $(N_1^2 + N_2^2)$, $N_1 N_2$, and $1$
\begin{dmath}
  \frac{c_{N_1^2}}{c_0}=-\zeta _2+\epsilon  \left(19 \zeta _3-24 \zeta _2 L_1\right)+\epsilon ^2 \left(-30 \zeta _4-48 \zeta _2 L_1^2+4 L_1^4+96 L_4\right)+\epsilon ^3 \left(\frac{41 \zeta _2 \zeta
   _3}{3}+\frac{975 \zeta _5}{2}-32 \zeta _2 L_1^3-1002 \zeta _4 L_1-\frac{16 L_1^5}{5}+384 L_5\right)+O\left(\epsilon ^4\right)
\end{dmath}
\begin{dmath}
\frac{c_{N_1N_2}}{c_0} =\epsilon  \left(36 \zeta _2 L_1-55 \zeta _3\right)+\epsilon ^2 \left(72 \zeta _2 L_1^2-\frac{463 \zeta _4}{2}\right)+\epsilon ^3 \left(139 \zeta _2 \zeta _3-2612 \zeta _5+96 \zeta _2
   L_1^3+564 \zeta _2^2 L_1\right)+O\left(\epsilon ^4\right)
\end{dmath}
\begin{dmath}
\frac{c_{1}}{c_0} = 2 \zeta _2+\epsilon  \left(17 \zeta _3+12 \zeta _2 L_1\right)+\epsilon ^2 \left(\frac{583 \zeta _4}{2}+24 \zeta _2 L_1^2-8 L_1^4-192 L_4\right)+\epsilon ^3 \left(-\frac{499 \zeta _2
   \zeta _3}{3}+1637 \zeta _5-32 \zeta _2 L_1^3+594 \zeta _4 L_1+\frac{32 L_1^5}{5}-768 L_5\right)+O\left(\epsilon ^4\right)
\end{dmath}
where $L_n\equiv \text{Li}_n\left(\frac{1}{2}\right)$, so for instance $L_1 = \log 2$.
Again, additional orders in the $\epsilon$ expansion can be found in \cite{Bianchi:2024nah}, 
but the main point is the empirical observation that these corrections exhibit uniform transcendentality.
Motivated by this result, in the following section we address the problem of defining a basis of uniformly transcendental integrals for these objects.


\section{Uniformly transcendental basis for two-point functions}\label{sec:bases}

We focus on two-point functions of operators built from a pair of scalar fields, both in four- and three-dimensional theories. 
Their perturbative expansion involves Feynman diagrams structurally identical to propagator corrections, which are ubiquitous in the renormalization of Quantum Field Theory.
We construct bases of integrals with manifestly uniform transcendental weight up to four loops in momentum space in four dimensions, extending the results of \cite{Lee:2011jt}. 
We perform an analogous analysis in three dimensions, building on our previous work \cite{Bianchi:2024nah}, and obtain several novel results.

\subsection{Four dimensions}

We begin with the four-dimensional case. 
We work up to transcendental weight twelve, for which explicit $\epsilon$-expansions are available from \cite{Lee:2011jt} through four loops. 
These expansions contain only rational combinations of multiple zeta values (MZVs). 

In the pictorial notation used below, solid dots on propagators denote squared propagators,
\[
\raisebox{-0.6mm}{\includegraphics[scale=0.15]{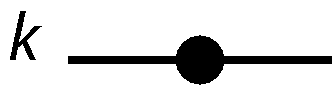}}=\frac{1}{k^2},
\]
while empty dots indicate inverse powers of propagators,
\[
\includegraphics[scale=0.15]{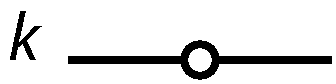}=k^2.
\]
Appropriate powers of the external momentum $p^2$ must be included so that all integrals exhibit the same overall scaling, but for convenience we set $p^2 = 1$ throughout.

\paragraph{One loop.}
At one loop only the bubble integral contributes. 
It becomes uniformly transcendental upon multiplication by the simple factor $(1 - 2\epsilon)$, or equivalently by squaring one of its propagators.
We normalize the integral as:
\begin{equation}\label{eq:basis1L}
    \begin{array}{l}
    M^{(1)}_{1} \equiv (1-2\epsilon)\, \epsilon\,\,\vcenter{\hbox{\includegraphics[scale=0.13]{figures/M1L1.png}}} \,\, =  -\epsilon\,\, \vcenter{\hbox{\includegraphics[scale=0.13]{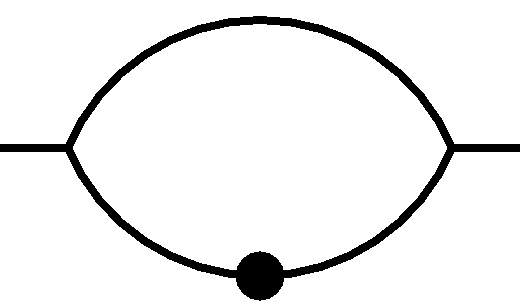}}}
    \end{array}
\end{equation}
This way the integral evaluates to the following function
\begin{equation}
    M^{(1)}_{1} = \frac{\Gamma (1-\epsilon )^2 \Gamma (1+\epsilon )}{\Gamma (1-2 \epsilon )} = 1-\frac{\zeta_2 \epsilon ^2}{2}-\frac{7 \zeta_3 \epsilon ^3}{3}-\frac{47 \zeta_4 \epsilon
   ^4}{16}+\epsilon ^5 \left(\frac{7 \zeta_2 \zeta_3}{6}-\frac{31
   \zeta_5}{5}\right)+O\left(\epsilon ^6\right)
\end{equation}
which will be needed later for reconstructing the expansions \eqref{eq:1L}--\eqref{eq:3L}.

\paragraph{Two loops.} At two loops the two master integrals can be made trivially uniformly transcendental. The specific choice we make is
\begin{equation}\label{eq:basis2L}
    \begin{array}{l}
    M^{(2)}_{1} \equiv  (1-2\epsilon)\,\,\vcenter{\hbox{\includegraphics[scale=0.13]{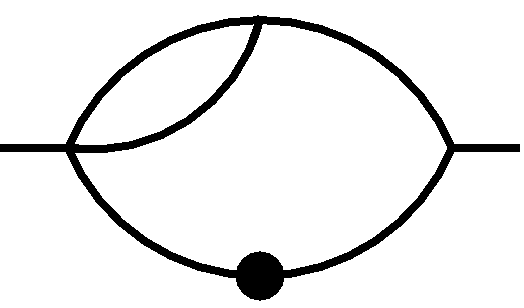}}} \,\, =  -\,\, \vcenter{\hbox{\includegraphics[scale=0.13]{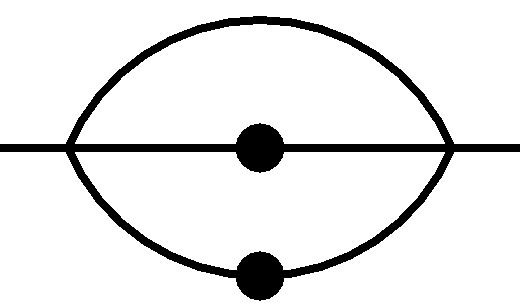}}} \\[1cm]
    M^{(2)}_{2} \equiv (1-2\epsilon)\,\,\vcenter{\hbox{\includegraphics[scale=0.13]{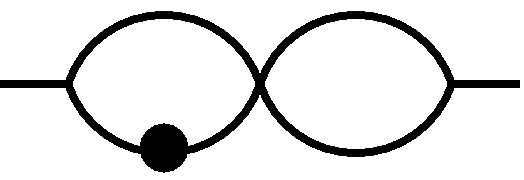}}} \,\, = - \,\,\vcenter{\hbox{\includegraphics[scale=0.13]{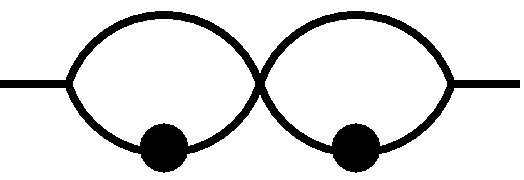}}}\\[0.5cm]
    \end{array}
\end{equation}
where we have again provided two alternative realizations, one with an explicit $(1-2\epsilon)$ factor which will appear ubiquitously across three and four loops integrals.

\paragraph{Three loops.} At three loops six master integrals are present, whose evaluation has been studied extensively \cite{Chetyrkin:1980pr,Tkachov:1981wb,Chetyrkin:1981qh,Kotikov:1995cw}. The first four can be expressed in terms of bubble integrals
\begin{equation*}
    \begin{array}{l}
    M^{(3)}_{1} \equiv  \dfrac{1-2\epsilon}{\epsilon}\,\,\vcenter{\hbox{\includegraphics[scale=0.13]{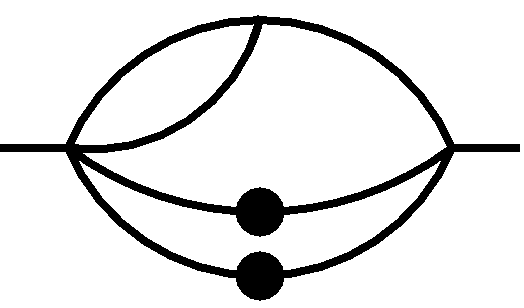}}} \,\, =  -\dfrac{1}{\epsilon}\,\, \vcenter{\hbox{\includegraphics[scale=0.13]{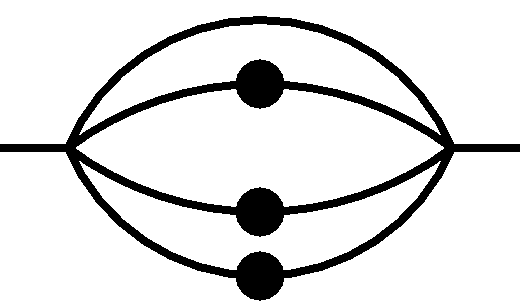}}}
    \end{array}
\end{equation*}
\begin{equation}\label{eq:basis3L1}
    \begin{array}{l}
    M^{(3)}_{2} \equiv  \dfrac{1-2\epsilon}{\epsilon}\,\,\vcenter{\hbox{\includegraphics[scale=0.13]{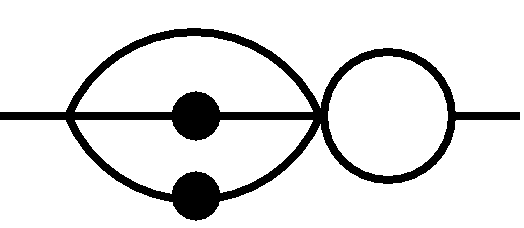}}} \,\, = -\dfrac{1}{\epsilon} \,\,\vcenter{\hbox{\includegraphics[scale=0.13]{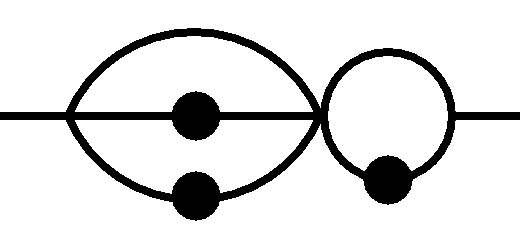}}} \\[1cm]
    M^{(3)}_{3} \equiv  \dfrac{1-2\epsilon}{\epsilon}\,\,\vcenter{\hbox{\includegraphics[scale=0.13]{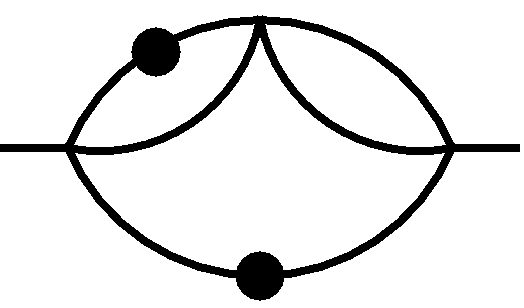}}}\,\, = -\dfrac{1}{\epsilon} \,\,
    \vcenter{\hbox{\includegraphics[scale=0.13]{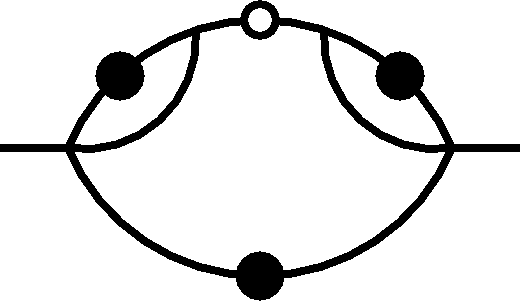}}} \\[1cm]
    M^{(3)}_{4} \equiv  \dfrac{1-2\epsilon}{\epsilon}\,\,\vcenter{\hbox{\includegraphics[scale=0.13]{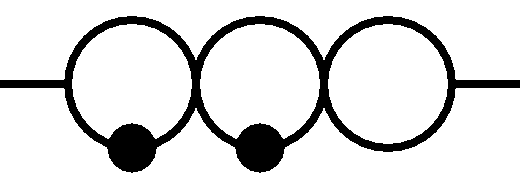}}} \,\, = -\dfrac{1}{\epsilon} \,\,
    \vcenter{\hbox{\includegraphics[scale=0.13]{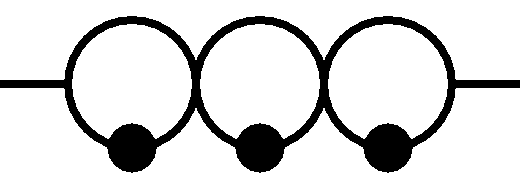}}} \\[0.5cm]
    \end{array}
\end{equation}
The rest of the basis is defined as:
\begin{equation}\label{eq:basis3L2}
    \begin{array}{ll}
    M^{(3)}_{5} \equiv (1-2\epsilon) \,\,\vcenter{\hbox{\includegraphics[scale=0.13]{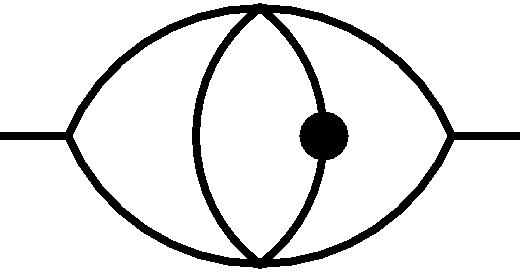}}} \quad &
    M^{(3)}_{6} \equiv (1-2\epsilon)\, \epsilon \,\,\vcenter{\hbox{\includegraphics[scale=0.13]{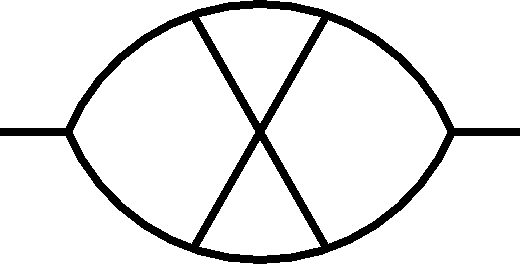}}} 
    \end{array}
\end{equation}
where the $\epsilon$ factor in the last definition is inserted to keep the same transcendental weight across all integrals.

\paragraph{Four loops.} In the four-dimensional case, a basis of master integrals relevant to this problem was expanded up to transcendental weight twelve in \cite{Lee:2011jt}. 
All of these integrals, except for three, were already presented in a form that makes uniform transcendentality manifest after factoring out simple rational combinations of the dimensional-regularization parameter $\epsilon$.

We build on these results and extend them by showing that the required rational prefactors naturally emerge from the IBP reductions of integrals with the same topology but with additional propagator powers or suitable numerator insertions, in the spirit of \cite{Henn:2013pwa}. 
By searching for such alternative representatives, we also construct uniformly transcendental versions of the three most complicated topologies of \cite{Lee:2011jt}.

We define the four-loop basis pictorially below. 
The first few integrals reduce entirely to products of bubble integrals whose powers can be chosen so that their expansions contain only Gamma functions with uniformly transcendental series. 
For convenience, we provide two equivalent versions of the basis: one in which the universal factor $(1-2\epsilon)$ appears explicitly, and another in which this factor is absorbed into squared or inverse propagators. The integral labeling is inherited from \cite{Lee:2011jt}.
\begin{equation*}
    \begin{array}{l}
    M^{(4)}_{01} \equiv  \dfrac{1-2\epsilon}{\epsilon^{2}}\,\,\vcenter{\hbox{\includegraphics[scale=0.13]{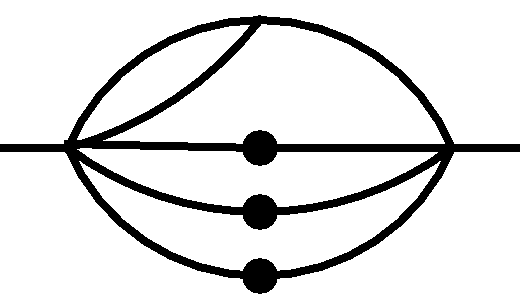}}} \,\, =  -\dfrac{1}{\epsilon^{2}}\,\, \vcenter{\hbox{\includegraphics[scale=0.13]{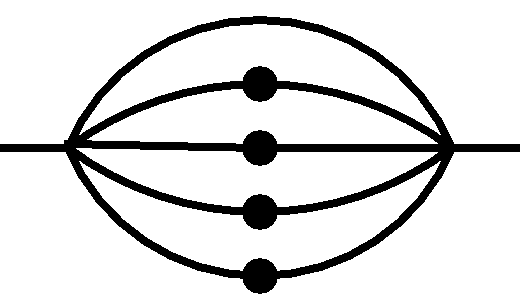}}} \\[1cm]
    M^{(4)}_{11} \equiv  \dfrac{1-2\epsilon}{\epsilon^{2}}\,\,\vcenter{\hbox{\includegraphics[scale=0.13]{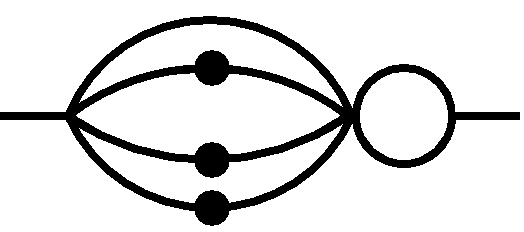}}} \,\, = -\dfrac{1}{\epsilon^{2}} \,\,\vcenter{\hbox{\includegraphics[scale=0.13]{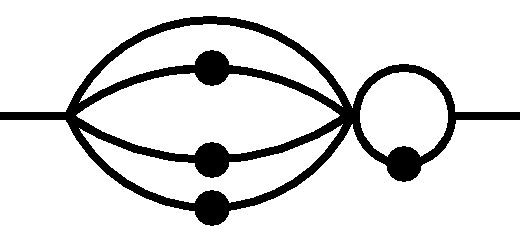}}}
    \end{array}
\end{equation*}
\begin{equation}\label{eq:basis4L1}
    \begin{array}{l}
    M^{(4)}_{12} \equiv  \dfrac{1-2\epsilon}{\epsilon^{2}}\,\,\vcenter{\hbox{\includegraphics[scale=0.13]{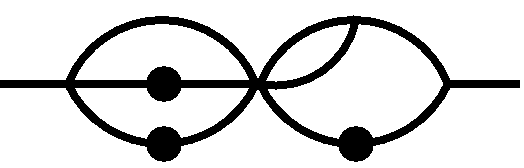}}}\,\, = -\dfrac{1}{\epsilon^{2}} \,\,
    \vcenter{\hbox{\includegraphics[scale=0.13]{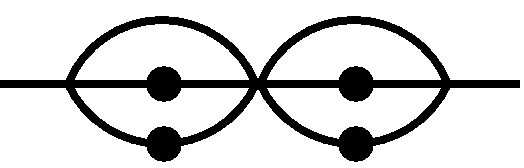}}} \\[1cm]
    M^{(4)}_{13} \equiv  \dfrac{1-2\epsilon}{\epsilon^{2}}\,\,\vcenter{\hbox{\includegraphics[scale=0.13]{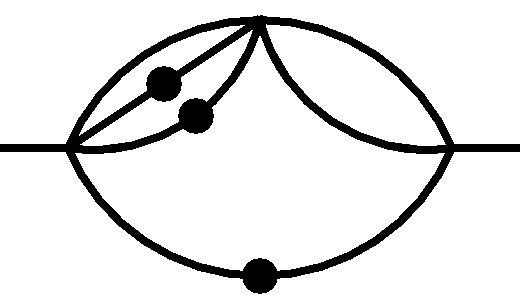}}} \,\, = -\dfrac{1}{\epsilon^{2}} \,\,
    \vcenter{\hbox{\includegraphics[scale=0.13]{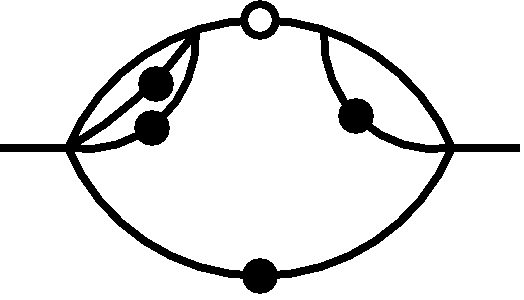}}}\\[1cm]   
    M^{(4)}_{14} \equiv \dfrac{1-2\epsilon}{\epsilon^{2}}\,\,\vcenter{\hbox{\includegraphics[scale=0.13]{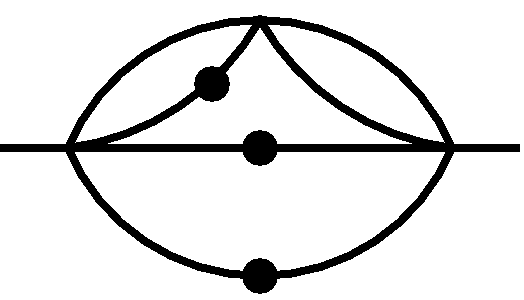}}} \,\, = -\dfrac{1}{\epsilon^{2}} \,\,
    \vcenter{\hbox{\includegraphics[scale=0.13]{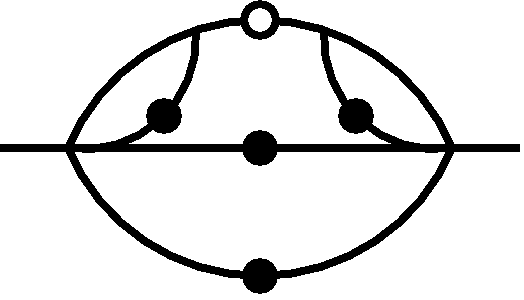}}}\\[1cm]   
    M^{(4)}_{23} \equiv \dfrac{1-2\epsilon}{\epsilon^{2}} \,\,\vcenter{\hbox{\includegraphics[scale=0.13]{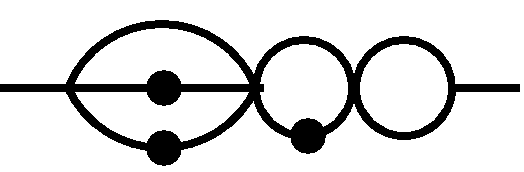}}}  \,\, =  -\dfrac{1}{\epsilon^{2}} \,\,\vcenter{\hbox{\includegraphics[scale=0.13]{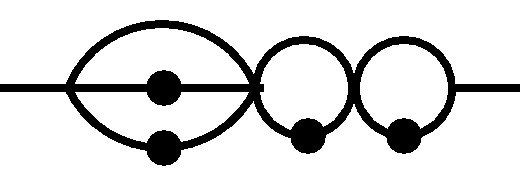}}}\\[1cm]
    M^{(4)}_{24} \equiv \dfrac{1-2\epsilon}{\epsilon^{2}} \,\,\vcenter{\hbox{\includegraphics[scale=0.13]{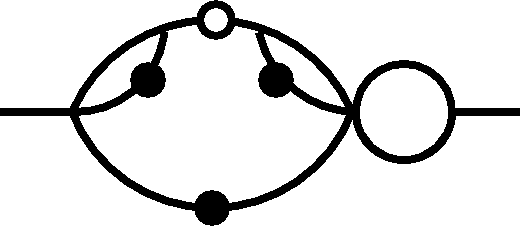}}}  \,\, =  -\dfrac{1}{\epsilon^{2}} \,\,
    \vcenter{\hbox{\includegraphics[scale=0.13]{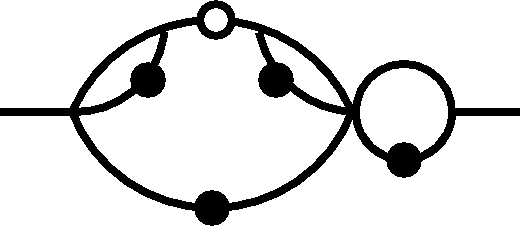}}}\\[1cm]
    M^{(4)}_{25} \equiv \dfrac{1-2\epsilon}{\epsilon^{2}} \,\,\vcenter{\hbox{\includegraphics[scale=0.13]{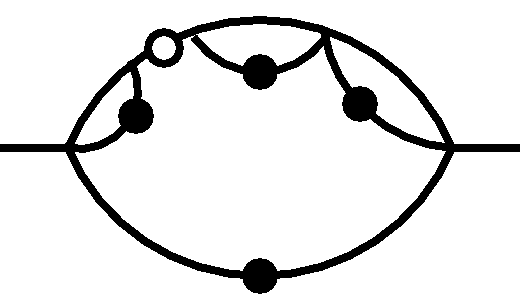}}} \,\, =  -\dfrac{1}{\epsilon^{2}} \,\,
    \vcenter{\hbox{\includegraphics[scale=0.13]{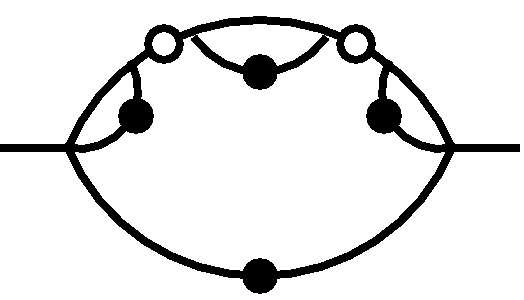}}}\\[1cm]
    M^{(4)}_{31} \equiv \dfrac{1-2\epsilon}{\epsilon^{2}} \,\,\vcenter{\hbox{\includegraphics[scale=0.13]{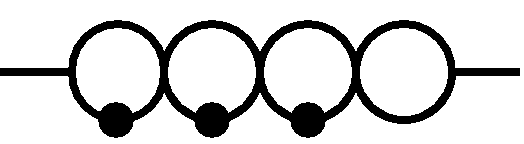}}} \,\, =  -\dfrac{1}{\epsilon^{2}} \,\,
    \vcenter{\hbox{\includegraphics[scale=0.13]{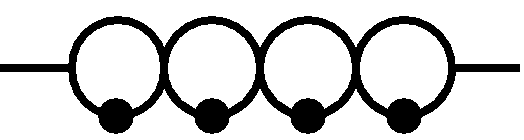}}}\\[0.5cm]
    \end{array}
\end{equation}
Again, the solid dot denotes a squared propagator, whereas the empty dot denotes an inverse power.
The rest of the basis is defined as:
\begin{equation*}
    \begin{array}{ll}
    M^{(4)}_{21} \equiv \dfrac{1-2\epsilon}{\epsilon} \,\,\vcenter{\hbox{\includegraphics[scale=0.13]{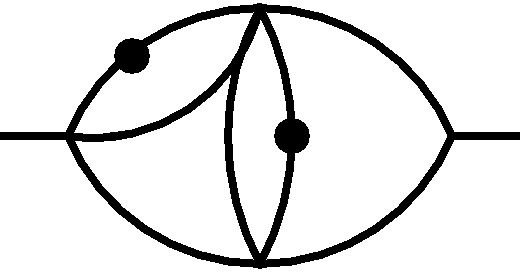}}} \quad
    & M^{(4)}_{22} \equiv \dfrac{1-2\epsilon}{\epsilon} \,\,\vcenter{\hbox{\includegraphics[scale=0.13]{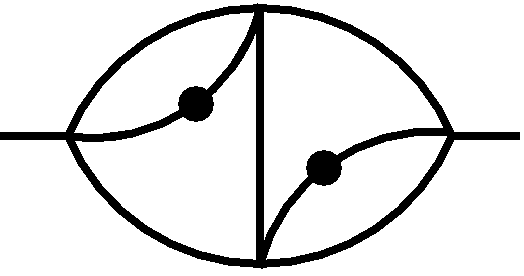}}} \\[1cm]
    M^{(4)}_{26} \equiv \dfrac{1-2\epsilon}{\epsilon} \,\,\vcenter{\hbox{\includegraphics[scale=0.13]{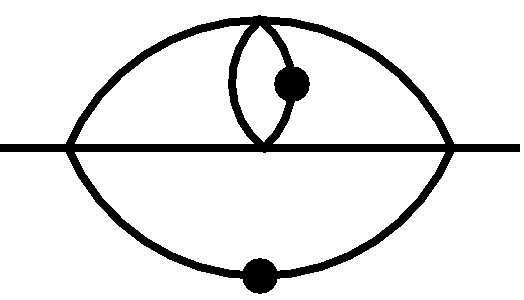}}}\quad &
    M^{(4)}_{27} \equiv \dfrac{1-2\epsilon}{\epsilon} \,\,\vcenter{\hbox{\includegraphics[scale=0.13]{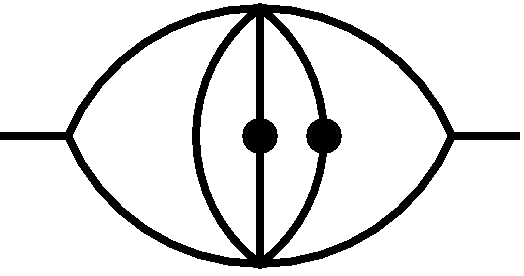}}} \\[1cm]
    M^{(4)}_{32} \equiv \dfrac{1-2\epsilon}{\epsilon} \,\,\vcenter{\hbox{\includegraphics[scale=0.13]{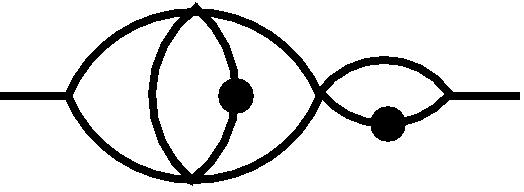}}} \quad
    & M^{(4)}_{33} \equiv \dfrac{1-2\epsilon}{\epsilon} \,\,\vcenter{\hbox{\includegraphics[scale=0.13]{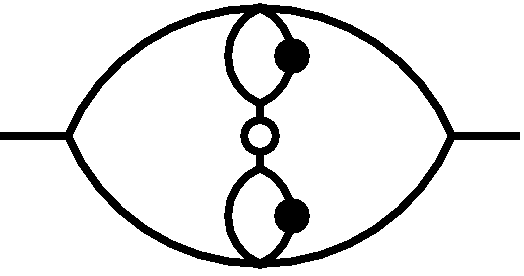}}} \\[1cm]
    M^{(4)}_{34} \equiv (1-2\epsilon) \,\,\vcenter{\hbox{\includegraphics[scale=0.13]{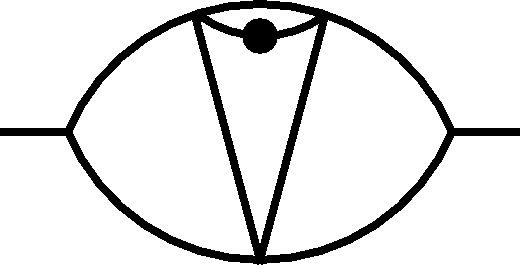}}} \quad
    & M^{(4)}_{36} \equiv (1-2\epsilon) \,\,\vcenter{\hbox{\includegraphics[scale=0.13]{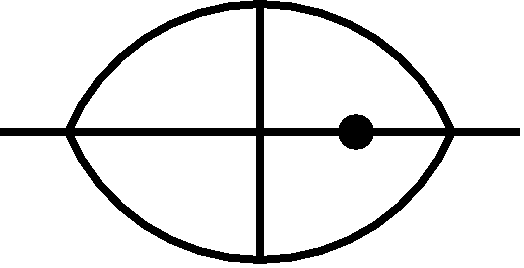}}} 
    \end{array}
\end{equation*}
\begin{equation*}
    \begin{array}{ll}
    M^{(4)}_{35} \equiv (1-2\epsilon) \,\,\vcenter{\hbox{\includegraphics[scale=0.13]{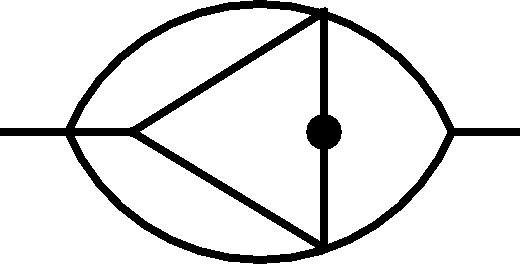}}} \quad &
    M^{(4)}_{41} \equiv (1-2\epsilon) \,\,\vcenter{\hbox{\includegraphics[scale=0.13]{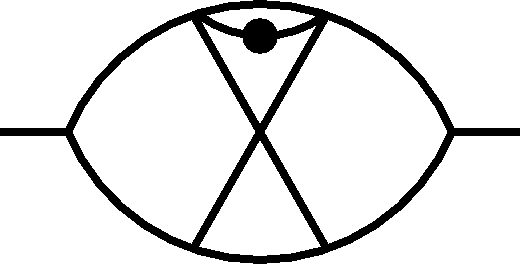}}} \\[1cm]
    M^{(4)}_{42} \equiv (1-2\epsilon) \,\,\vcenter{\hbox{\includegraphics[scale=0.13]{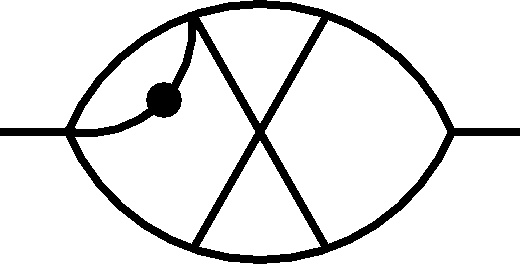}}} \quad
    & M^{(4)}_{43} \equiv (1-2\epsilon) \,\,\vcenter{\hbox{\includegraphics[scale=0.13]{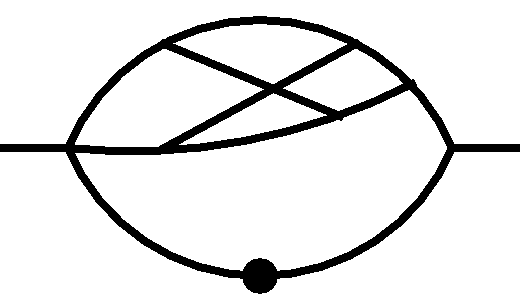}}} \\[1cm]
    M^{(4)}_{44} \equiv (1-2\epsilon)\, \epsilon \,\,\vcenter{\hbox{\includegraphics[scale=0.13]{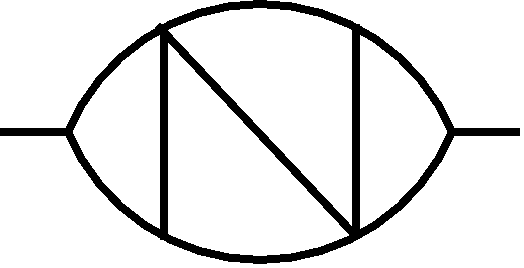}}} \quad
    & M^{(4)}_{52} \equiv (1-2\epsilon) \,\,\vcenter{\hbox{\includegraphics[scale=0.13]{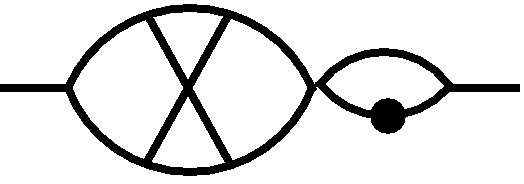}}} \\[1cm]
    M^{(4)}_{45} \equiv (1-2\epsilon)\, \epsilon \,\,\vcenter{\hbox{\includegraphics[scale=0.13]{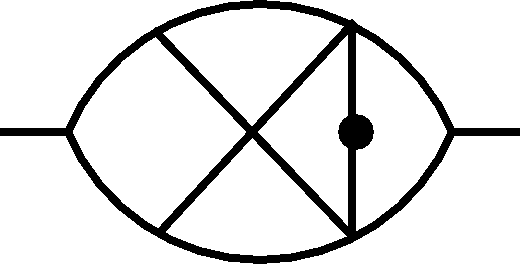}}} \quad &
    M^{(4)}_{51} \equiv (1-2\epsilon)\, \epsilon \,\,\vcenter{\hbox{\includegraphics[scale=0.13]{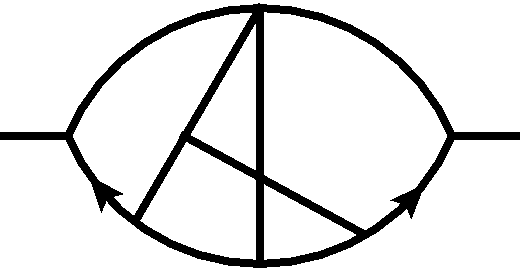}}} \\[1cm]
    M^{(4)}_{61} \equiv (1-2\epsilon)\, \epsilon \,\,\vcenter{\hbox{\includegraphics[scale=0.13]{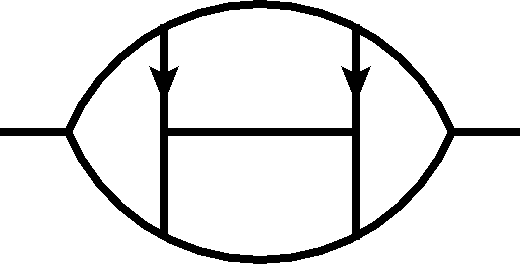}}} \quad
    & M^{(4)}_{62} \equiv (1-2\epsilon)\, \epsilon \,\,\vcenter{\hbox{\includegraphics[scale=0.13]{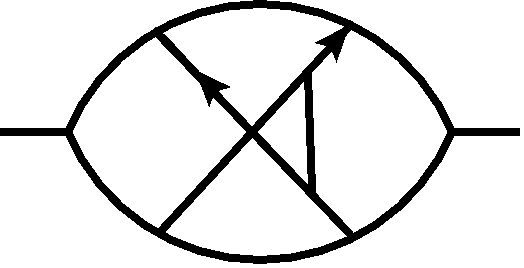}}}\\[0.5cm]
\end{array}
\end{equation*}
\begin{equation}\label{eq:basis4L2}
    \begin{array}{l}  
    M^{(4)}_{63} \equiv (1-2\epsilon)\, \epsilon \,\,\vcenter{\hbox{\includegraphics[scale=0.13]{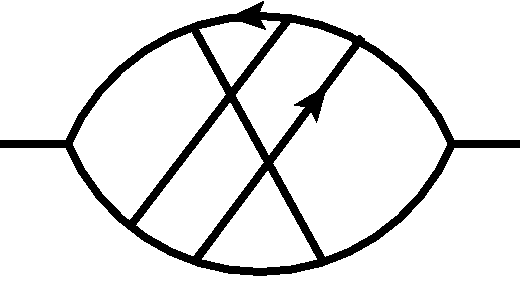}}}\\[0.5cm]
    \end{array}
\end{equation}
The arrows on propagators indicate a scalar product between the corresponding momenta, so, e.g.
\begin{equation}
\raisebox{-3mm}{\includegraphics[scale=0.15]{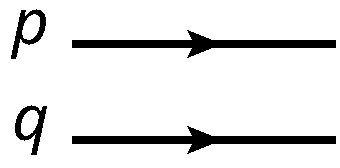}}\,\,=p\cdot q
\end{equation}
Instead of scalar products also inverse propagators may enforce uniform transcendental weight. Alternative basis elements with this property are showcased:
\begin{equation}\label{eq:basis4Lalt}
    \begin{array}{ll}
    \bar{M}^{(4)}_{51} \equiv (1-2\epsilon)\, \epsilon  \,\,\vcenter{\hbox{\includegraphics[scale=0.13]{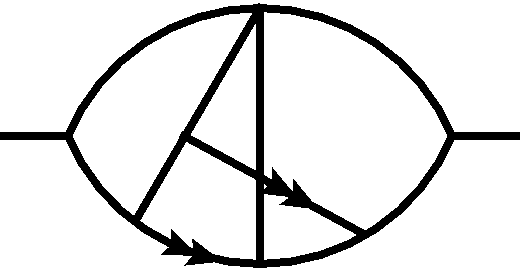}}} \quad
    & \bar{M}^{(4)}_{61} \equiv (1-2\epsilon)\, \epsilon  \,\,\vcenter{\hbox{\includegraphics[scale=0.13]{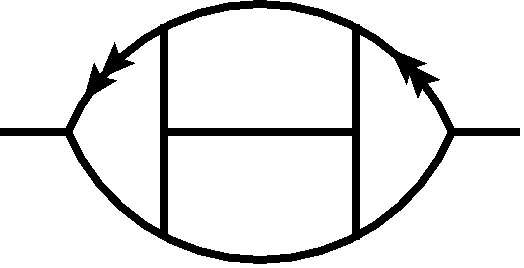}}} \\[1cm]
    \bar{M}^{(4)}_{62} \equiv (1-2\epsilon)\, \epsilon  \,\,\vcenter{\hbox{\includegraphics[scale=0.13]{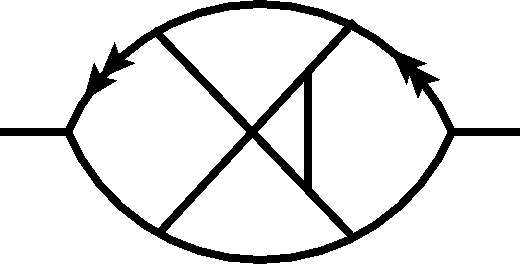}}} \quad
    & \bar{M}^{(4)}_{63} \equiv (1-2\epsilon)\, \epsilon  \,\,\vcenter{\hbox{\includegraphics[scale=0.13]{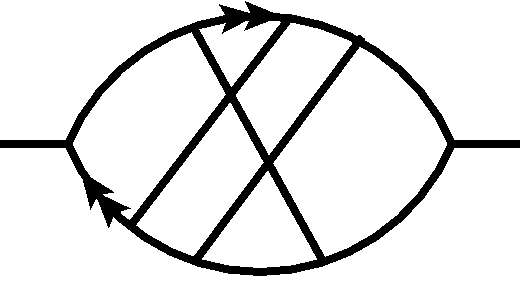}}} 
    \end{array}
\end{equation}
The double arrows on propagators with momenta $p_i$ and $p_j$ indicate a numerator of the form $(p_i-p_j)^2$.
So, for instance,
\begin{equation}
\raisebox{-3mm}{\includegraphics[scale=0.15]{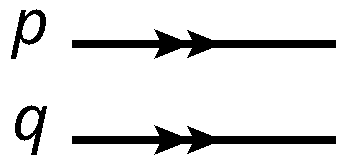}}\,\,=(p-q)^2
\end{equation}

The basis choice \eqref{eq:basis2L} through \eqref{eq:basis4L2} is not unique. 
Alternative assignments of propagator powers or numerator insertions can also ensure uniform transcendentality, as illustrated in \eqref{eq:basis4Lalt} for the most complicated topologies.

Our choice in \eqref{eq:basis4L2} is motivated by the fact that uniform transcendental weight is achieved with minimal deviations from the integrals of \cite{Lee:2011jt}; additionally, the rational coefficients of MZVs in their $\epsilon$-expansions appear simpler than in other possibilities, and the resulting coefficients in the two-point functions are correspondingly more compact.

The transformation from the basis of \cite{Lee:2011jt} to ours \eqref{eq:basis4L1}-\eqref{eq:basis4L2} is implemented by a matrix included in the Mathematica notebook \texttt{UT four loops.nb} submitted with this work, with $\epsilon$-expansions constructed using the ancillary files of \cite{Lee:2011jt}.

Our uniformly transcendental bases include higher propagator powers of propagators, which might introduce unphysical singularities at intermediate steps of the calculation. These higher powers should be understood as a compact way to produce $\epsilon$ factors enforcing uniform transcendentality after IBP reduction construction and do not correspond to new physical singularities. 
They merely reorganize the integral representation of the same physical quantities and do not introduce additional infrared or ultraviolet divergences. In fact, they can always be traded by rational functions of $\epsilon$, similarly to the construction of uniformly transcendental integrals in \cite{Lee:2011jt}. 
In the context of scattering amplitudes and form factors, uniform transcendentality is often associated with integrals admitting d-log representations at the integrand level. 
While it would be interesting to establish a direct connection between the present construction and possible d-log structures for these correlators, our focus here has been on a minimal modification of the integrated basis, leaving such an integrand-level analysis for future investigation.
In our analysis, simple trial and error proved sufficient to identify uniformly transcendental representatives. 
A more systematic strategy, however, would be to embed these integrals into a system of differential equations by introducing an auxiliary scale and then transforming the system to canonical form \cite{Henn:2013nsa}. 
This method has been employed, for example, in the study of the master integrals for four-loop Sudakov form factors \cite{Henn:2016men,Boels:2017ftb,Lee:2021lkc,Lee:2023dtc}. 
The uniformly transcendental combinations presented there already encompass the results we obtain, but we believe it is useful to provide an alternative and particularly simple realization of uniformly transcendental integrals based on minimal modifications of the original integrands.

\subsection{Three dimensions}
In three dimensions, the relevant master integrals were expanded to high orders in $\epsilon$ in \cite{Lee:2015eva}. 
Unlike the four-dimensional case, these expansions contain Euler sums in addition to MZVs, which makes the reconstruction of analytic coefficients from numerical data more cumbersome.
For the two-point functions in ABJM theory, however, only the one- and three-loop momentum integrals are required, and we focus on those here.

\paragraph{One loop.} The simple one-loop bubble integral in three dimensions is already uniformly transcendental with unit propagator powers \eqref{eq:1loop3d}
\begin{equation}\label{eq:basis1L3d}
    \begin{array}{l}
    N^{(1)}_{1} \equiv \,\,\vcenter{\hbox{\includegraphics[scale=0.13]{figures/M1L1.png}}}
    \end{array}
\end{equation}
Hence, in this setting there is no need to factor out the tree-level contribution to make uniform transcendentality manifest in the loop corrections to the two-point functions \cite{Bianchi:2024nah}.

\paragraph{Two loops.} We now identify a basis of two-loop integrals with uniform transcendental weight. 
Although these integrals are not required for the present computation, their structure is instructive for understanding the three-loop case.
We begin by considering the master integral
$\vcenter{\hbox{\includegraphics[scale=0.08]{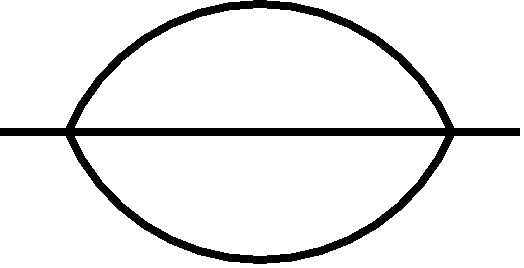}}}\,\,$.
Since it can be written entirely in terms of Gamma functions, one may easily render it uniformly transcendental by multiplying by an appropriate factor that compensates for the non-uniform Gamma function in the denominator.
So the rescaled integral
\begin{dgroup*}
\begin{dmath}\label{eq:2L3d}
    N^{(2)}_1=\frac{1-6\epsilon}{\epsilon}\,\, \vcenter{\hbox{\includegraphics[scale=0.13]{figures/M2L1.png}}} \raisebox{-3mm}{$\phantom{\Bigg|}$}
\end{dmath}
\begin{dmath*}
=
    \pi \left(\frac{1}{\epsilon ^2}-7 \zeta _2-\frac{176 \zeta _3 \epsilon }{3}-\frac{909 \zeta _4 \epsilon ^2}{4}+\left(\frac{1232 \zeta _2 \zeta _3}{3}-\frac{7472 \zeta
   _5}{5}\right) \epsilon ^3+\left(\frac{15488 \zeta _3^2}{9}-\frac{69253 \zeta _6}{16}\right) \epsilon ^4+O\left(\epsilon ^5\right)\right)
\end{dmath*}
\end{dgroup*}
is uniformly transcendental. 
An alternative way to tune the Gamma functions so as to obtain uniform transcendentality is to modify the powers of the propagators in the denominator.
This can be achieved by inserting a propagator with a $1/2$ power:
\begin{equation}\label{eq:2L3d2}
   \vcenter{\hbox{\includegraphics[scale=0.13]{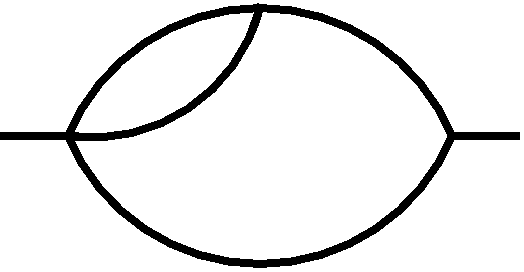}}} \raisebox{8mm}{\hspace{-7mm}$\frac12$}
\end{equation}
However, introducing a square root of a propagator is not ideal in this context, as it is incompatible with standard IBP reductions.

A further way of defining the same uniformly transcendental integral \eqref{eq:2L3d} is to trade the prefactor for a suitable combination of integer propagator powers and numerator insertions, as follows:
\begin{equation}\label{eq:2L3d3}
    N^{(2)}_1=\frac{1-6\epsilon}{\epsilon}\,\, \vcenter{\hbox{\includegraphics[scale=0.13]{figures/M2L1.png}}} = \frac{2}{\epsilon}\,p_1\cdot(p-p_2)\,\,\,\,\vcenter{\hbox{\includegraphics[scale=0.13]{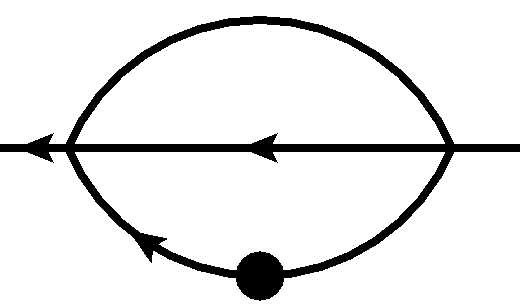}}} \raisebox{-5mm}{\hspace{-22mm}$p_1$}\raisebox{3.5mm}{\hspace{2mm}$p_2$}\raisebox{3.5mm}{\hspace{-12mm}$p$}\hspace{2.5cm}
\end{equation}
where the momenta and their flow directions are indicated by the arrows in the diagram. 
Upon IBP reduction, this representation is equivalent to \eqref{eq:2L3d}.

The product of two one-loop bubble integrals is already uniformly transcendental
\begin{dgroup*}
\begin{dmath}\label{eq:2L3dM2}
    N^{(2)}_2=\,\, \vcenter{\hbox{\includegraphics[scale=0.13]{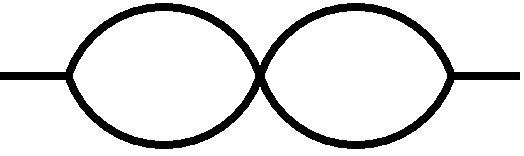}}} \raisebox{-3mm}{$\phantom{\Bigg|}$}
\end{dmath}
\begin{dmath*}
=
    \pi\, 2^{4\epsilon}\, \left(6 \zeta _2+75 \zeta _4 \epsilon ^2-4
   \zeta _2 \zeta _3 \epsilon ^3+\frac{3843 \zeta _6 \epsilon ^4}{8}+\left(-50 \zeta _3 \zeta _4-\frac{12 \zeta _2 \zeta _5}{5}\right) \epsilon
   ^5+O\left(\epsilon ^6\right)\right)
\end{dmath*}
\end{dgroup*}
The overall factors of $\epsilon$ in the definition of the integrals is a matter of convention, introduced solely to endow them with the same transcendental weight~3, in our normalization. 

\paragraph{Three loops.} By the same reasoning, the three-loop master integrals that factorize into products of bubble integrals can be rendered uniformly transcendental by multiplying them by suitable rational functions of $\epsilon$. 
Equivalently, one may obtain the same result through IBP reductions of integrals with appropriately modified propagator powers or numerator insertions.  
A third option would be to choose propagator powers that are not necessarily integer, but we do not pursue this possibility here. 
Instead, we also provide representations that avoid additional $\epsilon$ prefactors and achieve uniform transcendentality solely through adjusted denominator powers and numerator structures.
We define the basis elements as:
\begin{align}\label{eq:MIut}
    N^{(3)}_1 \equiv& \,\,
    \frac{(1-4\epsilon)(1-6\epsilon)}{\epsilon^2}\,\,\vcenter{\hbox{\includegraphics[scale=0.13]{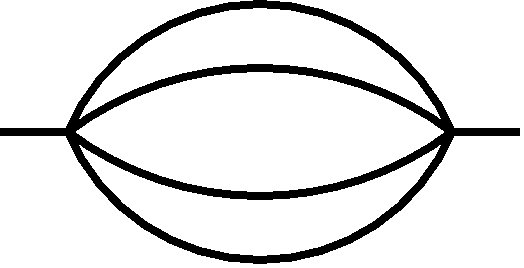}}}\,\, =\,\,\frac{2}{\epsilon^2}\,p_1\cdot p_2\,\,\,\,\vcenter{\hbox{\includegraphics[scale=0.13]{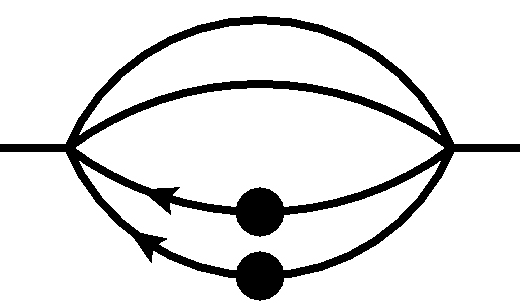}}} \nonumber\\[5mm]
    N^{(3)}_2 \equiv&\,\, \frac{1-6\epsilon}{\epsilon}\,\, \vcenter{\hbox{\includegraphics[scale=0.13]{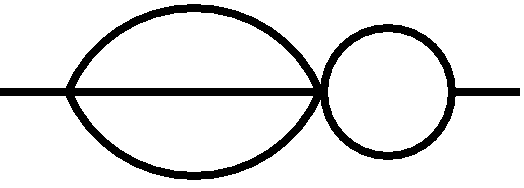}}}\,\, =\,\, \frac{2}{\epsilon}\, p_1\cdot(p-p_2)\,\,\,\,\vcenter{\hbox{\includegraphics[scale=0.13]{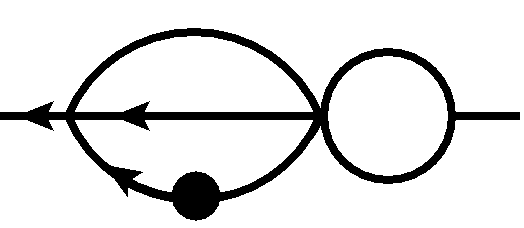}}} \nonumber\\[5mm]
    N^{(3)}_3 \equiv&\,\,\vcenter{\hbox{\includegraphics[scale=0.13]{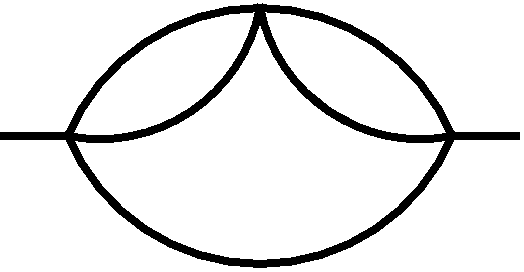}}}\nonumber\\[5mm]
    N^{(3)}_4 \equiv&\,\,\vcenter{\hbox{\includegraphics[scale=0.13]{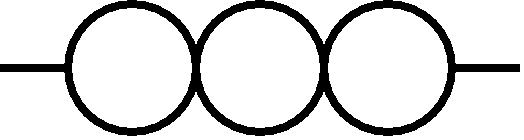}}} 
\end{align}
In the scalar product appearing in $N^{(3)}_1$, the momenta $p_1$ and $p_2$ are those indicated by arrows in the diagram and correspond to the propagators carrying squared powers. 
In contrast, the scalar product entering $N^{(3)}_2$ is the same as in $N^{(2)}_1$.

For the two more complicated master integrals, the planar topology already appears to be uniformly transcendental in three dimensions, as observed in \cite{Bianchi:2024nah},
\begin{equation}\label{eq:MIPut}
    N^{(3)}_5 \equiv \,\,\vcenter{\hbox{\includegraphics[scale=0.13]{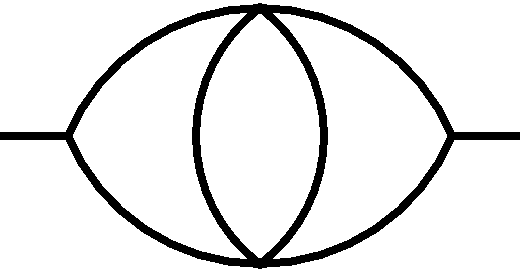}}}
\end{equation}

For the non-planar topology, a uniformly transcendental combination was identified empirically in \cite{Bianchi:2024nah}, arising naturally in the computation of the two-point function of the lowest-dimension BPS operators for a specific color structure.
\begin{align}\label{eq:MINPut}
    N^{(3)}_6 \equiv& \frac{1+2 \epsilon}{4(1+4 \epsilon)}\,\,\vcenter{\hbox{\includegraphics[scale=0.13]{figures/NUT3L6.png}}}   + \frac{5 (1+6 \epsilon) \epsilon }{(1+2 \epsilon) (1+4\epsilon)}\,\, \vcenter{\hbox{\includegraphics[scale=0.13]{figures/M3L5.png}}}\nonumber\\&
    +\frac{192 \epsilon ^2}{(1+2 \epsilon)^2}\,\,\vcenter{\hbox{\includegraphics[scale=0.13]{figures/M3L3.png}}} +\frac{14(1-6 \epsilon) (1+6 \epsilon) \epsilon }{(1+2 \epsilon)^2 (1+4 \epsilon)}\,\,\vcenter{\hbox{\includegraphics[scale=0.13]{figures/M3L2.png}}}
\nonumber\\&
   -\frac{2 (1-4 \epsilon) (1-6 \epsilon) \left(172 \epsilon^2+60 \epsilon +3\right)}{(1+2 \epsilon)^3 (1+4 \epsilon)}\,\,\vcenter{\hbox{\includegraphics[scale=0.13]{figures/M3L1.png}}}
\end{align}
We have searched for a simpler interpretation of this combination, for instance one arising directly from specific choices of numerators. 
Such an interpretation must exist, since \eqref{eq:MINPut} originates from a bona fide Feynman-diagram computation. 
After some trial and error, we found that \eqref{eq:MINPut} is equivalent to
\begin{align}
    N^{(3)}_6 &=\,\, 
    2\,\,\raisebox{-5.1mm}{\includegraphics[scale=0.13]{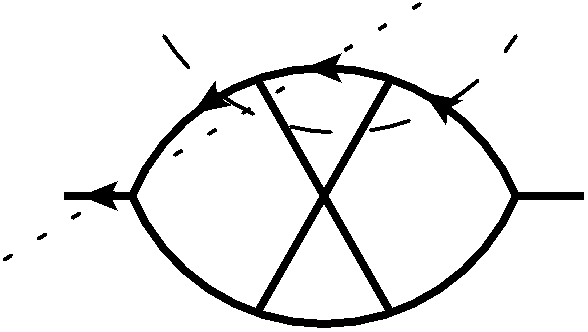}}
    +\vcenter{\hbox{\includegraphics[scale=0.13]{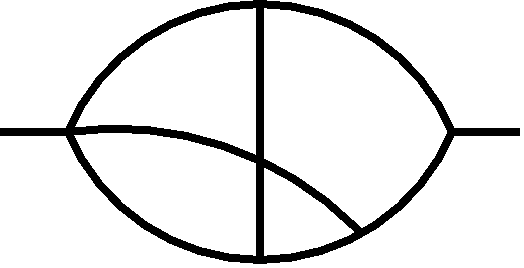}}}
    -\vcenter{\hbox{\includegraphics[scale=0.13]{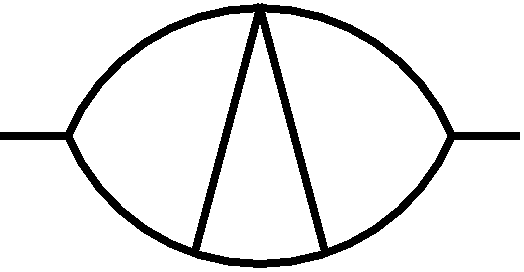}}}
    \nonumber\\[2mm]
    &
    +2\,\,\vcenter{\hbox{\includegraphics[scale=0.13]{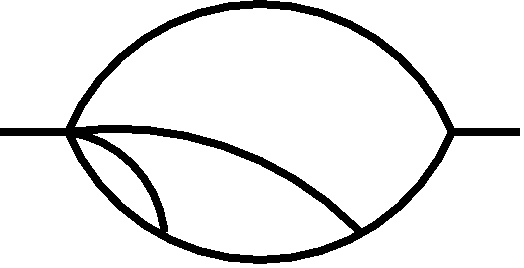}}}
    -2\,\,\vcenter{\hbox{\includegraphics[scale=0.13]{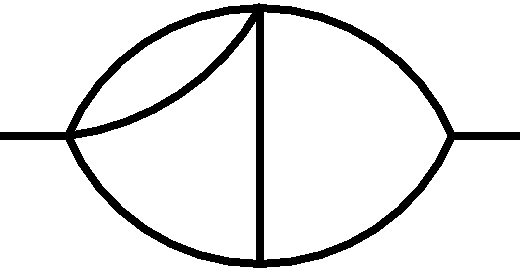}}}
    -\,\vcenter{\hbox{\includegraphics[scale=0.13]{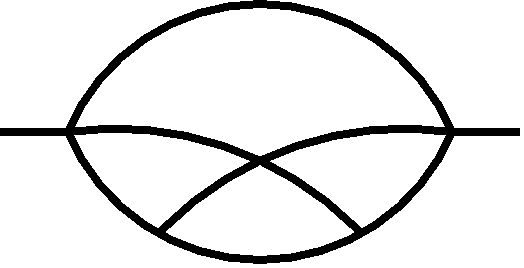}}}
\end{align}
where in the first integrand of the right-hand side two scalar products appear in the numerators, between the pairs of propagators joined by dashed lines in the picture.
At present we do not have a deeper understanding of why this particular linear combination yields a uniformly transcendental result, though it would certainly be interesting to uncover such an explanation.

The expansions of this basis of integrals are collected in Appendix~\ref{app:exp3d}.
Identifying a basis of four-loop uniformly transcendental integrals in three dimensions would also be of interest, but achieving this would require a more detailed analytic control over certain coefficients in their expansions \cite{Lee:2015eva}, a task that we leave for future work.


\section{Two-point functions in terms of uniformly transcendental integrals}

We present the perturbative corrections to the two-point functions of the lowest-dimension BPS operators, expressed in terms of the uniformly transcendental bases in four and three dimensions, respectively.

\paragraph{$\mathcal{N}=4$ SYM.}

The tree-level contribution reads
\begin{equation}
    n^{(0)} = \frac{1}{(1-2\epsilon)\epsilon}\, M^{(1)}_1
\end{equation}
where the additional factor $(1-2\epsilon)$ causes the departure from uniform transcendentality.   This invites redefining \eqref{eq:perturbative}, factoring out such a term
\begin{equation}\label{eq:perturbative2}
\left\langle O(p) \bar O(-p) \right\rangle 
= \frac{2\,(N^2-1)}{(4\pi)^{2-\epsilon}(1-2\epsilon)\epsilon}\,
\left( m^{(0)} + m^{(1)}\lambda + m^{(2)}\lambda^2 + m^{(3)}\lambda^3 + O(\lambda^4) \right)
\end{equation}
so that
\begin{equation}
    m^{(0)} = M^{(1)}_1
\end{equation}
This is partly why the representation of master integrals with explicit $(1-2\epsilon)$ factors in \eqref{eq:basis1L}-\eqref{eq:basis4L2} can be convenient for representing two-point functions. Such factors cancel naturally, leaving the two-point functions expressed in terms of the remaining integrals.

Once normalized by the tree-level contribution, the coefficients $m^{(L)}$ yield the same results as in \eqref{eq:1L}-\eqref{eq:3L}, that is 
\begin{equation}
    \frac{m^{(L)}}{m^{(0)}} = \frac{n^{(L)}}{n^{(0)}}
\end{equation}
Once the factorization in \eqref{eq:perturbative2} is performed, the coefficients $m^{(L)}$ are uniformly transcendental.
Moreover, this property becomes completely explicit when they are rewritten in terms of our basis of manifestly uniformly transcendental integrals.
We obtain for the one-loop correction to the two-point function
\begin{equation}
    m^{(1)} = -2\, M^{(2)}_{1} + 2\, M^{(2)}_{4}
\end{equation}
At two loops a more substantial simplification with respect to the expression in \cite{Bianchi:2023llc}, using a different integral basis, already appears
\begin{equation}
    m^{(2)} = \frac{8}{3}\,M^{(3)}_{1} + \frac{16}{3}\,M^{(3)}_{3}
    - 12\,M^{(3)}_{2} + 4\,M^{(3)}_{4} - 12\,M^{(3)}_{5}
    + M^{(3)}_{6}
\end{equation}
since the full contribution is expressed solely in terms of integrals with rational coefficients.
At three loops the simplification becomes impressive:
\begin{align}
m^{(3)} =&
-\frac{127}{6}\,M^{(4)}_{01}
+\frac{112}{9}\,M^{(4)}_{11}
+\frac{16}{3}\,M^{(4)}_{12}
+\frac{269}{6}\,M^{(4)}_{13}
+\frac{512}{3}\,M^{(4)}_{21}
+\frac{56}{3}\,M^{(4)}_{22} 
\nonumber\\ &
+\frac{176}{9}\,M^{(4)}_{24}
-29\,M^{(4)}_{14}
-40\,M^{(4)}_{23}
-15\,M^{(4)}_{26}
-58\,M^{(4)}_{27}
+8\,M^{(4)}_{31}
\nonumber\\ &
-48\,M^{(4)}_{32}
-48\,M^{(4)}_{34}
+240\,M^{(4)}_{35}
+66\,M^{(4)}_{36}
+12\,M^{(4)}_{41}
+8\,M^{(4)}_{44}
\nonumber\\ &
-6\,M^{(4)}_{45}
+4\,M^{(4)}_{52}
+32\,M^{(4)}_{51}
-4\,M^{(4)}_{61}
+8\,M^{(4)}_{63}
\end{align}
These results and their explicit $\epsilon$ expansions are available in electronic format in the Mathematica notebook \texttt{2-point function.nb}, included in the submission.

Alternatively, using the basis with inverse propagators \eqref{eq:basis4Lalt} we obtain analogously
\begin{align}
m^{(3)} =&
-\frac{211}{18}\,M^{(4)}_{01}
+\frac{59}{3}\,M^{(4)}_{11}
+\frac{19}{3}\,M^{(4)}_{12}
+\frac{53}{3}\,M^{(4)}_{13}
+\frac{448}{3}\,M^{(4)}_{21}
-\frac{28}{3}\,M^{(4)}_{22}
\nonumber\\ &
+\frac{200}{9}\,M^{(4)}_{24}
-\frac{133}{6}\,M^{(4)}_{14}
-40\,M^{(4)}_{23}
-\frac{15}{2}\,M^{(4)}_{26}
-23\,M^{(4)}_{27}
+8\,M^{(4)}_{31}
\nonumber\\ &
-48\,M^{(4)}_{32}
+40\,M^{(4)}_{34}
+148\,M^{(4)}_{35}
+28\,M^{(4)}_{36}
+4\,M^{(4)}_{41}
-4\,M^{(4)}_{44}
\nonumber\\ &
-4\,M^{(4)}_{42}
+4\,M^{(4)}_{52}
+16\,\bar{M}^{(4)}_{51}
-2\,\bar{M}^{(4)}_{61}
-2\,\bar{M}^{(4)}_{63}
\end{align}
In all such representations the resulting expressions simplify dramatically: the coefficients are purely rational numbers, in contrast with the rational functions of $\epsilon$ that appear when using a non-uniformly transcendental basis. For comparison, the expansion in the master-integral basis of \textsc{Forcer} reads \cite{Bianchi:2024nah}
\begin{dmath*}
\left\langle O(p) \bar O(-p) \right\rangle^{(3)}=
-\tfrac{4(3 \epsilon -1) (4 \epsilon -1) (5 \epsilon -1) \left(14484 \epsilon ^3+10379 \epsilon ^2+2433 \epsilon +190\right)}{9 \epsilon ^3 (3 \epsilon +1) (4 \epsilon +1) (5 \epsilon +1)}  \, \raisebox{-2.5mm}{\includegraphics[scale=0.1]{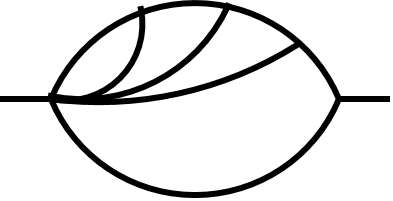}}+
+\tfrac{16 (4 \epsilon -1) \left(324 \epsilon ^2+157 \epsilon +18\right)(2 \epsilon -1)^2}{3 \epsilon ^3 (4 \epsilon +1) (5 \epsilon +1)}  \, \raisebox{-1.6mm}{\includegraphics[scale=0.1]{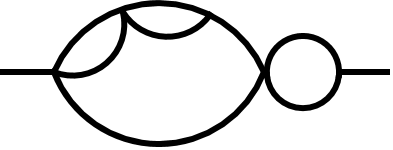}} 
-\tfrac{20 (5 \epsilon -1) (2 \epsilon -1)^2}{\epsilon ^3}  \, \raisebox{-2.5mm}{\includegraphics[scale=0.1]{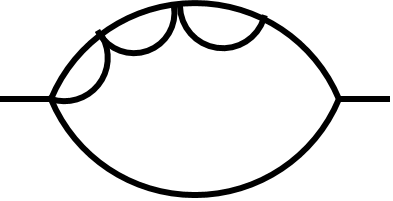}}
+\tfrac{16 (3 \epsilon -1) (4 \epsilon -1) \left(1257 \epsilon ^3+833 \epsilon ^2+165 \epsilon +10\right) (2 \epsilon -1)}{9 \epsilon ^3 (3 \epsilon +1) (4 \epsilon +1) (5 \epsilon +1)}  \, \raisebox{-1.6mm}{\includegraphics[scale=0.1]{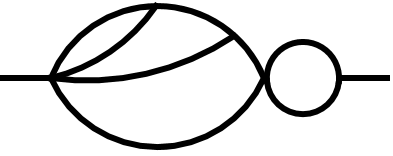}}
+\tfrac{8 (2 \epsilon -1)^3}{\epsilon ^3} \, \raisebox{-0.8mm}{\includegraphics[scale=0.1]{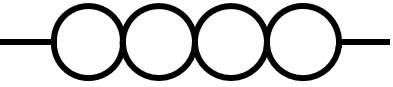}}
-\tfrac{4 (3 \epsilon -1) \left(408 \epsilon ^2+187 \epsilon +21\right) (2 \epsilon -1)^2}{\epsilon ^3 (4 \epsilon +1) (5 \epsilon +1)} \, \raisebox{-0.8mm}{\includegraphics[scale=0.1]{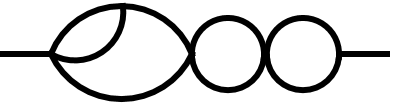}}
-\tfrac{4(3 \epsilon -1) (5 \epsilon -1) \left(3288 \epsilon ^3+2629 \epsilon ^2+579 \epsilon +38\right) (2 \epsilon -1)}{9 \epsilon ^3 (3 \epsilon +1) (4 \epsilon +1) (5 \epsilon +1)}  \, \raisebox{-2.5mm}{\includegraphics[scale=0.1]{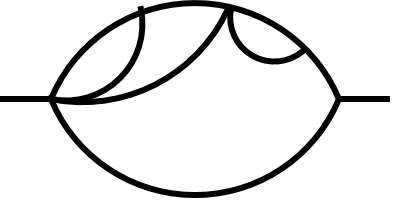}}
+\tfrac{4(4 \epsilon -1) (5 \epsilon -1) \left(3592 \epsilon ^3+2938 \epsilon ^2+761 \epsilon +63\right) (2 \epsilon -1)}{3 \epsilon ^3 (3 \epsilon +1) (4 \epsilon +1) (5 \epsilon +1)}  \, \raisebox{-2.5mm}{\includegraphics[scale=0.1]{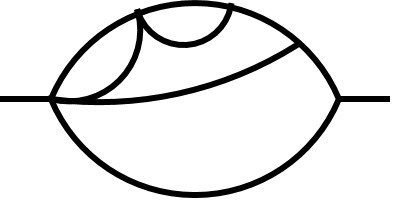}}
-\tfrac{48 (4 \epsilon +1) (2 \epsilon -1)}{\epsilon  (5 \epsilon +1)}  \, \raisebox{-2.5mm}{\includegraphics[scale=0.1]{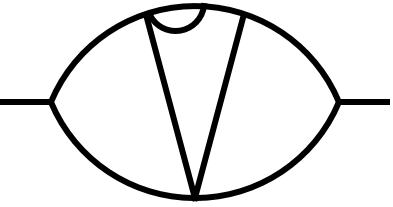}}
-\tfrac{48 (2 \epsilon -1)^2}{\epsilon ^2}  \, \raisebox{-1.6mm}{\includegraphics[scale=0.1]{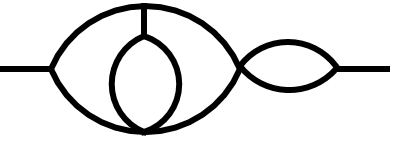}} 
+\tfrac{60 (5 \epsilon -1) (2 \epsilon -1)}{\epsilon  (3 \epsilon +1)} \, \raisebox{-2.5mm}{\includegraphics[scale=0.1]{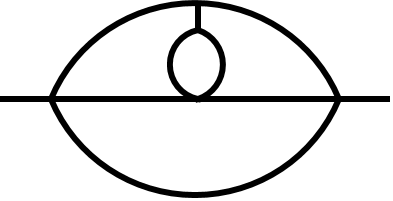}}
+\tfrac{16 (2 \epsilon -1)^2}{(3 \epsilon +1) (5 \epsilon +1)} \, \raisebox{-2.5mm}{\includegraphics[scale=0.1]{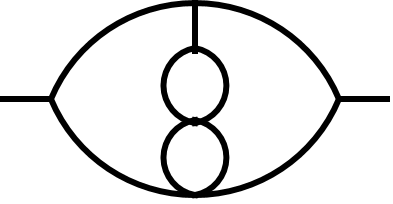}}
+\tfrac{8 (3 \epsilon -1)^2 \left(329 \epsilon ^3+223 \epsilon ^2+49 \epsilon +4\right) (2 \epsilon -1)}{3 \epsilon ^3 (3 \epsilon +1) (4 \epsilon +1) (5 \epsilon +1)}  \, \raisebox{-0.8mm}{\includegraphics[scale=0.1]{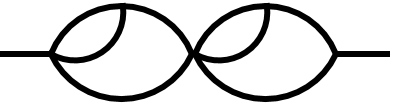}}
-\tfrac{12 (3 \epsilon -1) \left(128 \epsilon ^2+65 \epsilon +8\right) (2 \epsilon -1)}{\epsilon ^2 (4 \epsilon +1) (5 \epsilon +1)} \, \raisebox{-2.5mm}{\includegraphics[scale=0.1]{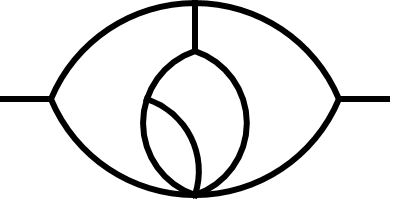}}
+\tfrac{32 \left(363 \epsilon ^2+205 \epsilon +26\right) (2 \epsilon -1)^2}{9 \epsilon ^2 (3 \epsilon +1) (5 \epsilon +1)} \, \raisebox{-2.5mm}{\includegraphics[scale=0.1]{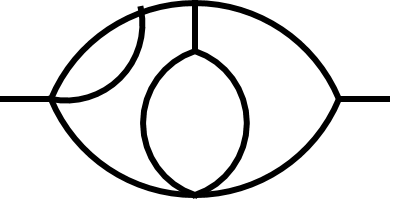}}
-\tfrac{8 \left(300 \epsilon ^2+142 \epsilon +15\right) (2 \epsilon -1)^2}{3 \epsilon ^2 (4 \epsilon +1) (5 \epsilon +1)} \, \raisebox{-2.5mm}{\includegraphics[scale=0.1]{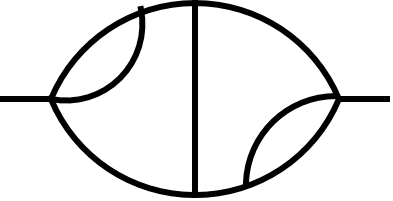}} 
+\tfrac{4 (2 \epsilon -1)}{\epsilon } \, \raisebox{-1.6mm}{\includegraphics[scale=0.1]{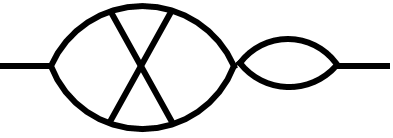}} 
+\tfrac{4 (2 \epsilon -1)}{\epsilon } \, \raisebox{-2.5mm}{\includegraphics[scale=0.1]{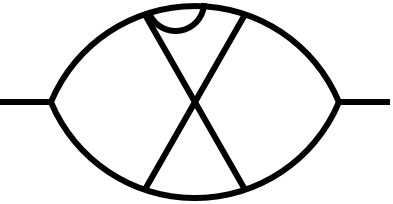}} 
-\tfrac{4 (6 \epsilon +1) (2 \epsilon -1)}{\epsilon  (4 \epsilon +1)} \, \raisebox{-2.5mm}{\includegraphics[scale=0.1]{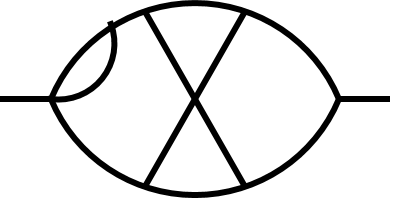}} 
-\tfrac{16 (4 \epsilon -1) (5 \epsilon -1) \left(104 \epsilon ^2+69 \epsilon +10\right) }{9 \epsilon ^2 (4 \epsilon +1) (5 \epsilon +1)} \, \raisebox{-2.5mm}{\includegraphics[scale=0.1]{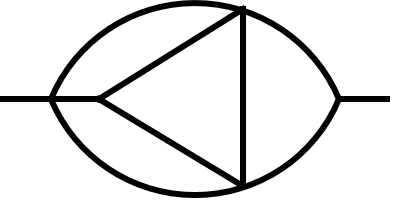}} 
+\tfrac{20 (5 \epsilon -1) \left(19 \epsilon ^2+13 \epsilon +2\right)}{3 \epsilon  (3 \epsilon +1) (5 \epsilon +1)} \, \raisebox{-2.5mm}{\includegraphics[scale=0.1]{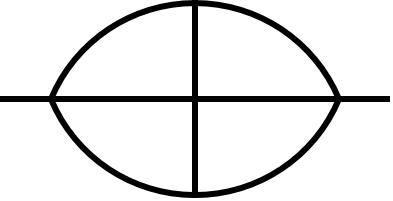}} 
-\tfrac{8 \epsilon}{(4 \epsilon +1) (5 \epsilon +1)} \, \raisebox{-2.5mm}{\includegraphics[scale=0.1]{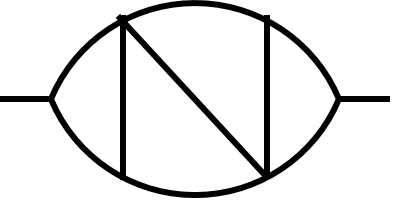}} 
-\tfrac{16 (6 \epsilon -1) (6 \epsilon +1)}{(4 \epsilon +1) (5 \epsilon +1)} \, \raisebox{-2.5mm}{\includegraphics[scale=0.1]{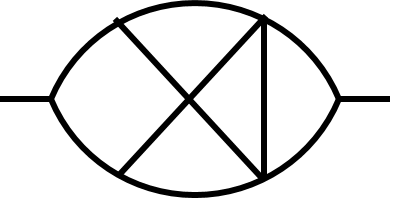}} 
-\tfrac{2 (3 \epsilon +1)}{5 \epsilon +1} \, \raisebox{-2.5mm}{\includegraphics[scale=0.1]{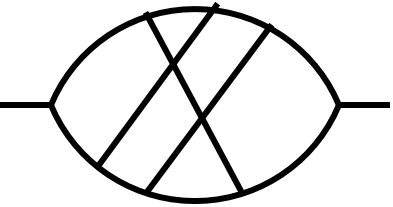}} 
-\tfrac{2 (3 \epsilon +1) (4 \epsilon +1)}{\epsilon  (5 \epsilon +1)} \, \raisebox{-2.5mm}{\includegraphics[scale=0.1]{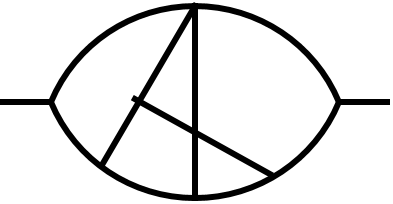}} 
+\tfrac{(3 \epsilon +1)^2 }{\epsilon  (5 \epsilon +1)} \, \raisebox{-2.5mm}{\includegraphics[scale=0.1]{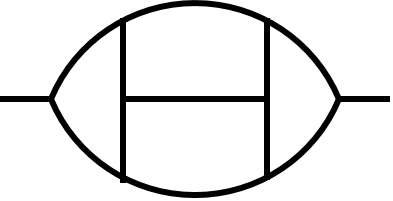}} 
\end{dmath*}
This comparison makes the simplification afforded by the uniformly transcendental basis completely transparent. 
Among the four-loop integrals, $M^{(4)}_{62}$ appears in individual diagrams but carries a vanishing coefficient in the final answer.  
The integrals $M^{(4)}_{25}$, $M^{(4)}_{33}$, and $M^{(4)}_{43}$ are also absent.

\paragraph{ABJM.} The same considerations as in the four-dimensional case apply here: ABJM two-point functions exhibit remarkable simplicity when expressed in terms of a basis of uniformly transcendental integrals.
At two loops we obtain the following contributions for each of the color structures:
\begin{equation}
    8\pi\, c_{N_1^2} = -\frac12N^{(3)}_1 - \, N^{(3)}_2 - 4\, N^{(3)}_3 - 2\, N^{(3)}_4 + 2\, N^{(3)}_5
\end{equation}
\begin{equation}
   8\pi\,  c_{N_1N_2} = N^{(3)}_1 + 2\, N^{(3)}_2 + 8\, N^{(3)}_3 + 4\, N^{(3)}_4 - 4\, N^{(3)}_5 + 2\, N^{(3)}_6
\end{equation}
\begin{equation}
   8\pi\,  c_{1} = - 2\, N^{(3)}_6
\end{equation}
where, once again, the factors of $8\pi$ arise solely from the normalization choices in \eqref{eq:norm3d} and \eqref{eq:1loop3d}, which we retain for consistency throughout this work. 
What truly matters is the appearance of simple rational numbers on the right-hand sides: this is the characteristic signature of the simplification achieved when expressing the result in a basis of manifestly uniformly transcendental integrals.


\section{Discussion and Outlook}

Our construction of uniformly transcendental bases for two-point functions provides a streamlined and remarkably compact representation of multi-loop correlators in both $\mathcal{N}=4$ SYM and ABJM. 
The analysis shows that the property of uniform transcendentality, long familiar from studies of scattering amplitudes, extends naturally to certain off-shell correlation functions as well.

Looking ahead, it would be interesting to extend this approach to higher-dimension protected operators. 
A natural next step is also to push the $\mathcal{N}=4$ SYM computation to five loops using the expansions of \cite{Georgoudis:2018olj,Georgoudis:2021onj} and to investigate whether a fully uniformly transcendental basis exists in that case. In three dimensions it would be valuable to identify the analogue of a four-loop uniformly transcendental basis and clarify its analytic structure \cite{Lee:2015eva}.

Also, it would be interesting to investigate whether the uniformly transcendental structure uncovered here admits a reformulation in terms of d-log integrands, as in the form-factor literature, and whether such a perspective becomes more natural in higher-point correlation functions where non-trivial kinematic structure is available \cite{Drummond:2013nda}.

At the present stage, however, the observations of uniform transcendentality remain empirical, and a deeper conceptual understanding of why these theories exhibit uniform transcendentality stands as an important goal for future research.

\acknowledgments

I thank Johannes Henn and Tobias Huber for suggesting to look for a representation of these two-point functions in terms of uniformly transcendental integrals.
This work was supported by Fondo Nacional de Desarrollo Cient\'ifico y Tecnol\'ogico, through Fondecyt Regular 1220240, Fondecyt Exploraci\'on 13220060 and Fondecyt Exploraci\'on 13250014.

\vfill
\newpage

\appendix

\section{Expansions of master integrals in four dimensions}\label{app:exp}
In this appendix we provide explicit expansions of the uniformly transcendental integrals defined in section \ref{sec:bases} up to transcendental order twelve. The results are also available in electronic format in auxiliary Mathematica files.
For two-loop integrals
\small
\begin{dmath}
M^{(2)}_{1}=-\frac{1}{\epsilon ^2}+\zeta _2+\frac{32 \zeta _3 \epsilon }{3}+\frac{57 \zeta _4 \epsilon
   ^2}{4}+\left(\frac{272 \zeta _5}{5}-\frac{32 \zeta _2 \zeta _3}{3}\right) \epsilon
   ^3+\left(\frac{1343 \zeta _6}{16}-\frac{512 \zeta _3^2}{9}\right) \epsilon ^4+\left(-152 \zeta _3
   \zeta _4-\frac{272 \zeta _2 \zeta _5}{5}+\frac{2312 \zeta _7}{7}\right) \epsilon
   ^5+\left(\frac{512}{9} \zeta _2 \zeta _3^2-\frac{8704 \zeta _5 \zeta _3}{15}+\frac{31105 \zeta
   _8}{64}\right) \epsilon ^6+\left(\frac{16384 \zeta _3^3}{81}-\frac{2686 \zeta _6 \zeta
   _3}{3}-\frac{3876 \zeta _4 \zeta _5}{5}-\frac{2312 \zeta _2 \zeta _7}{7}+\frac{20192 \zeta
   _9}{9}\right) \epsilon ^7+\left(\frac{2432}{3} \zeta _4 \zeta _3^2+\frac{8704}{15} \zeta _2 \zeta
   _5 \zeta _3-\frac{73984 \zeta _7 \zeta _3}{21}-\frac{36992 \zeta _5^2}{25}+\frac{183818639 \zeta
   _{10}}{61440}\right) \epsilon ^8+\left(-\frac{16384}{81} \zeta _2 \zeta _3^3+\frac{139264}{45}
   \zeta _5 \zeta _3^2-\frac{71667145 \zeta _8 \zeta _3}{13824}-\frac{22831 \zeta _5 \zeta
   _6}{5}-\frac{32946 \zeta _4 \zeta _7}{7}-\frac{20192 \zeta _2 \zeta _9}{9}+\frac{179192 \zeta
   _{11}}{11}\right) \epsilon ^9+\left(-\frac{131072 \zeta _3^4}{243}+\frac{16501069 \zeta _6 \zeta
   _3^2}{3456}+\frac{41344}{5} \zeta _4 \zeta _5 \zeta _3+\frac{73984}{21} \zeta _2 \zeta _7 \zeta
   _3-\frac{646144 \zeta _9 \zeta _3}{27}+\frac{36992}{25} \zeta _2 \zeta _5^2-\frac{628864 \zeta _5
   \zeta _7}{35}+\frac{3976724749325 \zeta _{12}}{203784192}\right) \epsilon ^{10}+O\left(\epsilon
   ^{11}\right)
\end{dmath}
\begin{dmath}
M^{(2)}_{2}=-\frac{1}{\epsilon ^2}+\zeta _2+\frac{14 \zeta _3 \epsilon }{3}+\frac{21 \zeta _4
   \epsilon ^2}{4}+\left(\frac{62 \zeta _5}{5}-\frac{14 \zeta _2 \zeta _3}{3}\right) \epsilon
   ^3+\left(\frac{155 \zeta _6}{16}-\frac{98 \zeta _3^2}{9}\right) \epsilon ^4+\left(-\frac{49}{2}
   \zeta _3 \zeta _4-\frac{62 \zeta _2 \zeta _5}{5}+\frac{254 \zeta _7}{7}\right) \epsilon
   ^5+\left(\frac{98}{9} \zeta _2 \zeta _3^2-\frac{868 \zeta _5 \zeta _3}{15}+\frac{889 \zeta
   _8}{64}\right) \epsilon ^6+\left(\frac{1372 \zeta _3^3}{81}-\frac{1085 \zeta _6 \zeta
   _3}{24}-\frac{651 \zeta _4 \zeta _5}{10}-\frac{254 \zeta _2 \zeta _7}{7}+\frac{1022 \zeta
   _9}{9}\right) \epsilon ^7+\left(\frac{343}{6} \zeta _4 \zeta _3^2+\frac{868}{15} \zeta _2 \zeta _5
   \zeta _3-\frac{508 \zeta _7 \zeta _3}{3}-\frac{1922 \zeta _5^2}{25}+\frac{220675 \zeta
   _{10}}{12288}\right) \epsilon ^8+\left(-\frac{1372}{81} \zeta _2 \zeta _3^3+\frac{6076}{45} \zeta
   _5 \zeta _3^2-\frac{897337 \zeta _8 \zeta _3}{13824}-\frac{961 \zeta _5 \zeta _6}{8}-\frac{381
   \zeta _4 \zeta _7}{2}-\frac{1022 \zeta _2 \zeta _9}{9}+\frac{4094 \zeta _{11}}{11}\right) \epsilon
   ^9+\left(-\frac{4802 \zeta _3^4}{243}+\frac{362845 \zeta _6 \zeta _3^2}{3456}+\frac{1519}{5} \zeta
   _4 \zeta _5 \zeta _3+\frac{508}{3} \zeta _2 \zeta _7 \zeta _3-\frac{14308 \zeta _9 \zeta
   _3}{27}+\frac{1922}{25} \zeta _2 \zeta _5^2-\frac{15748 \zeta _5 \zeta _7}{35}+\frac{4454962061
   \zeta _{12}}{203784192}\right) \epsilon ^{10}+O\left(\epsilon ^{11}\right)
\end{dmath}
For three-loop integrals
\begin{dmath}
M^{(3)}_{1}=\frac{1}{\epsilon
   ^4}-\frac{3 \zeta _2}{2 \epsilon ^2}-\frac{29 \zeta _3}{\epsilon }-\frac{639 \zeta
   _4}{16}+\left(\frac{87 \zeta _2 \zeta _3}{2}-\frac{1263 \zeta _5}{5}\right) \epsilon
   +\left(\frac{841 \zeta _3^2}{2}-\frac{57695 \zeta _6}{128}\right) \epsilon ^2+\left(\frac{18531
   \zeta _3 \zeta _4}{16}+\frac{3789 \zeta _2 \zeta _5}{10}-\frac{18567 \zeta _7}{7}\right) \epsilon
   ^3+\left(-\frac{2523}{4} \zeta _2 \zeta _3^2+\frac{36627 \zeta _5 \zeta _3}{5}-\frac{5163809 \zeta
   _8}{1024}\right) \epsilon ^4+\left(-\frac{24389 \zeta _3^3}{6}+\frac{1673155 \zeta _6 \zeta
   _3}{128}+\frac{807057 \zeta _4 \zeta _5}{80}+\frac{55701 \zeta _2 \zeta _7}{14}-\frac{93941 \zeta
   _9}{3}\right) \epsilon ^5+\left(-\frac{537399}{32} \zeta _4 \zeta _3^2-\frac{109881}{10} \zeta _2
   \zeta _5 \zeta _3+\frac{538443 \zeta _7 \zeta _3}{7}+\frac{1595169 \zeta _5^2}{50}-\frac{60014849
   \zeta _{10}}{1024}\right) \epsilon ^6+\left(\frac{24389}{4} \zeta _2 \zeta _3^3-\frac{1062183}{10}
   \zeta _5 \zeta _3^2+\frac{673877687 \zeta _8 \zeta _3}{4608}+\frac{14573757 \zeta _5 \zeta
   _6}{128}+\frac{11864313 \zeta _4 \zeta _7}{112}+\frac{93941 \zeta _2 \zeta _9}{2}-\frac{4371447
   \zeta _{11}}{11}\right) \epsilon ^7+\left(\frac{707281 \zeta _3^4}{24}-\frac{109172935}{576} \zeta
   _6 \zeta _3^2-\frac{23404653}{80} \zeta _4 \zeta _5 \zeta _3-\frac{1615329}{14} \zeta _2 \zeta _7
   \zeta _3+\frac{2724289 \zeta _9 \zeta _3}{3}-\frac{4785507}{100} \zeta _2 \zeta _5^2+\frac{23450121
   \zeta _5 \zeta _7}{35}-\frac{47825767580107 \zeta _{12}}{67928064}\right) \epsilon
   ^8+O\left(\epsilon^{9}\right)
\end{dmath}
\begin{dmath}
M^{(3)}_{2}=\frac{1}{\epsilon ^4}-\frac{3 \zeta _2}{2 \epsilon ^2}-\frac{13
   \zeta _3}{\epsilon }-\frac{255 \zeta _4}{16}+\left(\frac{39 \zeta _2 \zeta _3}{2}-\frac{303 \zeta
   _5}{5}\right) \epsilon +\left(\frac{169 \zeta _3^2}{2}-\frac{9439 \zeta _6}{128}\right) \epsilon
   ^2+\left(\frac{3315 \zeta _3 \zeta _4}{16}+\frac{909 \zeta _2 \zeta _5}{10}-\frac{2439 \zeta
   _7}{7}\right) \epsilon ^3+\left(-\frac{507}{4} \zeta _2 \zeta _3^2+\frac{3939 \zeta _5 \zeta
   _3}{5}-\frac{381985 \zeta _8}{1024}\right) \epsilon ^4+\left(-\frac{2197 \zeta
   _3^3}{6}+\frac{122707 \zeta _6 \zeta _3}{128}+\frac{15453 \zeta _4 \zeta _5}{16}+\frac{7317 \zeta
   _2 \zeta _7}{14}-\frac{6901 \zeta _9}{3}\right) \epsilon ^5+\left(-\frac{43095}{32} \zeta _4 \zeta
   _3^2-\frac{11817}{10} \zeta _2 \zeta _5 \zeta _3+\frac{31707 \zeta _7 \zeta _3}{7}+\frac{91809
   \zeta _5^2}{50}-\frac{11373541 \zeta _{10}}{5120}\right) \epsilon ^6+\left(\frac{2197}{4} \zeta _2
   \zeta _3^3-\frac{51207}{10} \zeta _5 \zeta _3^2+\frac{22346735 \zeta _8 \zeta
   _3}{4608}+\frac{2860017 \zeta _5 \zeta _6}{640}+\frac{621945 \zeta _4 \zeta _7}{112}+\frac{6901
   \zeta _2 \zeta _9}{2}-\frac{181239 \zeta _{11}}{11}\right) \epsilon ^7+\left(\frac{28561 \zeta
   _3^4}{24}-\frac{3588751}{576} \zeta _6 \zeta _3^2-\frac{200889}{16} \zeta _4 \zeta _5 \zeta
   _3-\frac{95121}{14} \zeta _2 \zeta _7 \zeta _3+\frac{89713 \zeta _9 \zeta _3}{3}-\frac{275427}{100}
   \zeta _2 \zeta _5^2+\frac{739017 \zeta _5 \zeta _7}{35}-\frac{994030450315 \zeta
   _{12}}{67928064}\right) \epsilon ^8+O\left(\epsilon^{9}\right)
\end{dmath}
\begin{dmath}
M^{(3)}_{3}=\frac{1}{\epsilon ^4}-\frac{3
   \zeta _2}{2 \epsilon ^2}-\frac{23 \zeta _3}{\epsilon }-\frac{495 \zeta _4}{16}+\left(\frac{69 \zeta
   _2 \zeta _3}{2}-\frac{1053 \zeta _5}{5}\right) \epsilon +\left(\frac{529 \zeta _3^2}{2}-\frac{49199
   \zeta _6}{128}\right) \epsilon ^2+\left(\frac{11385 \zeta _3 \zeta _4}{16}+\frac{3159 \zeta _2
   \zeta _5}{10}-\frac{16509 \zeta _7}{7}\right) \epsilon ^3+\left(-\frac{1587}{4} \zeta _2 \zeta
   _3^2+\frac{24219 \zeta _5 \zeta _3}{5}-\frac{5033345 \zeta _8}{1024}\right) \epsilon
   ^4+\left(-\frac{12167 \zeta _3^3}{6}+\frac{1131577 \zeta _6 \zeta _3}{128}+\frac{104247 \zeta _4
   \zeta _5}{16}+\frac{49527 \zeta _2 \zeta _7}{14}-\frac{87551 \zeta _9}{3}\right) \epsilon
   ^5+\left(-\frac{261855}{32} \zeta _4 \zeta _3^2-\frac{72657}{10} \zeta _2 \zeta _5 \zeta
   _3+\frac{379707 \zeta _7 \zeta _3}{7}+\frac{1108809 \zeta _5^2}{50}-\frac{319720921 \zeta
   _{10}}{5120}\right) \epsilon ^6+\left(\frac{12167}{4} \zeta _2 \zeta _3^3-\frac{557037}{10} \zeta
   _5 \zeta _3^2+\frac{130237955 \zeta _8 \zeta _3}{1152}+\frac{51806547 \zeta _5 \zeta
   _6}{640}+\frac{8171955 \zeta _4 \zeta _7}{112}+\frac{87551 \zeta _2 \zeta _9}{2}-\frac{4196349
   \zeta _{11}}{11}\right) \epsilon ^7+\left(\frac{279841 \zeta _3^4}{24}-\frac{58558681}{576} \zeta
   _6 \zeta _3^2-\frac{2397681}{16} \zeta _4 \zeta _5 \zeta _3-\frac{1139121}{14} \zeta _2 \zeta _7
   \zeta _3+\frac{2013673 \zeta _9 \zeta _3}{3}-\frac{3326427}{100} \zeta _2 \zeta _5^2+\frac{17383977
   \zeta _5 \zeta _7}{35}-\frac{53674581133075 \zeta _{12}}{67928064}\right) \epsilon
   ^8+O\left(\epsilon^{9}\right)
\end{dmath}
\begin{dmath}
M^{(3)}_{4}=\frac{1}{\epsilon ^4}-\frac{3 \zeta _2}{2 \epsilon ^2}-\frac{7
   \zeta _3}{\epsilon }-\frac{111 \zeta _4}{16}+\left(\frac{21 \zeta _2 \zeta _3}{2}-\frac{93 \zeta
   _5}{5}\right) \epsilon +\left(\frac{49 \zeta _3^2}{2}-\frac{943 \zeta _6}{128}\right) \epsilon
   ^2+\left(\frac{777 \zeta _3 \zeta _4}{16}+\frac{279 \zeta _2 \zeta _5}{10}-\frac{381 \zeta
   _7}{7}\right) \epsilon ^3+\left(-\frac{147}{4} \zeta _2 \zeta _3^2+\frac{651 \zeta _5 \zeta
   _3}{5}+\frac{6527 \zeta _8}{1024}\right) \epsilon ^4+\left(-\frac{343 \zeta _3^3}{6}+\frac{6601
   \zeta _6 \zeta _3}{128}+\frac{10323 \zeta _4 \zeta _5}{80}+\frac{1143 \zeta _2 \zeta
   _7}{14}-\frac{511 \zeta _9}{3}\right) \epsilon ^5+\left(-\frac{5439}{32} \zeta _4 \zeta
   _3^2-\frac{1953}{10} \zeta _2 \zeta _5 \zeta _3+381 \zeta _7 \zeta _3+\frac{8649 \zeta
   _5^2}{50}+\frac{254279 \zeta _{10}}{5120}\right) \epsilon ^6+\left(\frac{343}{4} \zeta _2 \zeta
   _3^3-\frac{4557}{10} \zeta _5 \zeta _3^2-\frac{51247 \zeta _8 \zeta _3}{1152}+\frac{87699 \zeta _5
   \zeta _6}{640}+\frac{42291 \zeta _4 \zeta _7}{112}+\frac{511 \zeta _2 \zeta _9}{2}-\frac{6141 \zeta
   _{11}}{11}\right) \epsilon ^7+\left(\frac{2401 \zeta _3^4}{24}-\frac{103537}{576} \zeta _6 \zeta
   _3^2-\frac{72261}{80} \zeta _4 \zeta _5 \zeta _3-\frac{1143}{2} \zeta _2 \zeta _7 \zeta
   _3+\frac{3577 \zeta _9 \zeta _3}{3}-\frac{25947}{100} \zeta _2 \zeta _5^2+\frac{35433 \zeta _5
   \zeta _7}{35}+\frac{10682301485 \zeta _{12}}{67928064}\right) \epsilon ^8+O\left(\epsilon
^{9}\right)
\end{dmath}
\begin{dmath}
M^{(3)}_{5}=-\frac{6 \zeta _3}{\epsilon }-9 \zeta _4+\left(9 \zeta _2 \zeta _3-102 \zeta _5\right)
   \epsilon +\left(120 \zeta _3^2-\frac{1731 \zeta _6}{8}\right) \epsilon ^2+\left(\frac{2709 \zeta _3
   \zeta _4}{8}+153 \zeta _2 \zeta _5-1413 \zeta _7\right) \epsilon ^3+\left(-180 \zeta _2 \zeta
   _3^2+\frac{13188 \zeta _5 \zeta _3}{5}-\frac{468033 \zeta _8}{160}-\frac{648 \zeta
   _{5,3}}{5}\right) \epsilon ^4+\left(-1281 \zeta _3^3+\frac{325413 \zeta _6 \zeta
   _3}{64}+\frac{182601 \zeta _4 \zeta _5}{40}+\frac{4239 \zeta _2 \zeta _7}{2}-18918 \zeta _9\right)
   \epsilon ^5+\left(-5337 \zeta _4 \zeta _3^2-\frac{19782}{5} \zeta _2 \zeta _5 \zeta _3+\frac{220203
   \zeta _7 \zeta _3}{7}+\frac{462087 \zeta _5^2}{35}-\frac{1606177521 \zeta
   _{10}}{44800}+\frac{972}{5} \zeta _2 \zeta _{5,3}-\frac{5751 \zeta _{7,3}}{7}\right) \epsilon
   ^6+\left(\frac{3843}{2} \zeta _2 \zeta _3^3-36483 \zeta _5 \zeta _3^2+\frac{159830943 \zeta _8
   \zeta _3}{2560}+\frac{7128}{5} \zeta _{5,3} \zeta _3+\frac{16626489 \zeta _5 \zeta
   _6}{320}+\frac{37186209 \zeta _4 \zeta _7}{560}+133353 \zeta _2 \zeta _9-\frac{8614689 \zeta
   _{11}}{20}+\frac{11664}{5} \zeta _{5,3,3}\right) \epsilon ^7+\left(9836 \zeta _3^4-\frac{292047}{4}
   \zeta _6 \zeta _3^2-\frac{2240811}{20} \zeta _4 \zeta _5 \zeta _3-\frac{297729}{14} \zeta _2 \zeta
   _7 \zeta _3+367216 \zeta _9 \zeta _3-\frac{751221}{70} \zeta _2 \zeta _5^2+\frac{11269773 \zeta _5
   \zeta _7}{35}-\frac{6782111909073 \zeta _{12}}{14151680}+\frac{90639}{10} \zeta _4 \zeta
   _{5,3}+\frac{126117}{14} \zeta _2 \zeta _{7,3}-16812 \zeta _{9,3}-2592 \zeta _{6,4,1,1}\right)
   \epsilon ^8+O\left(\epsilon ^{9}\right)
\end{dmath}
\begin{dmath}
M^{(3)}_{6}=20 \zeta _5 \epsilon +\left(68 \zeta _3^2+50 \zeta
   _6\right) \epsilon ^2+\left(204 \zeta _3 \zeta _4-30 \zeta _2 \zeta _5+450 \zeta _7\right) \epsilon
   ^3+\left(-102 \zeta _2 \zeta _3^2-2588 \zeta _5 \zeta _3+\frac{58383 \zeta _8}{10}-\frac{9072 \zeta
   _{5,3}}{5}\right) \epsilon ^4+\left(-\frac{6068 \zeta _3^3}{3}-\frac{14691 \zeta _6 \zeta
   _3}{2}+\frac{39189 \zeta _4 \zeta _5}{4}-675 \zeta _2 \zeta _7+\frac{88036 \zeta _9}{9}\right)
   \epsilon ^5+\left(-\frac{35439}{4} \zeta _4 \zeta _3^2+3882 \zeta _2 \zeta _5 \zeta _3-60302 \zeta
   _7 \zeta _3-\frac{173654 \zeta _5^2}{7}+\frac{88561889 \zeta _{10}}{700}+\frac{13608}{5} \zeta _2
   \zeta _{5,3}-\frac{85026 \zeta _{7,3}}{7}\right) \epsilon ^6+\left(3034 \zeta _2 \zeta
   _3^3+\frac{32806}{5} \zeta _5 \zeta _3^2-\frac{2006493 \zeta _8 \zeta _3}{8}+27216 \zeta _{5,3}
   \zeta _3-\frac{780395 \zeta _5 \zeta _6}{32}+\frac{8602149 \zeta _4 \zeta _7}{40}+\frac{2153134
   \zeta _2 \zeta _9}{3}-\frac{31418849 \zeta _{11}}{30}+\frac{81376}{5} \zeta _{5,3,3}\right)
   \epsilon ^7+\left(\frac{260342 \zeta _3^4}{9}+\frac{648401}{32} \zeta _6 \zeta
   _3^2-\frac{12083369}{60} \zeta _4 \zeta _5 \zeta _3+\frac{735359}{3} \zeta _2 \zeta _7 \zeta
   _3-\frac{31832092 \zeta _9 \zeta _3}{27}+\frac{1918243}{21} \zeta _2 \zeta _5^2-\frac{12010990
   \zeta _5 \zeta _7}{21}+\frac{152469976293 \zeta _{12}}{88448}+81973 \zeta _4 \zeta
   _{5,3}+\frac{452339}{7} \zeta _2 \zeta _{7,3}-\frac{1393864 \zeta _{9,3}}{9}-\frac{46400}{3} \zeta
   _{6,4,1,1}\right) \epsilon ^8+O\left(\epsilon ^{9}\right)
\end{dmath}
For four-loop integrals
\begin{dmath}
M^{(4)}_{01}=-\frac{1}{\epsilon ^6}+\frac{2 \zeta
   _2}{\epsilon ^4}+\frac{184 \zeta _3}{3 \epsilon ^3}+\frac{86 \zeta _4}{\epsilon
   ^2}+\frac{\frac{4144 \zeta _5}{5}-\frac{368 \zeta _2 \zeta _3}{3}}{\epsilon }+\left(1608 \zeta
   _6-\frac{16928 \zeta _3^2}{9}\right)+\left(-\frac{15824}{3} \zeta _3 \zeta _4-\frac{8288 \zeta _2
   \zeta _5}{5}+\frac{94504 \zeta _7}{7}\right) \epsilon +\left(\frac{33856}{9} \zeta _2 \zeta
   _3^2-\frac{762496 \zeta _5 \zeta _3}{15}+\frac{89786 \zeta _8}{3}\right) \epsilon
   ^2+\left(\frac{3114752 \zeta _3^3}{81}-98624 \zeta _6 \zeta _3-\frac{356384 \zeta _4 \zeta
   _5}{5}-\frac{189008 \zeta _2 \zeta _7}{7}+\frac{2215264 \zeta _9}{9}\right) \epsilon
   ^3+\left(\frac{1455808}{9} \zeta _4 \zeta _3^2+\frac{1524992}{15} \zeta _2 \zeta _5 \zeta
   _3-\frac{17388736 \zeta _7 \zeta _3}{21}-\frac{8586368 \zeta _5^2}{25}+\frac{17549487743 \zeta
   _{10}}{30720}\right) \epsilon ^4+\left(-\frac{6229504}{81} \zeta _2 \zeta _3^3+\frac{70149632}{45}
   \zeta _5 \zeta _3^2-\frac{12687840457 \zeta _8 \zeta _3}{6912}-\frac{6663552 \zeta _5 \zeta
   _6}{5}-\frac{8127344 \zeta _4 \zeta _7}{7}-\frac{4430528 \zeta _2 \zeta _9}{9}+\frac{53022424 \zeta
   _{11}}{11}\right) \epsilon ^5+\left(-\frac{143278592 \zeta _3^4}{243}+\frac{5226281293 \zeta _6
   \zeta _3^2}{1728}+\frac{65574656}{15} \zeta _4 \zeta _5 \zeta _3+\frac{34777472}{21} \zeta _2 \zeta
   _7 \zeta _3-\frac{407608576 \zeta _9 \zeta _3}{27}+\frac{17172736}{25} \zeta _2 \zeta
   _5^2-\frac{55946368 \zeta _5 \zeta _7}{5}+\frac{1133698129513985 \zeta _{12}}{101892096}\right)
   \epsilon ^6+O\left(\epsilon ^{7}\right)
\end{dmath}
\begin{dmath}
M^{(4)}_{11}=-\frac{1}{\epsilon ^6}+\frac{2 \zeta _2}{\epsilon
   ^4}+\frac{94 \zeta _3}{3 \epsilon ^3}+\frac{41 \zeta _4}{\epsilon ^2}+\frac{\frac{1294 \zeta
   _5}{5}-\frac{188 \zeta _2 \zeta _3}{3}}{\epsilon }+\left(\frac{831 \zeta _6}{2}-\frac{4418 \zeta
   _3^2}{9}\right)+\left(-\frac{3854}{3} \zeta _3 \zeta _4-\frac{2588 \zeta _2 \zeta
   _5}{5}+\frac{18694 \zeta _7}{7}\right) \epsilon +\left(\frac{8836}{9} \zeta _2 \zeta
   _3^2-\frac{121636 \zeta _5 \zeta _3}{15}+\frac{54359 \zeta _8}{12}\right) \epsilon
   ^2+\left(\frac{415292 \zeta _3^3}{81}-13019 \zeta _6 \zeta _3-\frac{53054 \zeta _4 \zeta
   _5}{5}-\frac{37388 \zeta _2 \zeta _7}{7}+\frac{282334 \zeta _9}{9}\right) \epsilon
   ^3+\left(\frac{181138}{9} \zeta _4 \zeta _3^2+\frac{243272}{15} \zeta _2 \zeta _5 \zeta
   _3-\frac{1757236 \zeta _7 \zeta _3}{21}-\frac{837218 \zeta _5^2}{25}+\frac{1617984383 \zeta
   _{10}}{30720}\right) \epsilon ^4+\left(-\frac{830584}{81} \zeta _2 \zeta _3^3+\frac{5716892}{45}
   \zeta _5 \zeta _3^2-\frac{981072457 \zeta _8 \zeta _3}{6912}-\frac{537657 \zeta _5 \zeta
   _6}{5}-\frac{766454 \zeta _4 \zeta _7}{7}-\frac{564668 \zeta _2 \zeta _9}{9}+\frac{4373494 \zeta
   _{11}}{11}\right) \epsilon ^5+\left(-\frac{9759362 \zeta _3^4}{243}+\frac{352448653 \zeta _6 \zeta
   _3^2}{1728}+\frac{4987076}{15} \zeta _4 \zeta _5 \zeta _3+\frac{3514472}{21} \zeta _2 \zeta _7
   \zeta _3-\frac{26539396 \zeta _9 \zeta _3}{27}+\frac{1674436}{25} \zeta _2 \zeta
   _5^2-\frac{24190036 \zeta _5 \zeta _7}{35}+\frac{64752848464385 \zeta _{12}}{101892096}\right)
   \epsilon ^6+O\left(\epsilon ^{7}\right)
\end{dmath}
\begin{dmath}
M^{(4)}_{12}=-\frac{1}{\epsilon ^6}+\frac{2 \zeta _2}{\epsilon
   ^4}+\frac{64 \zeta _3}{3 \epsilon ^3}+\frac{26 \zeta _4}{\epsilon ^2}+\frac{\frac{544 \zeta
   _5}{5}-\frac{128 \zeta _2 \zeta _3}{3}}{\epsilon }+\left(118 \zeta _6-\frac{2048 \zeta
   _3^2}{9}\right)+\left(-\frac{1664}{3} \zeta _3 \zeta _4-\frac{1088 \zeta _2 \zeta _5}{5}+\frac{4624
   \zeta _7}{7}\right) \epsilon +\left(\frac{4096}{9} \zeta _2 \zeta _3^2-\frac{34816 \zeta _5 \zeta
   _3}{15}+\frac{1366 \zeta _8}{3}\right) \epsilon ^2+\left(\frac{131072 \zeta _3^3}{81}-\frac{7552
   \zeta _6 \zeta _3}{3}-\frac{14144 \zeta _4 \zeta _5}{5}-\frac{9248 \zeta _2 \zeta
   _7}{7}+\frac{40384 \zeta _9}{9}\right) \epsilon ^3+\left(\frac{53248}{9} \zeta _4 \zeta
   _3^2+\frac{69632}{15} \zeta _2 \zeta _5 \zeta _3-\frac{295936 \zeta _7 \zeta _3}{21}-\frac{147968
   \zeta _5^2}{25}+\frac{53710463 \zeta _{10}}{30720}\right) \epsilon ^4+\left(-\frac{262144}{81}
   \zeta _2 \zeta _3^3+\frac{1114112}{45} \zeta _5 \zeta _3^2-\frac{67142857 \zeta _8 \zeta
   _3}{6912}-\frac{64192 \zeta _5 \zeta _6}{5}-\frac{120224 \zeta _4 \zeta _7}{7}-\frac{80768 \zeta _2
   \zeta _9}{9}+\frac{358384 \zeta _{11}}{11}\right) \epsilon ^5+\left(-\frac{2097152 \zeta
   _3^4}{243}+\frac{46397773 \zeta _6 \zeta _3^2}{1728}+\frac{905216}{15} \zeta _4 \zeta _5 \zeta
   _3+\frac{591872}{21} \zeta _2 \zeta _7 \zeta _3-\frac{2584576 \zeta _9 \zeta
   _3}{27}+\frac{295936}{25} \zeta _2 \zeta _5^2-\frac{2515456 \zeta _5 \zeta
   _7}{35}+\frac{696869277185 \zeta _{12}}{101892096}\right) \epsilon ^6+O\left(\epsilon
   ^{7}\right)
\end{dmath}
\begin{dmath}
M^{(4)}_{13}=-\frac{1}{\epsilon ^6}+\frac{2 \zeta _2}{\epsilon ^4}+\frac{136 \zeta _3}{3 \epsilon
   ^3}+\frac{62 \zeta _4}{\epsilon ^2}+\frac{\frac{3184 \zeta _5}{5}-\frac{272 \zeta _2 \zeta
   _3}{3}}{\epsilon }+\left(1252 \zeta _6-\frac{9248 \zeta _3^2}{9}\right)+\left(-\frac{8432}{3} \zeta
   _3 \zeta _4-\frac{6368 \zeta _2 \zeta _5}{5}+\frac{78376 \zeta _7}{7}\right) \epsilon
   +\left(\frac{18496}{9} \zeta _2 \zeta _3^2-\frac{433024 \zeta _5 \zeta _3}{15}+\frac{80746 \zeta
   _8}{3}\right) \epsilon ^2+\left(\frac{1257728 \zeta _3^3}{81}-\frac{170272 \zeta _6 \zeta
   _3}{3}-\frac{197408 \zeta _4 \zeta _5}{5}-\frac{156752 \zeta _2 \zeta _7}{7}+\frac{1954144 \zeta
   _9}{9}\right) \epsilon ^3+\left(\frac{573376}{9} \zeta _4 \zeta _3^2+\frac{866048}{15} \zeta _2
   \zeta _5 \zeta _3-\frac{10659136 \zeta _7 \zeta _3}{21}-\frac{5068928 \zeta
   _5^2}{25}+\frac{3523055539 \zeta _{10}}{6144}\right) \epsilon ^4+\left(-\frac{2515456}{81} \zeta _2
   \zeta _3^3+\frac{29445632}{45} \zeta _5 \zeta _3^2-\frac{8433759433 \zeta _8 \zeta
   _3}{6912}-\frac{3986368 \zeta _5 \zeta _6}{5}-\frac{4859312 \zeta _4 \zeta _7}{7}-\frac{3908288
   \zeta _2 \zeta _9}{9}+\frac{48832216 \zeta _{11}}{11}\right) \epsilon ^5+\left(-\frac{42762752
   \zeta _3^4}{243}+\frac{2223069517 \zeta _6 \zeta _3^2}{1728}+\frac{26847488}{15} \zeta _4 \zeta _5
   \zeta _3+\frac{21318272}{21} \zeta _2 \zeta _7 \zeta _3-\frac{265763584 \zeta _9 \zeta
   _3}{27}+\frac{10137856}{25} \zeta _2 \zeta _5^2-\frac{249549184 \zeta _5 \zeta
   _7}{35}+\frac{1227154944871937 \zeta _{12}}{101892096}\right) \epsilon ^6+O\left(\epsilon
   ^{7}\right)
\end{dmath}
\begin{dmath}
M^{(4)}_{14}=-\frac{1}{\epsilon ^6}+\frac{2 \zeta _2}{\epsilon ^4}+\frac{166 \zeta _3}{3 \epsilon
   ^3}+\frac{77 \zeta _4}{\epsilon ^2}+\frac{\frac{3934 \zeta _5}{5}-\frac{332 \zeta _2 \zeta
   _3}{3}}{\epsilon }+\left(\frac{3099 \zeta _6}{2}-\frac{13778 \zeta
   _3^2}{9}\right)+\left(-\frac{12782}{3} \zeta _3 \zeta _4-\frac{7868 \zeta _2 \zeta
   _5}{5}+\frac{92446 \zeta _7}{7}\right) \epsilon +\left(\frac{27556}{9} \zeta _2 \zeta
   _3^2-\frac{653044 \zeta _5 \zeta _3}{15}+\frac{364319 \zeta _8}{12}\right) \epsilon
   ^2+\left(\frac{2287148 \zeta _3^3}{81}-85739 \zeta _6 \zeta _3-\frac{302918 \zeta _4 \zeta
   _5}{5}-\frac{184892 \zeta _2 \zeta _7}{7}+\frac{2196094 \zeta _9}{9}\right) \epsilon
   ^3+\left(\frac{1060906}{9} \zeta _4 \zeta _3^2+\frac{1306088}{15} \zeta _2 \zeta _5 \zeta
   _3-\frac{15346036 \zeta _7 \zeta _3}{21}-\frac{7738178 \zeta _5^2}{25}+\frac{3635794099 \zeta
   _{10}}{6144}\right) \epsilon ^4+\left(-\frac{4574296}{81} \zeta _2 \zeta _3^3+\frac{54202652}{45}
   \zeta _5 \zeta _3^2-\frac{11611576393 \zeta _8 \zeta _3}{6912}-\frac{6095733 \zeta _5 \zeta
   _6}{5}-1016906 \zeta _4 \zeta _7-\frac{4392188 \zeta _2 \zeta _9}{9}+\frac{52847326 \zeta
   _{11}}{11}\right) \epsilon ^5+\left(-\frac{94916642 \zeta _3^4}{243}+\frac{4099008397 \zeta _6
   \zeta _3^2}{1728}+\frac{50284388}{15} \zeta _4 \zeta _5 \zeta _3+\frac{30692072}{21} \zeta _2 \zeta
   _7 \zeta _3-\frac{364551604 \zeta _9 \zeta _3}{27}+\frac{15476356}{25} \zeta _2 \zeta
   _5^2-\frac{51954652 \zeta _5 \zeta _7}{5}+\frac{1182726258489857 \zeta _{12}}{101892096}\right)
   \epsilon ^6+O\left(\epsilon ^{7}\right)
\end{dmath}
\begin{dmath}
M^{(4)}_{23}=-\frac{1}{\epsilon ^6}+\frac{2 \zeta _2}{\epsilon
   ^4}+\frac{46 \zeta _3}{3 \epsilon ^3}+\frac{17 \zeta _4}{\epsilon ^2}+\frac{\frac{334 \zeta
   _5}{5}-\frac{92 \zeta _2 \zeta _3}{3}}{\epsilon }+\left(\frac{119 \zeta _6}{2}-\frac{1058 \zeta
   _3^2}{9}\right)+\left(-\frac{782}{3} \zeta _3 \zeta _4-\frac{668 \zeta _2 \zeta _5}{5}+\frac{2566
   \zeta _7}{7}\right) \epsilon +\left(\frac{2116}{9} \zeta _2 \zeta _3^2-\frac{15364 \zeta _5 \zeta
   _3}{15}+\frac{3079 \zeta _8}{12}\right) \epsilon ^2+\left(\frac{48668 \zeta _3^3}{81}-\frac{2737
   \zeta _6 \zeta _3}{3}-\frac{5678 \zeta _4 \zeta _5}{5}-\frac{5132 \zeta _2 \zeta _7}{7}+\frac{21214
   \zeta _9}{9}\right) \epsilon ^3+\left(\frac{17986}{9} \zeta _4 \zeta _3^2+\frac{30728}{15} \zeta _2
   \zeta _5 \zeta _3-\frac{118036 \zeta _7 \zeta _3}{21}-\frac{55778 \zeta _5^2}{25}+\frac{9513139
   \zeta _{10}}{6144}\right) \epsilon ^4+\left(-\frac{97336}{81} \zeta _2 \zeta _3^3+\frac{353372}{45}
   \zeta _5 \zeta _3^2-\frac{27194953 \zeta _8 \zeta _3}{6912}-\frac{19873 \zeta _5 \zeta
   _6}{5}-\frac{43622 \zeta _4 \zeta _7}{7}-\frac{42428 \zeta _2 \zeta _9}{9}+\frac{183286 \zeta
   _{11}}{11}\right) \epsilon ^5+\left(-\frac{559682 \zeta _3^4}{243}+\frac{12084877 \zeta _6 \zeta
   _3^2}{1728}+\frac{261188}{15} \zeta _4 \zeta _5 \zeta _3+\frac{236072}{21} \zeta _2 \zeta _7 \zeta
   _3-\frac{975844 \zeta _9 \zeta _3}{27}+\frac{111556}{25} \zeta _2 \zeta _5^2-\frac{857044 \zeta _5
   \zeta _7}{35}+\frac{1095642712577 \zeta _{12}}{101892096}\right) \epsilon ^6+O\left(\epsilon
   ^{7}\right)
\end{dmath}
\begin{dmath}
M^{(4)}_{24}=-\frac{1}{\epsilon ^6}+\frac{2 \zeta _2}{\epsilon ^4}+\frac{76 \zeta _3}{3 \epsilon
   ^3}+\frac{32 \zeta _4}{\epsilon ^2}+\frac{\frac{1084 \zeta _5}{5}-\frac{152 \zeta _2 \zeta
   _3}{3}}{\epsilon }+\left(357 \zeta _6-\frac{2888 \zeta _3^2}{9}\right)+\left(-\frac{2432}{3} \zeta
   _3 \zeta _4-\frac{2168 \zeta _2 \zeta _5}{5}+\frac{16636 \zeta _7}{7}\right) \epsilon
   +\left(\frac{5776}{9} \zeta _2 \zeta _3^2-\frac{82384 \zeta _5 \zeta _3}{15}+\frac{13466 \zeta
   _8}{3}\right) \epsilon ^2+\left(\frac{219488 \zeta _3^3}{81}-9044 \zeta _6 \zeta _3-\frac{34688
   \zeta _4 \zeta _5}{5}-\frac{33272 \zeta _2 \zeta _7}{7}+\frac{263164 \zeta _9}{9}\right) \epsilon
   ^3+\left(\frac{92416}{9} \zeta _4 \zeta _3^2+\frac{164768}{15} \zeta _2 \zeta _5 \zeta
   _3-\frac{1264336 \zeta _7 \zeta _3}{21}-\frac{587528 \zeta _5^2}{25}+\frac{349587379 \zeta
   _{10}}{6144}\right) \epsilon ^4+\left(-\frac{438976}{81} \zeta _2 \zeta _3^3+\frac{3130592}{45}
   \zeta _5 \zeta _3^2-\frac{785984713 \zeta _8 \zeta _3}{6912}-\frac{386988 \zeta _5 \zeta
   _6}{5}-\frac{532352 \zeta _4 \zeta _7}{7}-\frac{526328 \zeta _2 \zeta _9}{9}+\frac{4198396 \zeta
   _{11}}{11}\right) \epsilon ^5+\left(-\frac{4170272 \zeta _3^4}{243}+\frac{197953357 \zeta _6 \zeta
   _3^2}{1728}+\frac{2636288}{15} \zeta _4 \zeta _5 \zeta _3+\frac{2528672}{21} \zeta _2 \zeta _7
   \zeta _3-\frac{20000464 \zeta _9 \zeta _3}{27}+\frac{1175056}{25} \zeta _2 \zeta
   _5^2-\frac{18033424 \zeta _5 \zeta _7}{35}+\frac{73315661303297 \zeta _{12}}{101892096}\right)
   \epsilon ^6+O\left(\epsilon ^{7}\right)
\end{dmath}
\begin{dmath}
M^{(4)}_{25}=-\frac{1}{\epsilon ^6}+\frac{2 \zeta _2}{\epsilon
   ^4}+\frac{118 \zeta _3}{3 \epsilon ^3}+\frac{53 \zeta _4}{\epsilon ^2}+\frac{\frac{2974 \zeta
   _5}{5}-\frac{236 \zeta _2 \zeta _3}{3}}{\epsilon }+\left(\frac{2387 \zeta _6}{2}-\frac{6962 \zeta
   _3^2}{9}\right)+\left(-\frac{6254}{3} \zeta _3 \zeta _4-\frac{5948 \zeta _2 \zeta
   _5}{5}+\frac{76318 \zeta _7}{7}\right) \epsilon +\left(\frac{13924}{9} \zeta _2 \zeta
   _3^2-\frac{350932 \zeta _5 \zeta _3}{15}+\frac{325135 \zeta _8}{12}\right) \epsilon
   ^2+\left(\frac{821516 \zeta _3^3}{81}-\frac{140833 \zeta _6 \zeta _3}{3}-\frac{157622 \zeta _4
   \zeta _5}{5}-\frac{152636 \zeta _2 \zeta _7}{7}+\frac{1934974 \zeta _9}{9}\right) \epsilon
   ^3+\left(\frac{368986}{9} \zeta _4 \zeta _3^2+\frac{701864}{15} \zeta _2 \zeta _5 \zeta
   _3-\frac{9005524 \zeta _7 \zeta _3}{21}-\frac{4422338 \zeta _5^2}{25}+\frac{18063500159 \zeta
   _{10}}{30720}\right) \epsilon ^4+\left(-\frac{1643032}{81} \zeta _2 \zeta _3^3+\frac{20704988}{45}
   \zeta _5 \zeta _3^2-\frac{7366259785 \zeta _8 \zeta _3}{6912}-\frac{3549469 \zeta _5 \zeta
   _6}{5}-\frac{4044854 \zeta _4 \zeta _7}{7}-\frac{3869948 \zeta _2 \zeta _9}{9}+\frac{48657118 \zeta
   _{11}}{11}\right) \epsilon ^5+\left(-\frac{24234722 \zeta _3^4}{243}+\frac{1595354509 \zeta _6
   \zeta _3^2}{1728}+\frac{18599396}{15} \zeta _4 \zeta _5 \zeta _3+\frac{18011048}{21} \zeta _2 \zeta
   _7 \zeta _3-\frac{228326932 \zeta _9 \zeta _3}{27}+\frac{8844676}{25} \zeta _2 \zeta
   _5^2-\frac{226969732 \zeta _5 \zeta _7}{35}+\frac{1267752375622145 \zeta _{12}}{101892096}\right)
   \epsilon ^6+O\left(\epsilon ^{7}\right)
\end{dmath}
\begin{dmath}
M^{(4)}_{31}=-\frac{1}{\epsilon ^6}+\frac{2 \zeta _2}{\epsilon
   ^4}+\frac{28 \zeta _3}{3 \epsilon ^3}+\frac{8 \zeta _4}{\epsilon ^2}+\frac{\frac{124 \zeta
   _5}{5}-\frac{56 \zeta _2 \zeta _3}{3}}{\epsilon }+\left(\zeta _6-\frac{392 \zeta
   _3^2}{9}\right)+\left(-\frac{224}{3} \zeta _3 \zeta _4-\frac{248 \zeta _2 \zeta _5}{5}+\frac{508
   \zeta _7}{7}\right) \epsilon +\left(\frac{784}{9} \zeta _2 \zeta _3^2-\frac{3472 \zeta _5 \zeta
   _3}{15}-\frac{110 \zeta _8}{3}\right) \epsilon ^2+\left(\frac{10976 \zeta _3^3}{81}-\frac{28 \zeta
   _6 \zeta _3}{3}-\frac{992 \zeta _4 \zeta _5}{5}-\frac{1016 \zeta _2 \zeta _7}{7}+\frac{2044 \zeta
   _9}{9}\right) \epsilon ^3+\left(\frac{3136}{9} \zeta _4 \zeta _3^2+\frac{6944}{15} \zeta _2 \zeta
   _5 \zeta _3-\frac{2032 \zeta _7 \zeta _3}{3}-\frac{7688 \zeta _5^2}{25}-\frac{3742081 \zeta
   _{10}}{30720}\right) \epsilon ^4+\left(-\frac{21952}{81} \zeta _2 \zeta _3^3+\frac{48608}{45} \zeta
   _5 \zeta _3^2+\frac{2364215 \zeta _8 \zeta _3}{6912}-\frac{124 \zeta _5 \zeta _6}{5}-\frac{4064
   \zeta _4 \zeta _7}{7}-\frac{4088 \zeta _2 \zeta _9}{9}+\frac{8188 \zeta _{11}}{11}\right) \epsilon
   ^5+\left(-\frac{76832 \zeta _3^4}{243}+\frac{73549 \zeta _6 \zeta _3^2}{1728}+\frac{27776}{15}
   \zeta _4 \zeta _5 \zeta _3+\frac{4064}{3} \zeta _2 \zeta _7 \zeta _3-\frac{57232 \zeta _9 \zeta
   _3}{27}+\frac{15376}{25} \zeta _2 \zeta _5^2-\frac{62992 \zeta _5 \zeta _7}{35}-\frac{27610067455
   \zeta _{12}}{101892096}\right) \epsilon ^6+O\left(\epsilon ^{7}\right)
\end{dmath}
\begin{dmath}
M^{(4)}_{21}=\frac{6 \zeta _3}{\epsilon
   ^3}+\frac{9 \zeta _4}{\epsilon ^2}+\frac{157 \zeta _5-12 \zeta _2 \zeta _3}{\epsilon }+\left(346
   \zeta _6-235 \zeta _3^2\right)+\left(-669 \zeta _3 \zeta _4-314 \zeta _2 \zeta _5+\frac{26657 \zeta
   _7}{8}\right) \epsilon +\left(470 \zeta _2 \zeta _3^2-\frac{125252 \zeta _5 \zeta
   _3}{15}+\frac{381707 \zeta _8}{48}+243 \zeta _{5,3}\right) \epsilon ^2+\left(\frac{14572 \zeta
   _3^3}{3}-\frac{103805 \zeta _6 \zeta _3}{6}-\frac{134057 \zeta _4 \zeta _5}{10}-\frac{26657 \zeta
   _2 \zeta _7}{4}+\frac{1657525 \zeta _9}{24}\right) \epsilon ^3+\left(20448 \zeta _4 \zeta
   _3^2+\frac{250504}{15} \zeta _2 \zeta _5 \zeta _3-\frac{6645125 \zeta _7 \zeta
   _3}{42}-\frac{37743301 \zeta _5^2}{560}+\frac{47544011 \zeta _{10}}{280}-486 \zeta _2 \zeta
   _{5,3}+\frac{242325 \zeta _{7,3}}{112}\right) \epsilon ^4+\left(-\frac{29144}{3} \zeta _2 \zeta
   _3^3+\frac{2015528}{9} \zeta _5 \zeta _3^2-\frac{13378025 \zeta _8 \zeta _3}{36}-6156 \zeta _{5,3}
   \zeta _3-\frac{5567551 \zeta _5 \zeta _6}{20}-\frac{31946105 \zeta _4 \zeta _7}{112}-\frac{5701045
   \zeta _2 \zeta _9}{12}+\frac{258270245 \zeta _{11}}{128}-7488 \zeta _{5,3,3}\right) \epsilon
   ^5+\left(-\frac{3885575 \zeta _3^4}{54}+\frac{8784121}{18} \zeta _6 \zeta _3^2+\frac{10146703}{15}
   \zeta _4 \zeta _5 \zeta _3+\frac{3060425}{21} \zeta _2 \zeta _7 \zeta _3-\frac{106068821 \zeta _9
   \zeta _3}{36}+\frac{21014701}{280} \zeta _2 \zeta _5^2-\frac{361112943 \zeta _5 \zeta
   _7}{140}+\frac{680114918205 \zeta _{12}}{176896}-44613 \zeta _4 \zeta _{5,3}-\frac{3110085}{56}
   \zeta _2 \zeta _{7,3}+\frac{748115 \zeta _{9,3}}{8}+17070 \zeta _{6,4,1,1}\right) \epsilon
   ^6+O\left(\epsilon ^{7}\right)
\end{dmath}
\begin{dmath}
M^{(4)}_{22}=\frac{6 \zeta _3}{\epsilon ^3}+\frac{9 \zeta _4}{\epsilon
   ^2}+\frac{127 \zeta _5-12 \zeta _2 \zeta _3}{\epsilon }+\left(271 \zeta _6-229 \zeta
   _3^2\right)+\left(-651 \zeta _3 \zeta _4-254 \zeta _2 \zeta _5+\frac{18989 \zeta _7}{8}\right)
   \epsilon +\left(458 \zeta _2 \zeta _3^2-\frac{111242 \zeta _5 \zeta _3}{15}+\frac{1342927 \zeta
   _8}{240}+\frac{243 \zeta _{5,3}}{5}\right) \epsilon ^2+\left(\frac{14182 \zeta _3^3}{3}-\frac{90353
   \zeta _6 \zeta _3}{6}-\frac{107267 \zeta _4 \zeta _5}{10}-\frac{18989 \zeta _2 \zeta
   _7}{4}+\frac{1084927 \zeta _9}{24}\right) \epsilon ^3+\left(19899 \zeta _4 \zeta
   _3^2+\frac{222484}{15} \zeta _2 \zeta _5 \zeta _3-\frac{10860403 \zeta _7 \zeta
   _3}{84}-\frac{31559011 \zeta _5^2}{560}+\frac{305940143 \zeta _{10}}{2800}-\frac{486}{5} \zeta _2
   \zeta _{5,3}+\frac{40851 \zeta _{7,3}}{112}\right) \epsilon ^4+\left(-\frac{28364}{3} \zeta _2
   \zeta _3^3+\frac{9454234}{45} \zeta _5 \zeta _3^2-\frac{21368395 \zeta _8 \zeta _3}{72}-2106 \zeta
   _{5,3} \zeta _3-\frac{4568941 \zeta _5 \zeta _6}{20}-\frac{105510709 \zeta _4 \zeta
   _7}{560}-\frac{1275439 \zeta _2 \zeta _9}{12}+\frac{591012973 \zeta _{11}}{640}-\frac{1764}{5}
   \zeta _{5,3,3}\right) \epsilon ^5+\left(-\frac{3772121 \zeta _3^4}{54}+\frac{3785216}{9} \zeta _6
   \zeta _3^2+\frac{9032572}{15} \zeta _4 \zeta _5 \zeta _3+\frac{10363963}{42} \zeta _2 \zeta _7
   \zeta _3-\frac{43120711 \zeta _9 \zeta _3}{18}+\frac{30400651}{280} \zeta _2 \zeta
   _5^2-\frac{67935699 \zeta _5 \zeta _7}{35}+\frac{1912275391257 \zeta _{12}}{884480}-\frac{25848}{5}
   \zeta _4 \zeta _{5,3}-\frac{239427}{56} \zeta _2 \zeta _{7,3}+\frac{64531 \zeta _{9,3}}{8}+1182
   \zeta _{6,4,1,1}\right) \epsilon ^6+O\left(\epsilon ^{7}\right)
\end{dmath}
\begin{dmath}
M^{(4)}_{26}=\frac{6 \zeta _3}{\epsilon
   ^3}+\frac{9 \zeta _4}{\epsilon ^2}+\frac{102 \zeta _5-12 \zeta _2 \zeta _3}{\epsilon
   }+\left(\frac{417 \zeta _6}{2}-314 \zeta _3^2\right)+\left(-906 \zeta _3 \zeta _4-204 \zeta _2
   \zeta _5+1413 \zeta _7\right) \epsilon +\left(628 \zeta _2 \zeta _3^2-\frac{46964 \zeta _5 \zeta
   _3}{5}+\frac{44831 \zeta _8}{20}+\frac{648 \zeta _{5,3}}{5}\right) \epsilon ^2+\left(\frac{24892
   \zeta _3^3}{3}-18683 \zeta _6 \zeta _3-\frac{72246 \zeta _4 \zeta _5}{5}-2826 \zeta _2 \zeta
   _7+18918 \zeta _9\right) \epsilon ^3+\left(35454 \zeta _4 \zeta _3^2+\frac{93928}{5} \zeta _2 \zeta
   _5 \zeta _3-\frac{995634 \zeta _7 \zeta _3}{7}-\frac{2519121 \zeta _5^2}{35}+\frac{12572799 \zeta
   _{10}}{1400}-\frac{1296}{5} \zeta _2 \zeta _{5,3}+\frac{5751 \zeta _{7,3}}{7}\right) \epsilon
   ^4+\left(-\frac{49784}{3} \zeta _2 \zeta _3^3+\frac{5340148}{15} \zeta _5 \zeta _3^2-\frac{1660235
   \zeta _8 \zeta _3}{6}-5616 \zeta _{5,3} \zeta _3-\frac{1448529 \zeta _5 \zeta _6}{5}-\frac{8019324
   \zeta _4 \zeta _7}{35}-142812 \zeta _2 \zeta _9+\frac{8614689 \zeta _{11}}{20}-\frac{11664}{5}
   \zeta _{5,3,3}\right) \epsilon ^5+\left(-\frac{3990178 \zeta _3^4}{27}+\frac{2119507}{3} \zeta _6
   \zeta _3^2+\frac{5275444}{5} \zeta _4 \zeta _5 \zeta _3+\frac{1809828}{7} \zeta _2 \zeta _7 \zeta
   _3-\frac{6803576 \zeta _9 \zeta _3}{3}+\frac{4720722}{35} \zeta _2 \zeta _5^2-\frac{78493614 \zeta
   _5 \zeta _7}{35}-\frac{16516924143 \zeta _{12}}{55280}-\frac{75168}{5} \zeta _4 \zeta
   _{5,3}-\frac{65934}{7} \zeta _2 \zeta _{7,3}+16812 \zeta _{9,3}+2592 \zeta _{6,4,1,1}\right)
   \epsilon ^6+O\left(\epsilon ^{7}\right)
\end{dmath}
\begin{dmath}
M^{(4)}_{27}=\frac{6 \zeta _3}{\epsilon ^3}+\frac{9 \zeta _4}{\epsilon
   ^2}+\frac{192 \zeta _5-12 \zeta _2 \zeta _3}{\epsilon }+\left(\frac{867 \zeta _6}{2}-260 \zeta
   _3^2\right)+\left(-744 \zeta _3 \zeta _4-384 \zeta _2 \zeta _5+4509 \zeta _7\right) \epsilon
   +\left(520 \zeta _2 \zeta _3^2-\frac{49184 \zeta _5 \zeta _3}{5}+10429 \zeta _8+648 \zeta
   _{5,3}\right) \epsilon ^2+\left(\frac{17872 \zeta _3^3}{3}-20630 \zeta _6 \zeta _3-\frac{92316
   \zeta _4 \zeta _5}{5}-9018 \zeta _2 \zeta _7+98490 \zeta _9\right) \epsilon ^3+\left(25248 \zeta _4
   \zeta _3^2+\frac{98368}{5} \zeta _2 \zeta _5 \zeta _3-\frac{1366620 \zeta _7 \zeta
   _3}{7}-\frac{2796366 \zeta _5^2}{35}+\frac{7796382 \zeta _{10}}{35}-1296 \zeta _2 \zeta
   _{5,3}+\frac{47790 \zeta _{7,3}}{7}\right) \epsilon ^4+\left(-\frac{35744}{3} \zeta _2 \zeta
   _3^3+\frac{831296}{3} \zeta _5 \zeta _3^2-\frac{1331320 \zeta _8 \zeta _3}{3}-16416 \zeta _{5,3}
   \zeta _3-\frac{1721994 \zeta _5 \zeta _6}{5}-\frac{3354030 \zeta _4 \zeta _7}{7}-1246740 \zeta _2
   \zeta _9+3887580 \zeta _{11}-23328 \zeta _{5,3,3}\right) \epsilon ^5+\left(-\frac{2634400 \zeta
   _3^4}{27}+\frac{1846516}{3} \zeta _6 \zeta _3^2+\frac{4541776}{5} \zeta _4 \zeta _5 \zeta
   _3+\frac{918840}{7} \zeta _2 \zeta _7 \zeta _3-\frac{11052848 \zeta _9 \zeta
   _3}{3}+\frac{2417532}{35} \zeta _2 \zeta _5^2-\frac{111405732 \zeta _5 \zeta
   _7}{35}+\frac{6991885065 \zeta _{12}}{1382}-94608 \zeta _4 \zeta _{5,3}-\frac{639900}{7} \zeta _2
   \zeta _{7,3}+189720 \zeta _{9,3}+25920 \zeta _{6,4,1,1}\right) \epsilon ^6+O\left(\epsilon
   ^{7}\right)
\end{dmath}
\begin{dmath}
M^{(4)}_{32}=\frac{6 \zeta _3}{\epsilon ^3}+\frac{9 \zeta _4}{\epsilon ^2}+\frac{102 \zeta _5-12
   \zeta _2 \zeta _3}{\epsilon }+\left(\frac{417 \zeta _6}{2}-134 \zeta _3^2\right)+\left(-366 \zeta
   _3 \zeta _4-204 \zeta _2 \zeta _5+1413 \zeta _7\right) \epsilon +\left(268 \zeta _2 \zeta
   _3^2-\frac{14564 \zeta _5 \zeta _3}{5}+\frac{54281 \zeta _8}{20}+\frac{648 \zeta _{5,3}}{5}\right)
   \epsilon ^2+\left(\frac{4732 \zeta _3^3}{3}-5273 \zeta _6 \zeta _3-\frac{23646 \zeta _4 \zeta
   _5}{5}-2826 \zeta _2 \zeta _7+18918 \zeta _9\right) \epsilon ^3+\left(6294 \zeta _4 \zeta
   _3^2+\frac{29128}{5} \zeta _2 \zeta _5 \zeta _3-\frac{244044 \zeta _7 \zeta _3}{7}-\frac{484221
   \zeta _5^2}{35}+\frac{45732849 \zeta _{10}}{1400}-\frac{1296}{5} \zeta _2 \zeta _{5,3}+\frac{5751
   \zeta _{7,3}}{7}\right) \epsilon ^4+\left(-\frac{9464}{3} \zeta _2 \zeta _3^3+\frac{656188}{15}
   \zeta _5 \zeta _3^2-\frac{380201 \zeta _8 \zeta _3}{6}-1728 \zeta _{5,3} \zeta _3-\frac{246129
   \zeta _5 \zeta _6}{5}-\frac{2382399 \zeta _4 \zeta _7}{35}-142812 \zeta _2 \zeta _9+\frac{8614689
   \zeta _{11}}{20}-\frac{11664}{5} \zeta _{5,3,3}\right) \epsilon ^5+\left(-\frac{355438 \zeta
   _3^4}{27}+\frac{240127}{3} \zeta _6 \zeta _3^2+\frac{640084}{5} \zeta _4 \zeta _5 \zeta
   _3+\frac{306648}{7} \zeta _2 \zeta _7 \zeta _3-\frac{1235096 \zeta _9 \zeta
   _3}{3}+\frac{650922}{35} \zeta _2 \zeta _5^2-\frac{11641164 \zeta _5 \zeta _7}{35}+\frac{35084937
   \zeta _{12}}{80}-\frac{46008}{5} \zeta _4 \zeta _{5,3}-\frac{65934}{7} \zeta _2 \zeta _{7,3}+16812
   \zeta _{9,3}+2592 \zeta _{6,4,1,1}\right) \epsilon ^6+O\left(\epsilon ^{7}\right)
\end{dmath}
\begin{dmath}
M^{(4)}_{33}=\frac{6 \zeta
   _3}{\epsilon ^3}+\frac{9 \zeta _4}{\epsilon ^2}+\frac{192 \zeta _5-12 \zeta _2 \zeta _3}{\epsilon
   }+\left(\frac{867 \zeta _6}{2}-224 \zeta _3^2\right)+\left(-636 \zeta _3 \zeta _4-384 \zeta _2
   \zeta _5+4509 \zeta _7\right) \epsilon +\left(448 \zeta _2 \zeta _3^2-\frac{42164 \zeta _5 \zeta
   _3}{5}+\frac{21047 \zeta _8}{2}+648 \zeta _{5,3}\right) \epsilon ^2+\left(\frac{13516 \zeta
   _3^3}{3}-17678 \zeta _6 \zeta _3-\frac{81786 \zeta _4 \zeta _5}{5}-9018 \zeta _2 \zeta _7+98490
   \zeta _9\right) \epsilon ^3+\left(18930 \zeta _4 \zeta _3^2+\frac{84328}{5} \zeta _2 \zeta _5 \zeta
   _3-\frac{1164894 \zeta _7 \zeta _3}{7}-\frac{2514126 \zeta _5^2}{35}+\frac{31911099 \zeta
   _{10}}{140}-1296 \zeta _2 \zeta _{5,3}+\frac{47790 \zeta _{7,3}}{7}\right) \epsilon
   ^4+\left(-\frac{27032}{3} \zeta _2 \zeta _3^3+\frac{3181888}{15} \zeta _5 \zeta _3^2-\frac{2313683
   \zeta _8 \zeta _3}{6}-12528 \zeta _{5,3} \zeta _3-\frac{1574799 \zeta _5 \zeta _6}{5}-\frac{3051441
   \zeta _4 \zeta _7}{7}-1246740 \zeta _2 \zeta _9+3887580 \zeta _{11}-23328 \zeta _{5,3,3}\right)
   \epsilon ^5+\left(-\frac{1789840 \zeta _3^4}{27}+\frac{1459273}{3} \zeta _6 \zeta
   _3^2+\frac{3463504}{5} \zeta _4 \zeta _5 \zeta _3+\frac{515388}{7} \zeta _2 \zeta _7 \zeta
   _3-\frac{9241688 \zeta _9 \zeta _3}{3}+\frac{1853052}{35} \zeta _2 \zeta _5^2-\frac{102801822 \zeta
   _5 \zeta _7}{35}+\frac{57592302801 \zeta _{12}}{11056}-88776 \zeta _4 \zeta _{5,3}-\frac{639900}{7}
   \zeta _2 \zeta _{7,3}+189720 \zeta _{9,3}+25920 \zeta _{6,4,1,1}\right) \epsilon ^6+O\left(\epsilon
   ^{7}\right)
\end{dmath}
\begin{dmath}
M^{(4)}_{34}=-\frac{20 \zeta _5}{\epsilon }+\left(4 \zeta _3^2-50 \zeta _6\right)+\left(12 \zeta _3
   \zeta _4+40 \zeta _2 \zeta _5-639 \zeta _7\right) \epsilon +\left(-8 \zeta _2 \zeta _3^2+\frac{1928
   \zeta _5 \zeta _3}{3}-\frac{23567 \zeta _8}{15}-\frac{648 \zeta _{5,3}}{5}\right) \epsilon
   ^2+\left(-\frac{440 \zeta _3^3}{3}+\frac{4634 \zeta _6 \zeta _3}{3}+1816 \zeta _4 \zeta _5+1278
   \zeta _2 \zeta _7-\frac{142462 \zeta _9}{9}\right) \epsilon ^3+\left(-636 \zeta _4 \zeta
   _3^2-\frac{3856}{3} \zeta _2 \zeta _5 \zeta _3+18422 \zeta _7 \zeta _3+\frac{56172 \zeta
   _5^2}{7}-\frac{13604091 \zeta _{10}}{350}+\frac{1296}{5} \zeta _2 \zeta _{5,3}-\frac{9276 \zeta
   _{7,3}}{7}\right) \epsilon ^4+\left(\frac{880}{3} \zeta _2 \zeta _3^3-\frac{603104}{45} \zeta _5
   \zeta _3^2+\frac{407428 \zeta _8 \zeta _3}{9}+3024 \zeta _{5,3} \zeta _3+39312 \zeta _5 \zeta
   _6+\frac{323607 \zeta _4 \zeta _7}{5}+\frac{2052668 \zeta _2 \zeta _9}{9}-\frac{10409948 \zeta
   _{11}}{15}+\frac{21824}{5} \zeta _{5,3,3}\right) \epsilon ^5+\left(\frac{31816 \zeta
   _3^4}{9}-\frac{454438}{9} \zeta _6 \zeta _3^2-\frac{317768}{5} \zeta _4 \zeta _5 \zeta
   _3+\frac{160828}{3} \zeta _2 \zeta _7 \zeta _3+\frac{9175892 \zeta _9 \zeta
   _3}{27}+\frac{327800}{21} \zeta _2 \zeta _5^2+\frac{44015786 \zeta _5 \zeta
   _7}{105}-\frac{13947004573 \zeta _{12}}{13820}+\frac{119368}{5} \zeta _4 \zeta
   _{5,3}+\frac{208504}{7} \zeta _2 \zeta _{7,3}-\frac{484088 \zeta _{9,3}}{9}-\frac{27136}{3} \zeta
   _{6,4,1,1}\right) \epsilon ^6+O\left(\epsilon ^{7}\right)
\end{dmath}
\begin{dmath}
M^{(4)}_{36}=-\frac{20 \zeta _5}{\epsilon }+\left(-32
   \zeta _3^2-50 \zeta _6\right)+\left(-96 \zeta _3 \zeta _4+40 \zeta _2 \zeta _5-569 \zeta _7\right)
   \epsilon +\left(64 \zeta _2 \zeta _3^2-\frac{1864 \zeta _5 \zeta _3}{3}-\frac{42619 \zeta
   _8}{30}-\frac{648 \zeta _{5,3}}{5}\right) \epsilon ^2+\left(1096 \zeta _3^3-\frac{3172 \zeta _6
   \zeta _3}{3}-80 \zeta _4 \zeta _5+1138 \zeta _2 \zeta _7-13198 \zeta _9\right) \epsilon
   ^3+\left(4740 \zeta _4 \zeta _3^2+\frac{3728}{3} \zeta _2 \zeta _5 \zeta _3-\frac{20800 \zeta _7
   \zeta _3}{3}-\frac{386 \zeta _5^2}{7}-\frac{24928237 \zeta _{10}}{700}+\frac{1296}{5} \zeta _2
   \zeta _{5,3}-\frac{7866 \zeta _{7,3}}{7}\right) \epsilon ^4+\left(-2192 \zeta _2 \zeta
   _3^3+\frac{1982152}{45} \zeta _5 \zeta _3^2-\frac{836611 \zeta _8 \zeta _3}{45}+\frac{24192}{5}
   \zeta _{5,3} \zeta _3+23848 \zeta _5 \zeta _6+\frac{5432 \zeta _4 \zeta _7}{5}-104068 \zeta _2
   \zeta _9-\frac{725897 \zeta _{11}}{10}-\frac{14496}{5} \zeta _{5,3,3}\right) \epsilon
   ^5+\left(-\frac{171236 \zeta _3^4}{9}+\frac{670394}{9} \zeta _6 \zeta _3^2+\frac{1346872}{15} \zeta
   _4 \zeta _5 \zeta _3+\frac{181760}{3} \zeta _2 \zeta _7 \zeta _3-\frac{1327808 \zeta _9 \zeta
   _3}{9}+\frac{115236}{7} \zeta _2 \zeta _5^2+\frac{4318112 \zeta _5 \zeta _7}{35}-\frac{24059482357
   \zeta _{12}}{27640}+\frac{45696}{5} \zeta _4 \zeta _{5,3}+\frac{113844}{7} \zeta _2 \zeta
   _{7,3}-\frac{95936 \zeta _{9,3}}{3}-4672 \zeta _{6,4,1,1}\right) \epsilon ^6+O\left(\epsilon
   ^{11}\right)
\end{dmath}
\begin{dmath}
M^{(4)}_{35}=-\frac{20 \zeta _5}{\epsilon }+\left(-32 \zeta _3^2-50 \zeta _6\right)+\left(-96 \zeta
   _3 \zeta _4+40 \zeta _2 \zeta _5-639 \zeta _7\right) \epsilon +\left(64 \zeta _2 \zeta
   _3^2-\frac{1744 \zeta _5 \zeta _3}{3}-\frac{49969 \zeta _8}{30}-\frac{648 \zeta _{5,3}}{5}\right)
   \epsilon ^2+\left(\frac{4168 \zeta _3^3}{3}-\frac{2872 \zeta _6 \zeta _3}{3}-20 \zeta _4 \zeta
   _5+1278 \zeta _2 \zeta _7-\frac{285749 \zeta _9}{18}\right) \epsilon ^3+\left(6060 \zeta _4 \zeta
   _3^2+\frac{3488}{3} \zeta _2 \zeta _5 \zeta _3-235 \zeta _7 \zeta _3-\frac{89679 \zeta
   _5^2}{28}-\frac{61813449 \zeta _{10}}{1400}+\frac{1296}{5} \zeta _2 \zeta _{5,3}-\frac{34989 \zeta
   _{7,3}}{28}\right) \epsilon ^4+\left(-\frac{8336}{3} \zeta _2 \zeta _3^3+\frac{2672752}{45} \zeta
   _5 \zeta _3^2+\frac{724093 \zeta _8 \zeta _3}{90}+\frac{20952}{5} \zeta _{5,3} \zeta _3-4452 \zeta
   _5 \zeta _6+\frac{606513 \zeta _4 \zeta _7}{20}+\frac{1402253 \zeta _2 \zeta _9}{9}-\frac{277291201
   \zeta _{11}}{480}+\frac{13784}{5} \zeta _{5,3,3}\right) \epsilon ^5+\left(-\frac{274466 \zeta
   _3^4}{9}+\frac{973484}{9} \zeta _6 \zeta _3^2+\frac{707024}{5} \zeta _4 \zeta _5 \zeta
   _3+\frac{113170}{3} \zeta _2 \zeta _7 \zeta _3+\frac{2840891 \zeta _9 \zeta
   _3}{27}+\frac{816661}{42} \zeta _2 \zeta _5^2+\frac{5585366 \zeta _5 \zeta
   _7}{105}-\frac{253292937191 \zeta _{12}}{221120}+\frac{76156}{5} \zeta _4 \zeta
   _{5,3}+\frac{191453}{14} \zeta _2 \zeta _{7,3}-\frac{515761 \zeta _{9,3}}{18}-\frac{11176}{3} \zeta
   _{6,4,1,1}\right) \epsilon ^6+O\left(\epsilon ^{7}\right)
\end{dmath}
\begin{dmath}
M^{(4)}_{41}=-\frac{20 \zeta _5}{\epsilon }+\left(-86
   \zeta _3^2-50 \zeta _6\right)+\left(-258 \zeta _3 \zeta _4+40 \zeta _2 \zeta _5-\frac{5777 \zeta
   _7}{8}\right) \epsilon +\left(172 \zeta _2 \zeta _3^2+\frac{20498 \zeta _5 \zeta
   _3}{3}-\frac{3244097 \zeta _8}{240}+\frac{22977 \zeta _{5,3}}{5}\right) \epsilon
   ^2+\left(\frac{15350 \zeta _3^3}{3}+\frac{55244 \zeta _6 \zeta _3}{3}-\frac{48673 \zeta _4 \zeta
   _5}{2}+\frac{5777 \zeta _2 \zeta _7}{4}-\frac{198659 \zeta _9}{8}\right) \epsilon ^3+\left(22509
   \zeta _4 \zeta _3^2-\frac{40996}{3} \zeta _2 \zeta _5 \zeta _3+\frac{3207559 \zeta _7 \zeta
   _3}{12}+\frac{1697411 \zeta _5^2}{16}-\frac{206969029 \zeta _{10}}{400}-\frac{45954}{5} \zeta _2
   \zeta _{5,3}+\frac{868257 \zeta _{7,3}}{16}\right) \epsilon ^4+\left(-\frac{30700}{3} \zeta _2
   \zeta _3^3-\frac{4794914}{45} \zeta _5 \zeta _3^2+\frac{89207513 \zeta _8 \zeta _3}{72}-159282
   \zeta _{5,3} \zeta _3-\frac{87707 \zeta _5 \zeta _6}{4}-\frac{72520703 \zeta _4 \zeta
   _7}{80}-\frac{9613357 \zeta _2 \zeta _9}{4}+\frac{2166134937 \zeta _{11}}{640}-\frac{272556}{5}
   \zeta _{5,3,3}\right) \epsilon ^5+\left(-\frac{2412083 \zeta _3^4}{18}-\frac{6372958}{9} \zeta _6
   \zeta _3^2+\frac{12660136}{15} \zeta _4 \zeta _5 \zeta _3+\frac{5425361}{6} \zeta _2 \zeta _7 \zeta
   _3+\frac{102364603 \zeta _9 \zeta _3}{18}+\frac{2331285}{8} \zeta _2 \zeta _5^2+\frac{778468933
   \zeta _5 \zeta _7}{140}-\frac{14101928102987 \zeta _{12}}{884480}-\frac{386472}{5} \zeta _4 \zeta
   _{5,3}+\frac{2584911}{8} \zeta _2 \zeta _{7,3}+\frac{2327797 \zeta _{9,3}}{24}-143882 \zeta
   _{6,4,1,1}\right) \epsilon ^6+O\left(\epsilon ^{7}\right)
\end{dmath}
\begin{dmath}
M^{(4)}_{42}=-\frac{20 \zeta _5}{\epsilon }+\left(-80
   \zeta _3^2-50 \zeta _6\right)+\left(-240 \zeta _3 \zeta _4+40 \zeta _2 \zeta _5-625 \zeta _7\right)
   \epsilon +\left(160 \zeta _2 \zeta _3^2+\frac{14000 \zeta _5 \zeta _3}{3}-\frac{59315 \zeta
   _8}{6}+3240 \zeta _{5,3}\right) \epsilon ^2+\left(\frac{11000 \zeta _3^3}{3}+\frac{38720 \zeta _6
   \zeta _3}{3}-17420 \zeta _4 \zeta _5+1250 \zeta _2 \zeta _7-\frac{109895 \zeta _9}{6}\right)
   \epsilon ^3+\left(16020 \zeta _4 \zeta _3^2-\frac{28000}{3} \zeta _2 \zeta _5 \zeta _3+\frac{455665
   \zeta _7 \zeta _3}{3}+\frac{1974593 \zeta _5^2}{28}-\frac{17704613 \zeta _{10}}{56}-6480 \zeta _2
   \zeta _{5,3}+\frac{907155 \zeta _{7,3}}{28}\right) \epsilon ^4+\left(-\frac{22000}{3} \zeta _2
   \zeta _3^3-\frac{747544}{9} \zeta _5 \zeta _3^2+\frac{12900791 \zeta _8 \zeta _3}{18}-99576 \zeta
   _{5,3} \zeta _3+70420 \zeta _5 \zeta _6-\frac{2373199 \zeta _4 \zeta _7}{4}-\frac{6055825 \zeta _2
   \zeta _9}{3}+\frac{95731961 \zeta _{11}}{32}-45672 \zeta _{5,3,3}\right) \epsilon
   ^5+\left(-\frac{918910 \zeta _3^4}{9}+\frac{1066460}{9} \zeta _6 \zeta _3^2+\frac{1850576}{3} \zeta
   _4 \zeta _5 \zeta _3-\frac{6384530}{3} \zeta _2 \zeta _7 \zeta _3+\frac{17142865 \zeta _9 \zeta
   _3}{3}-\frac{10914153}{14} \zeta _2 \zeta _5^2+\frac{23004085 \zeta _5 \zeta
   _7}{14}-\frac{198489001243 \zeta _{12}}{44224}-472092 \zeta _4 \zeta _{5,3}-\frac{8569635}{14}
   \zeta _2 \zeta _{7,3}+\frac{6730525 \zeta _{9,3}}{6}+182440 \zeta _{6,4,1,1}\right) \epsilon
   ^6+O\left(\epsilon ^{7}\right)
   \end{dmath}
   \begin{dmath}
M^{(4)}_{43}=-\frac{20 \zeta _5}{\epsilon }+\left(-68 \zeta _3^2-50 \zeta
   _6\right)+\left(-204 \zeta _3 \zeta _4+40 \zeta _2 \zeta _5-450 \zeta _7\right) \epsilon +\left(136
   \zeta _2 \zeta _3^2+\frac{9704 \zeta _5 \zeta _3}{3}-\frac{173899 \zeta _8}{30}+\frac{9072 \zeta
   _{5,3}}{5}\right) \epsilon ^2+\left(\frac{12664 \zeta _3^3}{3}+\frac{27422 \zeta _6 \zeta
   _3}{3}-8876 \zeta _4 \zeta _5+900 \zeta _2 \zeta _7-\frac{88036 \zeta _9}{9}\right) \epsilon
   ^3+\left(18588 \zeta _4 \zeta _3^2-\frac{19408}{3} \zeta _2 \zeta _5 \zeta _3+74852 \zeta _7 \zeta
   _3+\frac{254322 \zeta _5^2}{7}-\frac{83344677 \zeta _{10}}{700}-\frac{18144}{5} \zeta _2 \zeta
   _{5,3}+\frac{85026 \zeta _{7,3}}{7}\right) \epsilon ^4+\left(-\frac{25328}{3} \zeta _2 \zeta
   _3^3-\frac{2768072}{45} \zeta _5 \zeta _3^2+\frac{19958579 \zeta _8 \zeta _3}{45}-\frac{429408}{5}
   \zeta _{5,3} \zeta _3+85614 \zeta _5 \zeta _6-\frac{971628 \zeta _4 \zeta _7}{5}-\frac{6415384
   \zeta _2 \zeta _9}{9}+\frac{31418849 \zeta _{11}}{30}-\frac{81376}{5} \zeta _{5,3,3}\right)
   \epsilon ^5+\left(-\frac{1168844 \zeta _3^4}{9}-\frac{1947214}{9} \zeta _6 \zeta
   _3^2+\frac{2433576}{5} \zeta _4 \zeta _5 \zeta _3-\frac{913112}{3} \zeta _2 \zeta _7 \zeta
   _3+\frac{40371584 \zeta _9 \zeta _3}{27}-\frac{2662732}{21} \zeta _2 \zeta _5^2+\frac{22012300
   \zeta _5 \zeta _7}{21}-\frac{34479208947 \zeta _{12}}{27640}-\frac{827744}{5} \zeta _4 \zeta
   _{5,3}-\frac{494852}{7} \zeta _2 \zeta _{7,3}+\frac{1393864 \zeta _{9,3}}{9}+\frac{46400}{3} \zeta
   _{6,4,1,1}\right) \epsilon ^6+O\left(\epsilon ^{7}\right)
\end{dmath}
\begin{dmath}
M^{(4)}_{44}=\frac{441 \zeta _7 \epsilon }{8}+\left(-135 \zeta _3 \zeta _5+\frac{18567 \zeta _8}{80}-\frac{81 \zeta
   _{5,3}}{5}\right) \epsilon ^2+\left(-267 \zeta _3^3-\frac{675 \zeta _6 \zeta _3}{2}-81 \zeta _4
   \zeta _5-\frac{441 \zeta _2 \zeta _7}{4}+\frac{4583 \zeta _9}{2}\right) \epsilon
   ^3+\left(-\frac{2403}{2} \zeta _4 \zeta _3^2+270 \zeta _2 \zeta _5 \zeta _3-\frac{1293 \zeta _7
   \zeta _3}{4}-\frac{502287 \zeta _5^2}{56}+\frac{60113673 \zeta _{10}}{5600}+\frac{162}{5} \zeta _2
   \zeta _{5,3}-\frac{18441 \zeta _{7,3}}{56}\right) \epsilon ^4+\left(534 \zeta _2 \zeta _3^3-13422
   \zeta _5 \zeta _3^2-\frac{80673 \zeta _8 \zeta _3}{20}+\frac{6426}{5} \zeta _{5,3} \zeta _3-33198
   \zeta _5 \zeta _6-\frac{15759 \zeta _4 \zeta _7}{20}-117821 \zeta _2 \zeta _9+\frac{42903141 \zeta
   _{11}}{160}-\frac{12582}{5} \zeta _{5,3,3}\right) \epsilon ^5+\left(6729 \zeta _3^4-\frac{12945}{4}
   \zeta _6 \zeta _3^2-45357 \zeta _4 \zeta _5 \zeta _3-\frac{213027}{2} \zeta _2 \zeta _7 \zeta
   _3+159178 \zeta _9 \zeta _3-\frac{547881}{28} \zeta _2 \zeta _5^2-\frac{5290221 \zeta _5 \zeta
   _7}{10}+\frac{27703672977 \zeta _{12}}{44224}-18018 \zeta _4 \zeta _{5,3}-\frac{881703}{28} \zeta
   _2 \zeta _{7,3}+\frac{68177 \zeta _{9,3}}{2}+10716 \zeta _{6,4,1,1}\right) \epsilon
   ^6+O\left(\epsilon ^{7}\right)
\end{dmath}
\begin{dmath}
M^{(4)}_{52}=-\frac{20 \zeta _5}{\epsilon }+\left(-68 \zeta _3^2-50
   \zeta _6\right)+\left(-204 \zeta _3 \zeta _4+40 \zeta _2 \zeta _5-450 \zeta _7\right) \epsilon
   +\left(136 \zeta _2 \zeta _3^2+\frac{7904 \zeta _5 \zeta _3}{3}-\frac{173899 \zeta
   _8}{30}+\frac{9072 \zeta _{5,3}}{5}\right) \epsilon ^2+\left(\frac{6544 \zeta _3^3}{3}+\frac{22922
   \zeta _6 \zeta _3}{3}-9776 \zeta _4 \zeta _5+900 \zeta _2 \zeta _7-\frac{88036 \zeta _9}{9}\right)
   \epsilon ^3+\left(9408 \zeta _4 \zeta _3^2-\frac{15808}{3} \zeta _2 \zeta _5 \zeta _3+61352 \zeta
   _7 \zeta _3+\frac{174522 \zeta _5^2}{7}-\frac{85077177 \zeta _{10}}{700}-\frac{18144}{5} \zeta _2
   \zeta _{5,3}+\frac{85026 \zeta _{7,3}}{7}\right) \epsilon ^4+\left(-\frac{13088}{3} \zeta _2 \zeta
   _3^3-\frac{550472}{45} \zeta _5 \zeta _3^2+\frac{11651174 \zeta _8 \zeta _3}{45}-\frac{157248}{5}
   \zeta _{5,3} \zeta _3+33264 \zeta _5 \zeta _6-\frac{1072878 \zeta _4 \zeta _7}{5}-\frac{6415384
   \zeta _2 \zeta _9}{9}+\frac{31418849 \zeta _{11}}{30}-\frac{81376}{5} \zeta _{5,3,3}\right)
   \epsilon ^5+\left(-\frac{304484 \zeta _3^4}{9}-\frac{411544}{9} \zeta _6 \zeta
   _3^2+\frac{1113576}{5} \zeta _4 \zeta _5 \zeta _3-\frac{832112}{3} \zeta _2 \zeta _7 \zeta
   _3+\frac{32448344 \zeta _9 \zeta _3}{27}-\frac{2183932}{21} \zeta _2 \zeta _5^2+\frac{12077200
   \zeta _5 \zeta _7}{21}-\frac{22121494701 \zeta _{12}}{13820}-\frac{419504}{5} \zeta _4 \zeta
   _{5,3}-\frac{494852}{7} \zeta _2 \zeta _{7,3}+\frac{1393864 \zeta _{9,3}}{9}+\frac{46400}{3} \zeta
   _{6,4,1,1}\right) \epsilon ^6+O\left(\epsilon ^{7}\right)
\end{dmath}
\begin{dmath}
M^{(4)}_{45}=-\frac{20 \zeta _5}{\epsilon }+\left(-92
   \zeta _3^2-50 \zeta _6\right)+\left(-276 \zeta _3 \zeta _4+40 \zeta _2 \zeta _5-779 \zeta _7\right)
   \epsilon +\left(184 \zeta _2 \zeta _3^2+\frac{27296 \zeta _5 \zeta _3}{3}-\frac{253937 \zeta
   _8}{15}+\frac{29592 \zeta _{5,3}}{5}\right) \epsilon ^2+\left(5136 \zeta _3^3+\frac{72518 \zeta _6
   \zeta _3}{3}-30860 \zeta _4 \zeta _5+1558 \zeta _2 \zeta _7-\frac{246992 \zeta _9}{9}\right)
   \epsilon ^3+\left(22560 \zeta _4 \zeta _3^2-\frac{54592}{3} \zeta _2 \zeta _5 \zeta
   _3+\frac{1109690 \zeta _7 \zeta _3}{3}+\frac{1064089 \zeta _5^2}{7}-\frac{121454213 \zeta
   _{10}}{175}-\frac{59184}{5} \zeta _2 \zeta _{5,3}+\frac{526299 \zeta _{7,3}}{7}\right) \epsilon
   ^4+\left(-10272 \zeta _2 \zeta _3^3-\frac{12613688}{45} \zeta _5 \zeta _3^2+\frac{82366679 \zeta _8
   \zeta _3}{45}-\frac{1357488}{5} \zeta _{5,3} \zeta _3-373462 \zeta _5 \zeta _6-\frac{4154878 \zeta
   _4 \zeta _7}{5}+\frac{27190288 \zeta _2 \zeta _9}{9}-\frac{345840997 \zeta
   _{11}}{60}+\frac{329584}{5} \zeta _{5,3,3}\right) \epsilon ^5+\left(-\frac{322412 \zeta
   _3^4}{3}-\frac{17691136}{9} \zeta _6 \zeta _3^2+\frac{14926832}{15} \zeta _4 \zeta _5 \zeta
   _3+\frac{13067980}{3} \zeta _2 \zeta _7 \zeta _3+\frac{123344476 \zeta _9 \zeta
   _3}{27}+\frac{31069498}{21} \zeta _2 \zeta _5^2+\frac{1056218026 \zeta _5 \zeta
   _7}{105}-\frac{372374044931 \zeta _{12}}{13820}+\frac{2457536}{5} \zeta _4 \zeta
   _{5,3}+\frac{9648554}{7} \zeta _2 \zeta _{7,3}-\frac{9972628 \zeta _{9,3}}{9}-\frac{1528736}{3}
   \zeta _{6,4,1,1}\right) \epsilon ^6+O\left(\epsilon ^{7}\right)
\end{dmath}
\begin{dmath}
M^{(4)}_{51}=-\frac{21 \zeta _7 \epsilon }{2}+\left(420 \zeta _3 \zeta _5-\frac{12219 \zeta _8}{20}+\frac{1188
   \zeta _{5,3}}{5}\right) \epsilon ^2+\left(308 \zeta _3^3+1050 \zeta _6 \zeta _3-1152 \zeta _4 \zeta
   _5+21 \zeta _2 \zeta _7-\frac{2053 \zeta _9}{2}\right) \epsilon ^3+\left(1386 \zeta _4 \zeta
   _3^2-840 \zeta _2 \zeta _5 \zeta _3+\frac{18193 \zeta _7 \zeta _3}{2}+\frac{62373 \zeta
   _5^2}{4}-\frac{1505187 \zeta _{10}}{50}-\frac{2376}{5} \zeta _2 \zeta _{5,3}+\frac{13095 \zeta
   _{7,3}}{4}\right) \epsilon ^4+\left(-616 \zeta _2 \zeta _3^3-5942 \zeta _5 \zeta _3^2+\frac{1215841
   \zeta _8 \zeta _3}{20}-\frac{69732}{5} \zeta _{5,3} \zeta _3+25614 \zeta _5 \zeta _6-\frac{220647
   \zeta _4 \zeta _7}{5}+167617 \zeta _2 \zeta _9-\frac{25965659 \zeta _{11}}{80}+\frac{18396}{5}
   \zeta _{5,3,3}\right) \epsilon ^5+\left(-\frac{29635 \zeta _3^4}{3}-78590 \zeta _6 \zeta _3^2+76720
   \zeta _4 \zeta _5 \zeta _3+233247 \zeta _2 \zeta _7 \zeta _3-\frac{750950 \zeta _9 \zeta
   _3}{9}+\frac{113635}{2} \zeta _2 \zeta _5^2+\frac{3931247 \zeta _5 \zeta _7}{5}-\frac{17710326429
   \zeta _{12}}{13820}+\frac{118704}{5} \zeta _4 \zeta _{5,3}+\frac{137769}{2} \zeta _2 \zeta
   _{7,3}-\frac{179111 \zeta _{9,3}}{3}-25144 \zeta _{6,4,1,1}\right) \epsilon ^6+O\left(\epsilon ^{7}\right)
\end{dmath}
\begin{dmath}
M^{(4)}_{61}=-\frac{161 \zeta _7 \epsilon }{8}+\left(-50 \zeta _3 \zeta _5-\frac{8767 \zeta _8}{80}+\frac{81 \zeta
   _{5,3}}{5}\right) \epsilon ^2+\left(\frac{1402 \zeta _3^3}{3}-125 \zeta _6 \zeta _3-\frac{393 \zeta
   _4 \zeta _5}{2}+\frac{161 \zeta _2 \zeta _7}{4}-\frac{36131 \zeta _9}{24}\right) \epsilon
   ^3+\left(2103 \zeta _4 \zeta _3^2+100 \zeta _2 \zeta _5 \zeta _3+\frac{58175 \zeta _7 \zeta
   _3}{6}-\frac{778221 \zeta _5^2}{112}-\frac{17423337 \zeta _{10}}{1400}-\frac{162}{5} \zeta _2 \zeta
   _{5,3}+\frac{81297 \zeta _{7,3}}{112}\right) \epsilon ^4+\left(-\frac{2804}{3} \zeta _2 \zeta
   _3^3+\frac{128357}{3} \zeta _5 \zeta _3^2+\frac{191669 \zeta _8 \zeta _3}{6}+2700 \zeta _{5,3}
   \zeta _3-\frac{171343 \zeta _5 \zeta _6}{4}-\frac{236399 \zeta _4 \zeta _7}{80}-\frac{348133 \zeta
   _2 \zeta _9}{12}-\frac{45887177 \zeta _{11}}{1920}-\frac{3558}{5} \zeta _{5,3,3}\right) \epsilon
   ^5+\left(-\frac{42104 \zeta _3^4}{3}+\frac{317671}{6} \zeta _6 \zeta _3^2+\frac{294652}{3} \zeta _4
   \zeta _5 \zeta _3+\frac{517205}{3} \zeta _2 \zeta _7 \zeta _3+\frac{10621247 \zeta _9 \zeta
   _3}{54}+\frac{13612111}{168} \zeta _2 \zeta _5^2+\frac{3203981 \zeta _5 \zeta
   _7}{60}-\frac{2771942085673 \zeta _{12}}{2653440}+\frac{159244}{5} \zeta _4 \zeta
   _{5,3}+\frac{3140831}{56} \zeta _2 \zeta _{7,3}-\frac{4175797 \zeta _{9,3}}{72}-\frac{57538}{3}
   \zeta _{6,4,1,1}\right) \epsilon ^6+O\left(\epsilon ^{7}\right)
\end{dmath}
\begin{dmath}
M^{(4)}_{62}=-\frac{161 \zeta _7 \epsilon }{8}+\left(-470 \zeta _3 \zeta _5+\frac{32993 \zeta _8}{80}-\frac{999
   \zeta _{5,3}}{5}\right) \epsilon ^2+\left(-\frac{1238 \zeta _3^3}{3}-1175 \zeta _6 \zeta
   _3+\frac{1587 \zeta _4 \zeta _5}{2}+\frac{161 \zeta _2 \zeta _7}{4}+\frac{9077 \zeta _9}{72}\right)
   \epsilon ^3+\left(-1857 \zeta _4 \zeta _3^2+940 \zeta _2 \zeta _5 \zeta _3-\frac{348275 \zeta _7
   \zeta _3}{12}-\frac{1018651 \zeta _5^2}{112}+\frac{114458171 \zeta _{10}}{2800}+\frac{1998}{5}
   \zeta _2 \zeta _{5,3}-\frac{552873 \zeta _{7,3}}{112}\right) \epsilon ^4+\left(\frac{2476}{3} \zeta
   _2 \zeta _3^3-\frac{87808}{3} \zeta _5 \zeta _3^2-\frac{1960951 \zeta _8 \zeta _3}{24}-6246 \zeta
   _{5,3} \zeta _3-\frac{50663 \zeta _5 \zeta _6}{4}+\frac{8365741 \zeta _4 \zeta
   _7}{80}+\frac{24055051 \zeta _2 \zeta _9}{36}-\frac{2078568857 \zeta _{11}}{1920}+\frac{74272}{5}
   \zeta _{5,3,3}\right) \epsilon ^5+\left(-\frac{511969 \zeta _3^4}{18}+\frac{2564528}{3} \zeta _6
   \zeta _3^2+\frac{480932}{3} \zeta _4 \zeta _5 \zeta _3-\frac{8047655}{2} \zeta _2 \zeta _7 \zeta
   _3+\frac{195411407 \zeta _9 \zeta _3}{54}-\frac{236958199}{168} \zeta _2 \zeta _5^2-\frac{187321769
   \zeta _5 \zeta _7}{60}+\frac{7039832697729 \zeta _{12}}{884480}-\frac{3002836}{5} \zeta _4 \zeta
   _{5,3}-\frac{68022599}{56} \zeta _2 \zeta _{7,3}+\frac{114991963 \zeta
   _{9,3}}{72}+\frac{1224562}{3} \zeta _{6,4,1,1}\right) \epsilon ^6+O\left(\epsilon ^{7}\right)
\end{dmath}
\begin{dmath}
M^{(4)}_{63}=\frac{21 \zeta _7 \epsilon }{4}+\left(615 \zeta _3 \zeta _5-\frac{29541 \zeta _8}{40}+\frac{1566 \zeta
   _{5,3}}{5}\right) \epsilon ^2+\left(-149 \zeta _3^3+\frac{3075 \zeta _6 \zeta _3}{2}-\frac{2853
   \zeta _4 \zeta _5}{2}-\frac{21 \zeta _2 \zeta _7}{2}+\frac{12539 \zeta _9}{72}\right) \epsilon
   ^3+\left(-\frac{1341}{2} \zeta _4 \zeta _3^2-1230 \zeta _2 \zeta _5 \zeta _3+\frac{238499 \zeta _7
   \zeta _3}{8}+\frac{56439 \zeta _5^2}{16}-\frac{28515719 \zeta _{10}}{800}-\frac{3132}{5} \zeta _2
   \zeta _{5,3}+\frac{73725 \zeta _{7,3}}{16}\right) \epsilon ^4+\left(298 \zeta _2 \zeta
   _3^3-\frac{108463}{2} \zeta _5 \zeta _3^2+\frac{14418683 \zeta _8 \zeta _3}{80}-\frac{181179}{5}
   \zeta _{5,3} \zeta _3-\frac{752573 \zeta _5 \zeta _6}{4}+\frac{604131 \zeta _4 \zeta
   _7}{5}+\frac{93321169 \zeta _2 \zeta _9}{36}-\frac{4163459491 \zeta _{11}}{960}+\frac{288067}{5}
   \zeta _{5,3,3}\right) \epsilon ^5+\left(\frac{869195 \zeta _3^4}{36}-500720 \zeta _6 \zeta
   _3^2+\frac{92555}{3} \zeta _4 \zeta _5 \zeta _3+\frac{18943463}{12} \zeta _2 \zeta _7 \zeta
   _3-\frac{107580425 \zeta _9 \zeta _3}{108}+\frac{13591955}{24} \zeta _2 \zeta _5^2+\frac{165153311
   \zeta _5 \zeta _7}{120}-\frac{414702778929 \zeta _{12}}{110560}+\frac{1398413}{5} \zeta _4 \zeta
   _{5,3}+\frac{3858067}{8} \zeta _2 \zeta _{7,3}-\frac{21789949 \zeta _{9,3}}{36}-\frac{491474}{3}
   \zeta _{6,4,1,1}\right) \epsilon ^6+O\left(\epsilon ^{7}\right)
\end{dmath}
\begin{dmath}
\bar{M}^{(4)}_{51}=140 \zeta _7 \epsilon +\left(250 \zeta _3 \zeta _5+\frac{362 \zeta _8}{5}+\frac{864 \zeta
   _{5,3}}{5}\right) \epsilon ^2+\left(-\frac{290 \zeta _3^3}{3}+625 \zeta _6 \zeta _3-921 \zeta _4
   \zeta _5-280 \zeta _2 \zeta _7+\frac{185377 \zeta _9}{36}\right) \epsilon ^3+\left(-435 \zeta _4
   \zeta _3^2-500 \zeta _2 \zeta _5 \zeta _3-\frac{6619 \zeta _7 \zeta _3}{6}+\frac{604879 \zeta
   _5^2}{56}-\frac{304201 \zeta _{10}}{2800}-\frac{1728}{5} \zeta _2 \zeta _{5,3}+\frac{131421 \zeta
   _{7,3}}{56}\right) \epsilon ^4+\left(\frac{580}{3} \zeta _2 \zeta _3^3-\frac{60520}{3} \zeta _5
   \zeta _3^2+\frac{747451 \zeta _8 \zeta _3}{30}-\frac{69948}{5} \zeta _{5,3} \zeta _3-\frac{13381
   \zeta _5 \zeta _6}{2}-\frac{551713 \zeta _4 \zeta _7}{40}+\frac{9616919 \zeta _2 \zeta
   _9}{18}-\frac{747744919 \zeta _{11}}{960}+\frac{60508}{5} \zeta _{5,3,3}\right) \epsilon
   ^5+\left(\frac{58309 \zeta _3^4}{9}-\frac{580265}{6} \zeta _6 \zeta _3^2+\frac{105356}{3} \zeta _4
   \zeta _5 \zeta _3+217873 \zeta _2 \zeta _7 \zeta _3-\frac{19969957 \zeta _9 \zeta
   _3}{54}+\frac{4525963}{84} \zeta _2 \zeta _5^2+\frac{2920777 \zeta _5 \zeta
   _7}{6}-\frac{168884485933 \zeta _{12}}{442240}+\frac{152774}{5} \zeta _4 \zeta
   _{5,3}+\frac{1680179}{28} \zeta _2 \zeta _{7,3}-\frac{2091103 \zeta _{9,3}}{36}-\frac{64700}{3}
   \zeta _{6,4,1,1}\right) \epsilon ^6+O\left(\epsilon ^{7}\right)
\end{dmath}
\begin{dmath}
\bar{M}^{(4)}_{61}=70 \zeta _7 \epsilon +\left(350 \zeta _3 \zeta _5+245 \zeta _8\right) \epsilon ^2+\left(\frac{2402
   \zeta _3^3}{3}+875 \zeta _6 \zeta _3+525 \zeta _4 \zeta _5-140 \zeta _2 \zeta _7+\frac{51095 \zeta
   _9}{36}\right) \epsilon ^3+\left(3603 \zeta _4 \zeta _3^2-700 \zeta _2 \zeta _5 \zeta
   _3+\frac{150241 \zeta _7 \zeta _3}{6}-\frac{551055 \zeta _5^2}{56}-\frac{365447 \zeta
   _{10}}{112}+\frac{58515 \zeta _{7,3}}{56}\right) \epsilon ^4+\left(-\frac{4804}{3} \zeta _2 \zeta
   _3^3+\frac{185122}{3} \zeta _5 \zeta _3^2+\frac{913123 \zeta _8 \zeta _3}{15}+\frac{56052}{5} \zeta
   _{5,3} \zeta _3-\frac{123015 \zeta _5 \zeta _6}{2}+\frac{69929 \zeta _4 \zeta _7}{8}-\frac{2021015
   \zeta _2 \zeta _9}{18}+\frac{32391751 \zeta _{11}}{192}-2432 \zeta _{5,3,3}\right) \epsilon
   ^5+\left(-\frac{199112 \zeta _3^4}{9}+\frac{1035047}{6} \zeta _6 \zeta _3^2+111334 \zeta _4 \zeta
   _5 \zeta _3-\frac{641641}{3} \zeta _2 \zeta _7 \zeta _3+\frac{27280849 \zeta _9 \zeta
   _3}{27}-\frac{1054185}{28} \zeta _2 \zeta _5^2-\frac{491633 \zeta _5 \zeta
   _7}{2}-\frac{599907610933 \zeta _{12}}{1326720}-\frac{57012}{5} \zeta _4 \zeta
   _{5,3}-\frac{1434435}{28} \zeta _2 \zeta _{7,3}+\frac{395575 \zeta _{9,3}}{4}+16380 \zeta
   _{6,4,1,1}\right) \epsilon ^6+O\left(\epsilon ^{7}\right)
\end{dmath}
\begin{dmath}
\bar{M}^{(4)}_{62}=70 \zeta _7 \epsilon +\left(-400 \zeta _3 \zeta _5+1289 \zeta _8-432 \zeta _{5,3}\right) \epsilon
   ^2+\left(-\frac{4528 \zeta _3^3}{3}-1000 \zeta _6 \zeta _3+2640 \zeta _4 \zeta _5-140 \zeta _2
   \zeta _7+\frac{58460 \zeta _9}{9}\right) \epsilon ^3+\left(-6792 \zeta _4 \zeta _3^2+800 \zeta _2
   \zeta _5 \zeta _3-\frac{176527 \zeta _7 \zeta _3}{3}-\frac{143960 \zeta _5^2}{7}+\frac{9209341
   \zeta _{10}}{70}+864 \zeta _2 \zeta _{5,3}-\frac{88260 \zeta _{7,3}}{7}\right) \epsilon
   ^4+\left(\frac{9056}{3} \zeta _2 \zeta _3^3-\frac{612548}{3} \zeta _5 \zeta _3^2-\frac{1006669
   \zeta _8 \zeta _3}{30}-\frac{367128}{5} \zeta _{5,3} \zeta _3-337360 \zeta _5 \zeta
   _6+\frac{1337921 \zeta _4 \zeta _7}{2}+\frac{66306320 \zeta _2 \zeta _9}{9}-\frac{577508911 \zeta
   _{11}}{48}+164008 \zeta _{5,3,3}\right) \epsilon ^5+\left(-\frac{215302 \zeta
   _3^4}{9}+\frac{3247786}{3} \zeta _6 \zeta _3^2+\frac{845872}{3} \zeta _4 \zeta _5 \zeta _3-6806582
   \zeta _2 \zeta _7 \zeta _3+\frac{148596974 \zeta _9 \zeta _3}{27}-\frac{50029600}{21} \zeta _2
   \zeta _5^2-\frac{17557127 \zeta _5 \zeta _7}{3}+\frac{1771479707981 \zeta
   _{12}}{110560}-\frac{4526792}{5} \zeta _4 \zeta _{5,3}-\frac{14364440}{7} \zeta _2 \zeta
   _{7,3}+\frac{23626165 \zeta _{9,3}}{9}+\frac{2077280}{3} \zeta _{6,4,1,1}\right) \epsilon
   ^6+O\left(\epsilon ^{7}\right)
\end{dmath}
\begin{dmath}
   \bar{M}^{(4)}_{63}=\frac{161 \zeta _7 \epsilon }{2}+\left(1730 \zeta _3 \zeta _5-\frac{24641 \zeta _8}{20}+\frac{3132
   \zeta _{5,3}}{5}\right) \epsilon ^2+\left(\frac{1526 \zeta _3^3}{3}+4325 \zeta _6 \zeta _3-2103
   \zeta _4 \zeta _5-161 \zeta _2 \zeta _7+\frac{111709 \zeta _9}{36}\right) \epsilon ^3+\left(2289
   \zeta _4 \zeta _3^2-3460 \zeta _2 \zeta _5 \zeta _3+\frac{350821 \zeta _7 \zeta
   _3}{6}+\frac{1341143 \zeta _5^2}{56}-\frac{158963083 \zeta _{10}}{2800}-\frac{6264}{5} \zeta _2
   \zeta _{5,3}+\frac{500565 \zeta _{7,3}}{56}\right) \epsilon ^4+\left(-\frac{3052}{3} \zeta _2 \zeta
   _3^3-\frac{437114}{3} \zeta _5 \zeta _3^2+\frac{12651161 \zeta _8 \zeta _3}{30}-\frac{501948}{5}
   \zeta _{5,3} \zeta _3-\frac{601363 \zeta _5 \zeta _6}{2}+\frac{10285231 \zeta _4 \zeta
   _7}{40}+\frac{96184979 \zeta _2 \zeta _9}{18}-\frac{8522762527 \zeta _{11}}{960}+\frac{594424}{5}
   \zeta _{5,3,3}\right) \epsilon ^5+\left(-12280 \zeta _3^4-\frac{7997035}{6} \zeta _6 \zeta
   _3^2+\frac{340730}{3} \zeta _4 \zeta _5 \zeta _3+\frac{12405859}{3} \zeta _2 \zeta _7 \zeta
   _3-\frac{83251955 \zeta _9 \zeta _3}{27}+\frac{120992035}{84} \zeta _2 \zeta _5^2+\frac{52550644
   \zeta _5 \zeta _7}{15}-\frac{3624073372357 \zeta _{12}}{442240}+\frac{3346676}{5} \zeta _4 \zeta
   _{5,3}+\frac{35218139}{28} \zeta _2 \zeta _{7,3}-\frac{58741243 \zeta _{9,3}}{36}-\frac{1275668}{3}
   \zeta _{6,4,1,1}\right) \epsilon ^6++O\left(\epsilon ^{7}\right)
\end{dmath}
\normalsize

\section{Expansions of master integrals in three dimensions}\label{app:exp3d}
\small
\begin{dmath*}
N^{(3)}_1= \pi^{5/2}\,2^{2\epsilon}\,\left(-\frac{2}{\epsilon ^2}-23 \zeta _2+150 \zeta _3 \epsilon -\frac{6721 \zeta _4 \epsilon
   ^2}{8}+\left(1725 \zeta _2 \zeta _3+\frac{16866 \zeta _5}{5}\right) \epsilon ^3+\left(-5625 \zeta
   _3^2-\frac{1695955 \zeta _6}{64}\right) \epsilon ^4+\left(\frac{504075 \zeta _3 \zeta
   _4}{8}+\frac{193959 \zeta _2 \zeta _5}{5}+\frac{587250 \zeta _7}{7}\right) \epsilon
   ^5+\left(-\frac{129375}{2} \zeta _2 \zeta _3^2-252990 \zeta _5 \zeta _3-\frac{1314208253 \zeta
   _8}{1536}\right) \epsilon ^6+\left(140625 \zeta _3^3+\frac{127196625 \zeta _6 \zeta
   _3}{64}+\frac{56678193 \zeta _4 \zeta _5}{40}+\frac{6753375 \zeta _2 \zeta _7}{7}+2292914 \zeta
   _9\right) \epsilon ^7+\left(-\frac{37805625}{16} \zeta _4 \zeta _3^2-2909385 \zeta _2 \zeta _5
   \zeta _3-\frac{44043750 \zeta _7 \zeta _3}{7}-\frac{71115489 \zeta _5^2}{25}-\frac{115029223713
   \zeta _{10}}{4096}\right) \epsilon ^8+O\left(\epsilon ^9\right)\right)
\end{dmath*}
\begin{dmath*}
N^{(3)}_2 = \pi^{5/2}\,2^{2\epsilon}\,\left(  
   \frac{1}{\epsilon ^2}-\frac{9
   \zeta _2}{2}-59 \zeta _3 \epsilon -\frac{4095 \zeta _4 \epsilon ^2}{16}+\left(\frac{531 \zeta _2
   \zeta _3}{2}-\frac{7473 \zeta _5}{5}\right) \epsilon ^3+\left(\frac{3481 \zeta
   _3^2}{2}-\frac{696717 \zeta _6}{128}\right) \epsilon ^4+\left(\frac{241605 \zeta _3 \zeta
   _4}{16}+\frac{67257 \zeta _2 \zeta _5}{10}-\frac{277497 \zeta _7}{7}\right) \epsilon
   ^5+\left(-\frac{31329}{4} \zeta _2 \zeta _3^2+\frac{440907 \zeta _5 \zeta _3}{5}-\frac{116779905
   \zeta _8}{1024}\right) \epsilon ^6+\left(-\frac{205379 \zeta _3^3}{6}+\frac{41106303 \zeta _6
   \zeta _3}{128}+\frac{6120387 \zeta _4 \zeta _5}{16}+\frac{2497473 \zeta _2 \zeta
   _7}{14}-\frac{3352331 \zeta _9}{3}\right) \epsilon ^7+\left(-\frac{14254695}{32} \zeta _4 \zeta
   _3^2-\frac{3968163}{10} \zeta _2 \zeta _5 \zeta _3+\frac{16372323 \zeta _7 \zeta
   _3}{7}+\frac{55845729 \zeta _5^2}{50}-\frac{109515055419 \zeta _{10}}{40960}\right) \epsilon
   ^8+O\left(\epsilon ^9\right)\right)
\end{dmath*}
\begin{dmath*}
N^{(3)}_3 = \pi^{5/2}\,2^{6\epsilon}\,\left(  6 \zeta _2+\frac{705 \zeta _4 \epsilon ^2}{2}-102 \left(\zeta _2
   \zeta _3\right) \epsilon ^3+\frac{419349 \zeta _6 \epsilon ^4}{32}+\left(-\frac{11985}{2} \zeta _3
   \zeta _4-\frac{5778 \zeta _2 \zeta _5}{5}\right) \epsilon ^5+\left(867 \zeta _2 \zeta
   _3^2+\frac{30123735 \zeta _8}{64}\right) \epsilon ^6+\left(-\frac{135783}{2} \zeta _4 \zeta
   _5-\frac{7128933 \zeta _3 \zeta _6}{32}-\frac{96786 \zeta _2 \zeta _7}{7}\right) \epsilon
   ^7+\left(\frac{203745}{4} \zeta _4 \zeta _3^2+\frac{98226}{5} \zeta _2 \zeta _5 \zeta
   _3+\frac{173150843613 \zeta _{10}}{10240}\right) \epsilon ^8+O\left(\epsilon ^9\right)\right)
\end{dmath*}
\begin{dmath*}
N^{(3)}_4 = \pi^{5/2}\,2^{6\epsilon}\,\left(  6 \zeta
   _2+\frac{225 \zeta _4 \epsilon ^2}{2}-6 \left(\zeta _2 \zeta _3\right) \epsilon ^3+\frac{30933
   \zeta _6 \epsilon ^4}{32}+\left(-\frac{225}{2} \zeta _3 \zeta _4-\frac{18 \zeta _2 \zeta
   _5}{5}\right) \epsilon ^5+\left(3 \zeta _2 \zeta _3^2+\frac{419575 \zeta _8}{64}\right) \epsilon
   ^6+\left(-\frac{135}{2} \zeta _4 \zeta _5-\frac{30933 \zeta _3 \zeta _6}{32}-\frac{18 \zeta _2
   \zeta _7}{7}\right) \epsilon ^7+\left(\frac{225}{4} \zeta _4 \zeta _3^2+\frac{18}{5} \zeta _2
   \zeta _5 \zeta _3+\frac{404851293 \zeta _{10}}{10240}\right) \epsilon ^8+O\left(\epsilon
   ^9\right)\right)
\end{dmath*}
\begin{dmath*}
N^{(3)}_5 = \pi^{5/2}\,2^{-2\epsilon}\,\bigg(  6 \zeta _2+84 \zeta _3 \epsilon +\left(16 L_1^4-96 \zeta _2 L_1^2+336 \zeta _3 L_1+384
   L_4+\frac{669 \zeta _4}{2}\right) \epsilon ^2+\left(\frac{256 L_1^5}{5}-256 \zeta _2 L_1^3+672
   \zeta _3 L_1^2+1536 L_4 L_1+1536 L_5+600 \zeta _2 \zeta _3+2046 \zeta _5\right) \epsilon
   ^3+\left(\frac{256 L_1^6}{3}-200 \zeta _2 L_1^4+896 \zeta _3 L_1^3+3072 L_4 L_1^2-2760 \zeta _4
   L_1^2+6144 L_5 L_1+3864 \zeta _2 \zeta _3 L_1+8184 \zeta _5 L_1-4128 \zeta _3^2+6144 L_6+4416 L_4
   \zeta _2+\frac{493773 \zeta _6}{32}-4224 \zeta _{-5,-1}\right) \epsilon ^4+\left(\frac{2048
   L_1^7}{21}+\frac{896}{5} \zeta _2 L_1^5-304 \zeta _3 L_1^4+4096 L_4 L_1^3-7360 \zeta _4
   L_1^3+12288 L_5 L_1^2+14928 \zeta _2 \zeta _3 L_1^2-\frac{192384}{7} \zeta _3^2 L_1+24576 L_6
   L_1+17664 L_4 \zeta _2 L_1+\frac{98568 \zeta _6 L_1}{7}-\frac{56832}{7} \zeta _{-5,-1} L_1+24576
   L_7+17664 L_5 \zeta _2-28800 L_4 \zeta _3+\frac{1234581 \zeta _3 \zeta _4}{28}+\frac{237027 \zeta
   _2 \zeta _5}{5}+12000 \zeta _7-\frac{61440}{7} \zeta _{-5,1,1}+\frac{56832}{7} \zeta
   _{5,-1,-1}\right) \epsilon ^5+\left(\frac{256 L_1^8}{3}+640 \zeta _2 L_1^6-\frac{129792}{35} \zeta
   _3 L_1^5+4096 L_4 L_1^4+801 \zeta _4 L_1^4+16384 L_5 L_1^3+\frac{247488}{7} \zeta _2 \zeta _3
   L_1^3-43648 \zeta _5 L_1^3-\frac{506400}{7} \zeta _3^2 L_1^2+49152 L_6 L_1^2+35328 L_4 \zeta _2
   L_1^2-\frac{1924947}{14} \zeta _6 L_1^2+\frac{122880}{7} \zeta _{-5,-1} L_1^2+98304 L_7 L_1+70656
   L_5 \zeta _2 L_1-115200 L_4 \zeta _3 L_1+\frac{1262451}{7} \zeta _3 \zeta _4 L_1+\frac{897468}{7}
   \zeta _2 \zeta _5 L_1+48000 \zeta _7 L_1-\frac{245760}{7} \zeta _{-5,1,1} L_1+\frac{227328}{7}
   \zeta _{5,-1,-1} L_1-\frac{518067}{7} \zeta _2 \zeta _3^2+98304 L_8+70656 L_6 \zeta
   _2-\frac{314880 L_5 \zeta _3}{7}+284184 L_4 \zeta _4-\frac{2619654 \zeta _3 \zeta
   _5}{35}+\frac{1736740617 \zeta _8}{2240}+\frac{1628160}{7} \zeta _{-7,-1}-\frac{1068096}{7} \zeta
   _2 \zeta _{-5,-1}+\frac{634236 \zeta _{5,3}}{35}-\frac{227328}{7} \zeta
   _{-5,-1,-1,-1}-\frac{245760}{7} \zeta _{-5,-1,1,1}\right) \epsilon ^6
\end{dmath*}
\vspace{-5mm}
\begin{dmath*}   
   +\bigg(\frac{368768
   L_1^9}{12285}+\frac{1881088 \zeta _2 L_1^7}{1365}-\frac{10055488 \zeta _3
   L_1^6}{1365}+\frac{1156096}{455} L_4 L_1^5+\frac{9274448}{455} \zeta _4 L_1^5 
   +\frac{1825792}{91}
   L_5 L_1^4+\frac{2908552}{91} \zeta _2 \zeta _3 L_1^4-\frac{72827568}{455} \zeta _5
   L_1^4-\frac{7092928}{91} \zeta _3^2 L_1^3+65536 L_6 L_1^3+\frac{4956160}{91} L_4 \zeta _2
   L_1^3-\frac{68477828}{91} \zeta _6 L_1^3+\frac{14246912}{91} \zeta _{-5,-1} L_1^3+196608 L_7
   L_1^2+\frac{10850304}{91} L_5 \zeta _2 L_1^2-\frac{2492928}{13} L_4 \zeta _3
   L_1^2+\frac{25576662}{91} \zeta _3 \zeta _4 L_1^2-\frac{103700832}{455} \zeta _2 \zeta _5
   L_1^2+\frac{15340272}{91} \zeta _7 L_1^2-\frac{5385216}{91} \zeta _{-5,1,1}
   L_1^2+\frac{10933248}{91} \zeta _{5,-1,-1} L_1^2-\frac{63126240}{91} \zeta _2 \zeta _3^2
   L_1+393216 L_8 L_1+282624 L_6 \zeta _2 L_1-\frac{1047552}{13} L_5 \zeta _3 L_1+\frac{6701472}{7}
   L_4 \zeta _4 L_1+\frac{25641432}{455} \zeta _3 \zeta _5 L_1-\frac{30566168 \zeta _8
   L_1}{65}+\frac{24158208}{13} \zeta _{-7,-1} L_1-\frac{29009664}{91} \zeta _2 \zeta _{-5,-1}
   L_1+\frac{6139176}{35} \zeta _{5,3} L_1-\frac{23875584}{91} \zeta _{-5,-1,-1,-1}
   L_1-\frac{964608}{13} \zeta _{-5,-1,1,1} L_1
\end{dmath*}
\vspace{-5mm}
\begin{dmath*}
   +\frac{9562478 \zeta _3^3}{91}+\frac{8036352 L_4
   L_5}{91}+393216 L_9+282624 L_7 \zeta _2-\frac{34455552 L_6 \zeta _3}{91}-\frac{59963712}{91} L_4
   \zeta _2 \zeta _3+\frac{34187808 L_5 \zeta _4}{91}-\frac{12736896 L_4 \zeta
   _5}{65}+\frac{163930677 \zeta _4 \zeta _5}{3640}+\frac{683051001 \zeta _3 \zeta
   _6}{1456}+\frac{21203436 \zeta _2 \zeta _7}{91}+\frac{1156507903 \zeta
   _9}{182}+\frac{18374016}{91} \zeta _3 \zeta _{-5,-1}-33792 \zeta _{-7,1,1}-\frac{20562432}{91}
   \zeta _2 \zeta _{-5,1,1}-\frac{3077376}{91} \zeta _2 \zeta _{5,-1,-1}-\frac{22023168}{91} \zeta
   _{7,-1,-1}+\frac{116736}{91} \zeta _{-5,-1,-1,-1,1}+\frac{6027264}{91} \zeta
   _{-5,-1,-1,1,1}-\frac{2009088}{91} \zeta _{-5,-1,1,-1,-1}-\frac{528384}{7} \zeta
   _{-5,-1,1,1,1}\bigg) \epsilon ^7
\end{dmath*}
\vspace{-4mm}
\begin{dmath*}
   +\bigg(-\frac{267119204608
   L_1^{10}}{734868225}+\frac{1059510917824 \zeta _2 L_1^8}{146973645}-\frac{4354879826944 \zeta _3
   L_1^7}{342938505}-\frac{367120162816 L_4 L_1^6}{48991215}-\frac{3187135993072 \zeta _4
   L_1^6}{48991215}+\frac{2973421170688 L_5 L_1^5}{48991215}-\frac{17549910811648 \zeta _2 \zeta _3
   L_1^5}{114312835}-\frac{19449684288224 \zeta _5 L_1^5}{244956075}+\frac{711998403000 \zeta _3^2
   L_1^4}{3266081}+65536 L_6 L_1^4+\frac{1399964646400 L_4 \zeta _2
   L_1^4}{9798243}-\frac{8060197839419 \zeta _6 L_1^4}{3732664}+\frac{2586103604736 \zeta _{-5,-1}
   L_1^4}{3266081}+262144 L_7 L_1^3-\frac{16137699328 L_5 \zeta _2 L_1^3}{171899}-\frac{47425642496
   L_4 \zeta _3 L_1^3}{466583}+\frac{1620347298952 \zeta _3 \zeta _4
   L_1^3}{22862567}-\frac{106278735184512 \zeta _2 \zeta _5 L_1^3}{16330405}+\frac{21031848624000
   \zeta _7 L_1^3}{3266081}+\frac{19638038528 \zeta _{-5,1,1} L_1^3}{3266081}+\frac{212118097920
   \zeta _{5,-1,-1} L_1^3}{466583}-\frac{126365863260288 \zeta _2 \zeta _3^2 L_1^2}{22862567}+786432
   L_8 L_1^2+565248 L_6 \zeta _2 L_1^2+\frac{2412561063936 L_5 \zeta _3
   L_1^2}{3266081}-\frac{4820338624 L_4 \zeta _4 L_1^2}{35891}+\frac{241774624089504 \zeta _3 \zeta
   _5 L_1^2}{16330405}-\frac{530779417673327 \zeta _8 L_1^2}{24065860}+\frac{21908593603584 \zeta
   _{-7,-1} L_1^2}{3266081}+\frac{71393813455872 \zeta _2 \zeta _{-5,-1}
   L_1^2}{22862567}+\frac{1329019555776 \zeta _{5,3} L_1^2}{1256185}-\frac{669510500352 \zeta
   _{-5,-1,-1,-1} L_1^2}{466583}+\frac{646989496320 \zeta _{-5,-1,1,1}
   L_1^2}{3266081}+\frac{103101626784584 \zeta _3^3 L_1}{68587701}+\frac{3793144840192 L_4 L_5
   L_1}{3266081}+1572864 L_9 L_1+1130496 L_7 \zeta _2 L_1-\frac{7293084819456 L_6 \zeta _3
   L_1}{3266081}-\frac{22416580808448 L_4 \zeta _2 \zeta _3 L_1}{3266081}-\frac{2385271188096 L_5
   \zeta _4 L_1}{3266081}+\frac{19750103021056 L_4 \zeta _5 L_1}{2332915}
\end{dmath*}
\begin{dmath*}
   -\frac{1167482364324503
   \zeta _4 \zeta _5 L_1}{45725134}-\frac{1244921435967255 \zeta _3 \zeta _6
   L_1}{182900536}+\frac{5629010967512 \zeta _2 \zeta _7 L_1}{3266081}+\frac{513240571732220 \zeta _9
   L_1}{9798243}-\frac{37451775451136 \zeta _3 \zeta _{-5,-1} L_1}{22862567}+\frac{141637036032 \zeta
   _{-7,1,1} L_1}{35891}-\frac{10015471073280 \zeta _2 \zeta _{-5,1,1}
   L_1}{3266081}-\frac{6893160328192 \zeta _2 \zeta _{5,-1,-1} L_1}{3266081}-\frac{579441825792 \zeta
   _{7,-1,-1} L_1}{466583}+\frac{1097286746112 \zeta _{-5,-1,-1,-1,1}
   L_1}{3266081}+\frac{1919017230336 \zeta _{-5,-1,-1,1,1} L_1}{3266081}-\frac{101308825600 \zeta
   _{-5,-1,1,-1,-1} L_1}{3266081}-\frac{240707223552 \zeta _{-5,-1,1,1,1}
   L_1}{251237}+\frac{12132925440 L_5^2}{251237}+\frac{44960410176 L_4 \zeta
   _3^2}{251237}-\frac{8710146527772 \zeta _5^2}{8793295}+1572864 L_{10}-\frac{3033231360 L_4^2 \zeta
   _2}{251237}+1130496 L_8 \zeta _2-\frac{18358272 L_7 \zeta _3}{7}-\frac{347925518592 L_5 \zeta _2
   \zeta _3}{1758659}-\frac{18883236593385 \zeta _3^2 \zeta _4}{7034636}+\frac{708732514176 L_6 \zeta
   _4}{251237}+\frac{2987132699904 L_5 \zeta _5}{1256185}-\frac{6696293713551 \zeta _2 \zeta _3 \zeta
   _5}{1758659}+\frac{4048529317437 L_4 \zeta _6}{251237}-\frac{15491203452 \zeta _3 \zeta
   _7}{251237}+\frac{634335474560110119 \zeta _{10}}{18008668160}+\frac{6123951126528 \zeta
   _{-9,-1}}{251237}-\frac{5175869932032 \zeta _2 \zeta _{-7,-1}}{1758659}-\frac{23190147072 L_4
   \zeta _{-5,-1}}{35891}-\frac{19484093971224 \zeta _4 \zeta _{-5,-1}}{1758659}+\frac{478123252842
   \zeta _2 \zeta _{5,3}}{1758659}+\frac{698630689767 \zeta _{7,3}}{1758659}-\frac{1467554595840
   \zeta _3 \zeta _{-5,1,1}}{1758659}+\frac{3954511872 \zeta _3 \zeta
   _{5,-1,-1}}{1758659}-\frac{296420997120 \zeta _{-7,-1,-1,-1}}{251237}-\frac{465218684928 \zeta
   _{-7,-1,1,1}}{251237}-\frac{554638015488 \zeta _2 \zeta
   _{-5,-1,-1,-1}}{1758659}-\frac{1242786920448 \zeta _2 \zeta
   _{-5,-1,1,1}}{1758659}-\frac{65152106496 \zeta _{-5,1,-1,-1,-1,-1}}{251237}-\frac{6066462720 \zeta
   _{-5,1,1,-1,1,-1}}{251237}+\frac{51077111808 \zeta _{-5,1,1,1,-1,-1}}{251237}-\frac{54567419904
   \zeta _{-5,1,1,1,1,1}}{251237}\bigg) \epsilon ^8+O\left(\epsilon ^9\right)\bigg)
\end{dmath*}
\begin{dmath*}
N^{(3)}_6 = \pi^{5/2}\,2^{8\epsilon}\,\bigg(  -8 \zeta _2-68 \zeta
   _3 \epsilon +\left(32 L_1^4+48 \zeta _2 L_1^2+408 \zeta _3 L_1+768 L_4-1216 \zeta _4\right)
   \epsilon ^2+\left(-\frac{1088 L_1^5}{5}+128 \zeta _2 L_1^3-1224 \zeta _3 L_1^2-4608 L_4 L_1+4620
   \zeta _4 L_1+3072 L_5+498 \zeta _2 \zeta _3-6548 \zeta _5\right) \epsilon ^3+\left(\frac{2240
   L_1^6}{3}-832 \zeta _2 L_1^4+2448 \zeta _3 L_1^3+13824 L_4 L_1^2-7920 \zeta _4 L_1^2-18432 L_5
   L_1+2628 \zeta _2 \zeta _3 L_1+39288 \zeta _5 L_1+7732 \zeta _3^2+12288 L_6+11904 L_4 \zeta
   _2-\frac{1689739 \zeta _6}{24}-8448 \zeta _{-5,-1}\right) \epsilon ^4+\left(-\frac{36224
   L_1^7}{21}+\frac{12576}{5} \zeta _2 L_1^5-7864 \zeta _3 L_1^4-27648 L_4 L_1^3-7000 \zeta _4
   L_1^3+55296 L_5 L_1^2+10692 \zeta _2 \zeta _3 L_1^2-150600 \zeta _5 L_1^2-\frac{693384}{7} \zeta
   _3^2 L_1-73728 L_6 L_1-34560 L_4 \zeta _2 L_1+\frac{3983417 \zeta _6 L_1}{14}+\frac{649728}{7}
   \zeta _{-5,-1} L_1+49152 L_7+72192 L_5 \zeta _2-100608 L_4 \zeta _3+\frac{4964081 \zeta _3 \zeta
   _4}{28}-\frac{358286 \zeta _2 \zeta _5}{5}-231632 \zeta _7-\frac{294912}{7} \zeta
   _{-5,1,1}-\frac{58368}{7} \zeta _{5,-1,-1}\right) \epsilon ^5+\left(\frac{59840
   L_1^8}{21}-\frac{59072}{15} \zeta _2 L_1^6+\frac{1010608}{35} \zeta _3 L_1^5+33280 L_4 L_1^4+92226
   \zeta _4 L_1^4-110592 L_5 L_1^3-\frac{752984}{7} \zeta _2 \zeta _3 L_1^3+432144 \zeta _5
   L_1^3+\frac{3250008}{7} \zeta _3^2 L_1^2+221184 L_6 L_1^2+66816 L_4 \zeta _2
   L_1^2-\frac{18581403}{28} \zeta _6 L_1^2-\frac{2606592}{7} \zeta _{-5,-1} L_1^2-294912 L_7
   L_1-384000 L_5 \zeta _2 L_1+603648 L_4 \zeta _3 L_1-\frac{12456987}{14} \zeta _3 \zeta _4
   L_1+\frac{1358220}{7} \zeta _2 \zeta _5 L_1+1389792 \zeta _7 L_1+\frac{1769472}{7} \zeta _{-5,1,1}
   L_1+\frac{350208}{7} \zeta _{5,-1,-1} L_1-98304 L_4^2-\frac{180258}{7} \zeta _2 \zeta _3^2+196608
   L_8+190464 L_6 \zeta _2+\frac{574464 L_5 \zeta _3}{7}+758064 L_4 \zeta _4+\frac{23949208 \zeta _3
   \zeta _5}{35}-\frac{2232450923 \zeta _8}{840}+\frac{3256320}{7} \zeta _{-7,-1}-\frac{1727616}{7}
   \zeta _2 \zeta _{-5,-1}-\frac{107784 \zeta _{5,3}}{35}-\frac{454656}{7} \zeta
   _{-5,-1,-1,-1}-\frac{491520}{7} \zeta _{-5,-1,1,1}\right) \epsilon ^6
\end{dmath*}
\vspace{-4mm}
\begin{dmath*}   
   +\bigg(-\frac{38033024
   L_1^9}{12285}-\frac{6236864 \zeta _2 L_1^7}{4095}-\frac{92357968 \zeta _3
   L_1^6}{1365}+\frac{4320256 L_4 L_1^5}{1365}-\frac{113395476}{455} \zeta _4
   L_1^5+\frac{40128512}{273} L_5 L_1^4+\frac{13954132}{91} \zeta _2 \zeta _3 L_1^4-\frac{1112466808
   \zeta _5 L_1^4}{1365}-\frac{116459120}{91} \zeta _3^2 L_1^3-442368 L_6 L_1^3-\frac{40090112}{273}
   L_4 \zeta _2 L_1^3+\frac{140704457}{91} \zeta _6 L_1^3+\frac{58762240}{91} \zeta _{-5,-1}
   L_1^3+884736 L_7 L_1^2+\frac{128568320}{91} L_5 \zeta _2 L_1^2-\frac{21648896}{13} L_4 \zeta _3
   L_1^2+\frac{1054300677}{182} \zeta _3 \zeta _4 L_1^2-\frac{435730244}{455} \zeta _2 \zeta _5
   L_1^2-\frac{539341440}{91} \zeta _7 L_1^2-\frac{65222656}{91} \zeta _{-5,1,1}
   L_1^2-\frac{1802240}{13} \zeta _{5,-1,-1} L_1^2+589824 L_4^2 L_1-\frac{134714796}{91} \zeta _2
   \zeta _3^2 L_1-1179648 L_8 L_1-552960 L_6 \zeta _2 L_1-\frac{64401408}{91} L_5 \zeta _3
   L_1
\end{dmath*}
\begin{dmath*}     
   -\frac{12036064}{7} L_4 \zeta _4 L_1-\frac{2000343984}{455} \zeta _3 \zeta _5
   L_1+\frac{60661266409 \zeta _8 L_1}{3640}-\frac{453255168}{91} \zeta _{-7,-1}
   L_1+\frac{35241216}{13} \zeta _2 \zeta _{-5,-1} L_1-\frac{7159968}{35} \zeta _{5,3}
   L_1+\frac{8798208}{13} \zeta _{-5,-1,-1,-1} L_1+\frac{25276416}{91} \zeta _{-5,-1,1,1}
   L_1-\frac{117389666 \zeta _3^3}{273}-\frac{41271296 L_4 L_5}{91}+786432 L_9+1155072 L_7 \zeta
   _2+\frac{15384576 L_6 \zeta _3}{91}-\frac{292518528}{91} L_4 \zeta _2 \zeta _3+\frac{583116864 L_5
   \zeta _4}{91}+\frac{153864448 L_4 \zeta _5}{65}+\frac{5110832357 \zeta _4 \zeta
   _5}{1820}+\frac{116047009093 \zeta _3 \zeta _6}{8736}-\frac{5604052 \zeta _2 \zeta
   _7}{13}-\frac{23121935561 \zeta _9}{819}-\frac{44766464}{91} \zeta _3 \zeta _{-5,-1}+251904 \zeta
   _{-7,1,1}-\frac{138273792}{91} \zeta _2 \zeta _{-5,1,1}+\frac{17606144}{91} \zeta _2 \zeta
   _{5,-1,-1}+\frac{49022976}{91} \zeta _{7,-1,-1}-\frac{6991872}{91} \zeta
   _{-5,-1,-1,-1,1}-\frac{13062144}{91} \zeta _{-5,-1,-1,1,1}-\frac{7573504}{91} \zeta
   _{-5,-1,1,-1,-1}-\frac{2433024}{7} \zeta _{-5,-1,1,1,1}\bigg) \epsilon
   ^7
\end{dmath*}
\vspace{-4mm}
\begin{dmath*}   
   +\bigg(\frac{11230669702528 L_1^{10}}{11023023375}+\frac{17304076971248 \zeta _2
   L_1^8}{734868225}+\frac{54050307038816 \zeta _3 L_1^7}{571564175}-\frac{30310186720256 L_4
   L_1^6}{244956075}+\frac{4465423135196 \zeta _4 L_1^6}{11664575}-\frac{26287796740096 L_5
   L_1^5}{244956075}+\frac{127227862249704 \zeta _2 \zeta _3 L_1^5}{571564175}+\frac{330783445199152
   \zeta _5 L_1^5}{244956075}+\frac{41167786245128 \zeta _3^2 L_1^4}{16330405}+532480 L_6
   L_1^4+\frac{13321238419712 L_4 \zeta _2 L_1^4}{48991215}-\frac{34719652951336 \zeta _6
   L_1^4}{16330405}-\frac{71144187392 \zeta _{-5,-1} L_1^4}{466583}-1769472 L_7
   L_1^3-\frac{463017445376 L_5 \zeta _2 L_1^3}{122785}+\frac{6380264805376 L_4 \zeta _3
   L_1^3}{2332915}-\frac{2359283757414677 \zeta _3 \zeta _4 L_1^3}{114312835}+\frac{3590171688712
   \zeta _2 \zeta _5 L_1^3}{2332915}+\frac{49254868205312 \zeta _7
   L_1^3}{2332915}+\frac{21249064349696 \zeta _{-5,1,1} L_1^3}{16330405}+\frac{2423795785728 \zeta
   _{5,-1,-1} L_1^3}{16330405}-\frac{8650752}{5} L_4^2 L_1^2+\frac{691630461902412 \zeta _2 \zeta
   _3^2 L_1^2}{114312835}+3538944 L_8 L_1^2+1069056 L_6 \zeta _2 L_1^2+\frac{5949623580672 L_5 \zeta
   _3 L_1^2}{2332915}+\frac{314208494176 L_4 \zeta _4 L_1^2}{1256185}+\frac{370570556547696 \zeta _3
   \zeta _5 L_1^2}{16330405}-\frac{1391398158187773 \zeta _8 L_1^2}{24065860}+\frac{279790816733184
   \zeta _{-7,-1} L_1^2}{16330405}-\frac{1134919234420992 \zeta _2 \zeta _{-5,-1}
   L_1^2}{114312835}+\frac{267910432560 \zeta _{5,3} L_1^2}{251237}-\frac{42242552291328 \zeta
   _{-5,-1,-1,-1} L_1^2}{16330405}-\frac{1467051466752 \zeta _{-5,-1,1,1}
   L_1^2}{2332915}
\end{dmath*}
\vspace{-4mm}
\begin{dmath*}      
   +\frac{1228763001975148 \zeta _3^3 L_1}{342938505}+\frac{48903747043328 L_4 L_5
   L_1}{16330405}-4718592 L_9 L_1-6144000 L_7 \zeta _2 L_1-\frac{845356843008 L_6 \zeta _3
   L_1}{466583}+\frac{220477732229376 L_4 \zeta _2 \zeta _3 L_1}{16330405}-\frac{527625447364224 L_5
   \zeta _4 L_1}{16330405}-\frac{8671088124416 L_4 \zeta _5 L_1}{2332915}-\frac{9685282337878051
   \zeta _4 \zeta _5 L_1}{228625670}-\frac{128870369716187473 \zeta _3 \zeta _6
   L_1}{1829005360}+\frac{4157102564152 \zeta _2 \zeta _7 L_1}{2332915}+\frac{1018916433083702 \zeta
   _9 L_1}{6998745}-\frac{21333917746688 \zeta _3 \zeta _{-5,-1} L_1}{22862567}+\frac{507058520064
   \zeta _{-7,1,1} L_1}{179455}+\frac{112316599412736 \zeta _2 \zeta _{-5,1,1}
   L_1}{16330405}-\frac{39954053037056 \zeta _2 \zeta _{5,-1,-1} L_1}{16330405}-\frac{13844259618816
   \zeta _{7,-1,-1} L_1}{16330405}+\frac{10287179440128 \zeta _{-5,-1,-1,-1,1}
   L_1}{16330405}+\frac{12217349578752 \zeta _{-5,-1,-1,1,1} L_1}{16330405}+\frac{254190166016 \zeta
   _{-5,-1,1,-1,-1} L_1}{466583}+\frac{1710195818496 \zeta _{-5,-1,1,1,1}
   L_1}{1256185}+\frac{2489326338048 L_5^2}{1256185}+\frac{313669015680 L_4 \zeta
   _3^2}{251237}+\frac{19228507858032 \zeta _5^2}{1758659}-3145728 L_4 L_6+3145728
   L_{10}-\frac{1647282069504 L_4^2 \zeta _2}{1256185}+3047424 L_8 \zeta _2-\frac{106512384 L_7 \zeta
   _3}{35}-\frac{5918951594496 L_5 \zeta _2 \zeta _3}{8793295}+\frac{18434723332163 \zeta _3^2 \zeta
   _4}{35173180}+\frac{10944636365568 L_6 \zeta _4}{1256185}+\frac{5608912955904 L_5 \zeta
   _5}{1256185}+\frac{51367321749244 \zeta _2 \zeta _3 \zeta _5}{8793295}+\frac{42287248089522 L_4
   \zeta _6}{1256185}+\frac{10429832868908 \zeta _3 \zeta _7}{1256185}-\frac{9442052851614329921
   \zeta _{10}}{112554176000}+\frac{13177232191488 \zeta _{-9,-1}}{251237}-\frac{115711045764096
   \zeta _2 \zeta _{-7,-1}}{8793295}+\frac{340590944256 L_4 \zeta
   _{-5,-1}}{179455}-\frac{106046751995376 \zeta _4 \zeta _{-5,-1}}{8793295}-\frac{11868938016428
   \zeta _2 \zeta _{5,3}}{43966475}+\frac{1799886540902 \zeta _{7,3}}{8793295}-\frac{5059305867264
   \zeta _3 \zeta _{-5,1,1}}{8793295}+\frac{25333532909568 \zeta _3 \zeta
   _{5,-1,-1}}{8793295}-\frac{3416942456832 \zeta _{-7,-1,-1,-1}}{1256185}-\frac{4282629083136 \zeta
   _{-7,-1,1,1}}{1256185}+\frac{81731708928 \zeta _2 \zeta
   _{-5,-1,-1,-1}}{8793295}
\end{dmath*}
\begin{dmath*}
   -\frac{4028426072064 \zeta _2 \zeta
   _{-5,-1,1,1}}{8793295}-\frac{142958690304 \zeta _{-5,1,-1,-1,-1,-1}}{1256185}-\frac{196608}{5}
   \zeta _{-5,1,-1,1,-1,-1}+\frac{393216}{5} \zeta _{-5,1,1,-1,-1,1}-\frac{157968678912 \zeta
   _{-5,1,1,-1,1,-1}}{1256185}+\frac{180067811328 \zeta _{-5,1,1,1,-1,-1}}{1256185}-\frac{37111824384
   \zeta _{-5,1,1,1,1,1}}{1256185}\bigg) \epsilon ^8+O\left(\epsilon ^9\right)\bigg)
\end{dmath*}
\normalsize

\bibliographystyle{JHEP}

\bibliography{biblio3}

\end{document}